\documentclass[12pt,a4paper]{article}
\usepackage[T1]{fontenc}
\usepackage[utf8]{inputenc}
\usepackage{amsfonts}
\usepackage{verbatim}
\usepackage{graphicx}
\usepackage{geometry}
\usepackage{slashed}
\usepackage{hyperref}
\usepackage{subfig}
\usepackage{amssymb}

\usepackage{cite}
\usepackage{cancel}
\usepackage{caption}
\usepackage{epstopdf}
\usepackage{amsthm}
\usepackage{amsmath}
\usepackage{color}

\usepackage{float}
\usepackage{numprint}

\newcommand{\newc}{\newcommand*}

\long\def\begincomment#1\endcomment{%
        \begingroup\sf\baselineskip12pt#1\endgroup}

\newc{\etal}{\textrm{et al.}} 
\newc{\eg}{\textrm{e.g.}} 
\newc{\ie}{\textrm{i.e.}}
\newc{\etc}{\textrm{etc}}
\newc\vs{\textrm{vs.}}
\newc{\cl}{\rm {C.L.}}

\newc{\ev}{\ensuremath{\,\mathrm{eV}}}
\newc{\kev}{\ensuremath{\,\mathrm{keV}}}
\newc{\mev}{\ensuremath{\,\mathrm{MeV}}}
\newc{\gev}{\ensuremath{\,\mathrm{GeV}}}
\newc{\tev}{\ensuremath{\,\mathrm{TeV}}}
\newc{\MeV}{\mev} 
\newc{\TeV}{\tev}
\newc{\invpb}{\ensuremath{/\text{pb}}}
\newc{\invfb}{\ensuremath{/\text{fb}}}
\newc\nb{\ensuremath{\,\mathrm{nb}}} \newc\pb{\ensuremath{\,\mathrm{pb}}} \newc\fb{\ensuremath{\,\mathrm{fb}}}
\newc\pc{\ensuremath{\,\mathrm{pc}}}
\newc\kpc{\ensuremath{\,\mathrm{kpc}}}
\newc\mpc{\ensuremath{\,\mathrm{Mpc}}}
\newc\ps{\ensuremath{\,\mathrm{ps}}} 
\newc\cmeter{\ensuremath{\,\mathrm{cm}}} 
\newc\meter{\ensuremath{\,\mathrm{m}}} 
\newc\kmeter{\ensuremath{\,\mathrm{km}}}
\newc\second{\ensuremath{\,\mathrm{s}}}
\newc\msecond{\ensuremath{\,\mathrm{ms}}}
\newc\nsecond{\ensuremath{\,\mathrm{ns}}}
\newc\psecond{\ensuremath{\,\mathrm{ps}}}

\newc{\chisqmin}{\ensuremath{\chi^2_{\mathrm{min}}}}
\newc{\Delchisq}{\ensuremath{\Delta\chi^2}}
\newc{\chisq}{\ensuremath{\chi^2}}
\newc{\like}{\ensuremath{\mathcal{L}}}

\newc\lsim{\ensuremath{\mathrel{\rlap{\lower4pt\hbox{\hskip1pt$\sim$}}\raise1pt\hbox{$<$}}}}
\newc\gsim{\ensuremath{\mathrel{\rlap{\lower4pt\hbox{\hskip1pt$\sim$}}\raise1pt\hbox{$>$}}}}
\newc{\VEV}[1]{\ensuremath{\langle #1 \rangle}}
\newc{\dl}{\ensuremath{\stackrel{\leftarrow}{D}}}
\newc{\dr}{\ensuremath{\stackrel{\rightarrow}{D}}}

\newc{\bcenter}{\begin{center}}    \newc{\ecenter}{\end{center}}
\newc{\bfl}{\begin{flushleft}}    \newc{\efl}{\end{flushleft}}
\newc{\bfr}{\begin{flushright}}    \newc{\efr}{\end{flushright}}

\newc{\bi}{\begin{itemize}}
\newc{\ei}{\end{itemize}}
\newc{\bed}{\begin{description}}
\newc{\eed}{\end{description}}
\newc{\ben}{\begin{enumerate}}
\newc{\een}{\end{enumerate}}

\newc{\be}{\begin{equation}}
\newc{\ee}{\end{equation}}
\newc{\bea}{\begin{eqnarray}}
\newc{\eea}{\end{eqnarray}}
\newc{\bfle}{\begin{flalign}}
\newc{\efle}{\end{flalign}}
\newc{\ra}{\rightarrow}

\newc{\alphas}{\ensuremath{\alpha_s}}
\newc{\alphatwo}{\ensuremath{\alpha_2}}
\newc{\alphaone}{\ensuremath{\alpha_1}}
\newc{\alphai}[1]{\ensuremath{\alpha_{#1}}}
\newc{\alphaem}{\ensuremath{\alpha_{\mathrm{em}}}}
\newc{\alphaeff}{\ensuremath{\alpha_{\mathrm{eff}}}}
\newc{\sineff}{\ensuremath{\sin^2 \theta_{\mathrm{eff}}}}
\newc{\sinsqeff}{\ensuremath{\sin^2 \theta_{\mathrm{eff}}}}
\newc{\dalphahad}{\ensuremath{\Delta \alpha_{\mathrm{had}}}}
\newc{\yt}{\ensuremath{h_t}} \newc{\yb}{\ensuremath{h_b}} \newc{\ytau}{\ensuremath{h_{\tau}}}
\newc\mz{\ensuremath{m_Z}} 
\newc\mw{\ensuremath{m_W}}
\newc\mZ{\mz}        \newc\mW{\mw}
\newc\mhsm{\ensuremath{ m_{H_{\mathrm{SM}}}}}
\newc{\mtop}{\ensuremath{ m_t}}               
\newc{\mbottom}{\ensuremath{ m_b}} 
\newc{\mtau}{\ensuremath{ m_{\tau}}}
\newc{\mt}{\mtpole}
\newc{\mb}{\mbottom} 
\newc{\rtwogg}{\ensuremath{R_{h_2}(\gamma\gamma)}}
\newc{\rtwozz}{\ensuremath{R_{h_2}(ZZ)}}
\newc{\ronegg}{\ensuremath{R_{h_1}(\gamma\gamma)}}
\newc{\ronezz}{\ensuremath{R_{h_1}(ZZ)}}
\newc{\rsiggg}{\ensuremath{R_{h_\textrm{sig}}(\gamma\gamma)}}
\newc{\rsigzz}{\ensuremath{R_{h_\textrm{sig}}(ZZ)}}
\newc{\llbar}{\ensuremath{\ell\bar{\ell}}}
\newc{\tauptaum}{\ensuremath{ \tau^+\tau^-}}
\newc{\qqbar}{\ensuremath{ q\bar{q}}} \newc{\ppbar}{\ensuremath{ p\bar{p}}}
\newc{\bbbar}{\ensuremath{ b\bar{b}}} \newc{\ttbar}{\ensuremath{ t\bar{t}}}
\newc{\ffbar}{\ensuremath{ f\bar{f}}} \newc{\tautaubar}{\ensuremath{ \tau\bar{\tau}}}

\newc{\mchi}{\ensuremath{m_\neutone}}
\newc{\squark}{\ensuremath{\tilde{q}}}
\newc{\slepton}{\ensuremath{\tilde{l}}}
\newc{\gluino}{\ensuremath{\tilde{g}}} 
\newc{\mgluino}{\ensuremath{{m_{\gluino}}}}
\newc{\wino}{\ensuremath{\tilde{W}}} 
\newc{\mwino}{\ensuremath{{m_{\wino}}}}
\newc{\tone}{\ensuremath{{\tilde{t}_1}}}
\newc{\Hone}{\ensuremath{{\tilde{H}_{1}}}}
\newc{\Htwo}{\ensuremath{{\tilde{H}_{2}}}}
\newc{\Hhtwo}{\ensuremath{{H_{2}}}}
\newc{\qli}{\ensuremath{{\tilde{Q}_{i}}}}
\newc{\uri}{\ensuremath{{\tilde{u}_{i}}}}
\newc{\dri}{\ensuremath{{\tilde{d}_{i}}}}
\newc{\lli}{\ensuremath{{\tilde{L}_{i}}}}
\newc{\eri}{\ensuremath{{\tilde{e}_{i}}}}

\newc{\sthw}{\ensuremath{ \sin\theta_W}}              \newc{\cthw}{\ensuremath{\cos\theta_W}}
\newc{\tanthw}{\ensuremath{ \tan\theta_W}}              \newc{\cotthw}{\ensuremath{\cot\theta_W}}
\newc{\ssqthw}{\ensuremath{\sin^2 \theta_W}}
\newc{\msbar}{\ensuremath{\overline{MS}}} \newc{\drbar}{\ensuremath{\overline{DR}}}
\newc{\mtmtsmmsbar}{\ensuremath{ m_t(m_t)^{\msbar}_{{\mathrm{SM}}}}}
\newc{\mtmtsmdrbar}{\ensuremath{ m_t(m_t)^{\drbar}_{{\mathrm{SM}}}}}
\newc{\mtmtmssmdrbar}{\ensuremath{ m_t(m_t)^{\drbar}_{{\mathrm{SUSY}}}}}
\newc{\mbmbmsbar}{\ensuremath{ m_b^{\msbar}(m_b)}}
\newc{\mcmbmsbar}{\ensuremath{ m_c^{\msbar}(m_c)}}
\newc{\msmbmsbar}{\ensuremath{ m_s^{\msbar}}}
\newc{\mdmbmsbar}{\ensuremath{ m_d^{\msbar}}}
\newc{\mumbmsbar}{\ensuremath{ m_u^{\msbar}}}
\newc{\mtaupole}{\ensuremath{m_\tau^{\rm pole}}}
\newc{\mmupole}{\ensuremath{m_\mu^{\rm pole}}}
\newc{\mepole}{\ensuremath{m_e^{\rm pole}}}
\newc{\mzpole}{\ensuremath{M_Z^{\rm pole}}}
\newc{\mbmbsmmsbar}{\ensuremath{ m_b(m_b)^{\msbar}_{{\mathrm{SM}}}}}
\newc{\mbmzsmmsbar}{\ensuremath{ m_b(\mz)^{\msbar}_{{\mathrm{SM}}}}}
\newc{\mbmzsmdrbar}{\ensuremath{ m_b(\mz)^{\drbar}_{{\mathrm{SM}}}}}
\newc{\mbmzmssmdrbar}{\ensuremath{ m_b(\mz)^{\drbar}_{{\mathrm{SUSY}}}}}
\newc{\mtaumzsmmsbar}{\ensuremath{ m_{\tau}(\mz)^{\msbar}_{{\mathrm{SM}}}}}
\newc{\mtaumzsmdrbar}{\ensuremath{ m_{\tau}(\mz)^{\drbar}_{{\mathrm{SM}}}}}
\newc{\mtaumzmssmdrbar}{\ensuremath{ m_{\tau}(\mz)^{\drbar}_{{\mathrm{SUSY}}}}}
\newc{\alphasmzms}{\ensuremath{\alpha_s^{\overline{MS}}(M_Z)}}
\newc{\alphaimzms}[1]{\ensuremath{\alpha_{#1}(M_Z)^{\overline{MS}}}}
\newc{\alphaemmz}{\ensuremath{\alpha_{\mathrm{em}}^{-1}(M_Z)}}

\newc{\mzero}{\ensuremath{{m_0}}}
\newc{\mhalf}{\ensuremath{ M_{1/2}}}
\newc{\tanb}{\ensuremath{\tan\beta}}
\newc{\azero}{\ensuremath{ A_0}}
\newc{\signmu}{\ensuremath{\rm{sgn}\,\mu}}
\newc{\atau}{\ensuremath{{A_{\tau}}}}
\newc{\mueff}{\ensuremath{\mu_{\rm{eff}}}}
\newc{\lam}{\ensuremath{{\lambda}}}
\newc{\kap}{\ensuremath{{\kappa}}}
\newc{\alam}{\ensuremath{{A_{\lambda}}}}
\newc{\akap}{\ensuremath{{A_{\kappa}}}}
\newc{\hs}{\ensuremath{ H_s}}      
\newc{\mhs}{\ensuremath{ m_{H_s}}} 
\newc{\mgut}{\ensuremath{M_{\textrm{GUT}}}}
\newc{\gut}{\ensuremath{{\rm GUT}}}
\newc{\mplanck}{\ensuremath{ M_{\rm P}}}      \newc{\mpl}{\ensuremath{ M_{\rm Pl}}}
\newc{\msusy}{\ensuremath{ M_{\rm SUSY}}}      \newc{\ms}{\ensuremath{ M_{\rm S}}}
\newc{\mew}{\ensuremath{ M_{\rm EW}}}  
 \newc{\hu}{\ensuremath{ H_u}}       \newc{\hd}{\ensuremath{ H_d}}
 \newc{\mhu}{\ensuremath{ m_{H_u}}}       \newc{\mhd}{\ensuremath{ m_{H_d}}}
 \newc{\mhuew}{\ensuremath{ m^{\ast}_{H_u}}}       \newc{\mhdew}{\ensuremath{ m^{\ast}_{H_d}}}
 \newc{\mhuewsq}{\ensuremath{ m^{\ast\, 2}_{H_u}}}       \newc{\mhdewsq}{\ensuremath{ m^{\ast\, 2}_{H_d}}}
 \newc{\mhl}{\ensuremath{m_\hl}} 
 \newc{\mhone}{\ensuremath{m_{h_1}}} 
 \newc{\mhtwo}{\ensuremath{m_{h_2}}} 
 \newc{\mhi}{\ensuremath{m_{\tilde{h}}}} 
 \newc{\mul}{\ensuremath{m_{\tilde{u}_L}}} 
 \newc{\mtone}{\ensuremath{m_{\tilde{t}_1}}} 
 \newc{\ma}{\ensuremath{m_A}} 
 \newc{\mH}{\ensuremath{m_H}} 
 \newc{\maone}{\ensuremath{m_{a_1}}} 
 \newc{\matwo}{\ensuremath{m_{a_2}}}
 \newc{\hone}{\ensuremath{h_1}}
 \newc{\htwo}{\ensuremath{h_2}}
 \newc{\aone}{\ensuremath{a_1}}
 \newc{\atwo}{\ensuremath{a_2}}
 \newc{\mqthree}{\ensuremath{m_{\tilde{q}_3}^2}}
 \newc{\muthree}{\ensuremath{m_{\tilde{u}_3}^2}}
 \newc{\mql}{\ensuremath{m_{\tilde{q}}}}
 \newc{\mqlij}{\ensuremath{(m^2_{\tilde{q}})_{ij}}}
 \newc{\mur}{\ensuremath{m_{\tilde{u}}}}
 \newc{\mdr}{\ensuremath{m_{\tilde{d}}}}
 \newc{\murij}{\ensuremath{(m^2_{\tilde{u}})_{ij}}}
 \newc{\md}{\ensuremath{m_{\tilde{D}}}}
 \newc{\me}{\ensuremath{m_{\tilde{E}}}}
 \newc{\muu}{\ensuremath{m_{\tilde{U}}}}
 \newc{\mdrij}{\ensuremath{(m^2_{\tilde{d}})_{ij}}}
 \newc{\mll}{\ensuremath{m_{\tilde{l}}}}
 \newc{\mllij}{\ensuremath{(m^2_{\tilde{l}})_{ij}}}
 \newc{\mdlij}{\ensuremath{(m^2_{dl})_{ij}}}
 \newc{\mer}{\ensuremath{m_{\tilde{e}}}}
 \newc{\merij}{\ensuremath{(m^2_{\tilde{e}})_{ij}}}
 \newc{\ts}{\ensuremath{T_{SUSY}}}

\newc{\sigsip}{\ensuremath{\sigma^{\rm SI}_{p}}}	\newc{\sigsin}{\ensuremath{\sigma^{\rm SI}_{n}}}
\newc{\sigsdp}{\ensuremath{\sigma^{\rm SD}_{p}}}	\newc{\sigsdn}{\ensuremath{\sigma^{\rm SD}_{n}}}
\newc{\sigsi}{\ensuremath{\sigma^{\rm SI}}}	\newc{\sigsd}{\ensuremath{\sigma^{\rm SD}}}
\newc{\abund}{\ensuremath{ \Omega h^2}}
\newc{\omegadm}{\ensuremath{ \Omega_{{\rm DM}}}}     \newc{\abunddm}{\ensuremath{ \Omega_{{\rm DM}} h^2}} 
\newc{\omegam}{\ensuremath{ \Omega_{{\rm m}}}}       \newc{\abundm}{\ensuremath{ \Omega_{{\rm m}} h^2}}
\newc{\omegab}{\ensuremath{ \Omega_{{\rm b}}}}	\newc{\abundb}{\ensuremath{ \Omega_{{\rm b}} h^2}}
\newc{\omegatot}{\ensuremath{ \Omega_{{\rm TOT}}}}
\newc{\omegacdm}{\ensuremath{ \Omega_{{\rm CDM}}}}   \newc{\abundcdm}{\ensuremath{ \Omega_{{\rm CDM}} h^2}}
\newc{\omegalambda}{\ensuremath{ \Omega_{\Lambda}}} \newc{\abundlambda}{\ensuremath{ \Omega_{\Lambda} h^2}}
\newc{\omegarad}{\ensuremath{ \Omega_{{\rm rad}}}}  \newc{\abundrad}{\ensuremath{ \Omega_{{\rm rad}} h^2}}
\newc{\rhocrit}{\ensuremath{ \rho_{\rm crit}}}
\newc{\rhochi}{\ensuremath{ \rho_{\chi}}}
\newc{\abunchi}{\ensuremath{\Omega_\chi h^2}}
\newc{\abundlsp}{\ensuremath{\Omega_{\rm LSP}h^2}}
\newcommand*{\abundchi}{\ensuremath{\Omega_\chi h^2}}

\newc{\amu}{\ensuremath{ a_{\mu}}}        \newc{\amususy}{\ensuremath{ a_{\mu}^{\mathrm{SUSY}}}}
\newc{\amuexpt}{\ensuremath{ a_{\mu}^{\mathrm{expt}}}}        \newc{\amusm}{\ensuremath{ a_{\mu}^{\mathrm{SM}}}}
\newc\deltaamu{\ensuremath{\Delta a_{\mu}}} \newc{\deltaamususy}{\ensuremath{\delta a_{\mu}^{\mathrm{SUSY}}}}
\newc\gmtwo{\ensuremath{ (g-2)_{\mu}}} 
\newc{\deltagmtwomususy}{\ensuremath{\delta\left(g-2\right)_{\mu}^{\mathrm{SUSY}}}}
\newc{\deltagmtwomu}{\ensuremath{\delta\left(g-2\right)_{\mu}}}
\newc\BR{\ensuremath{\rm BR}}
\newc\bsgamma{\ensuremath{ b\rightarrow s \gamma }}
\newc\bxsgamma{\ensuremath{\overline{B}\rightarrow X_{s}\gamma}}
\newc\brbsgamma{\ensuremath{\BR\left(\bsgamma\right)}}
\newc\brbxsgamma{\ensuremath{\BR\left(\bxsgamma\right)}}
\newc\bsmumu{\ensuremath{B_s\to\mu^+\mu^-}}
\newc\bdmumu{\ensuremath{B_d\to\mu^+\mu^-}}
\newc\brbsmumu{\ensuremath{\BR\left(B_s\to\mu^+\mu^-\right)}}
\newc\brbdmumu{\ensuremath{\BR\left(B_d\to\mu^+\mu^-\right)}}
\newc\bdmmumu{\ensuremath{\overline{B}_d\to\mu^+\mu^-}}
\newc\bbbarmix{\ensuremath{\overline{B}_s\mbox{-}B_s}}      
\newc\delmbs{\ensuremath{\Delta M_{B_s}}}
\newc\thc{\ensuremath{t\to h c}}
\newc\thu{\ensuremath{t\to h u}}
\newc{\butaunu}{\ensuremath{B_u \rightarrow \tau \nu}}
\newc{\brbutaunu}{\ensuremath{\BR\left(B_u \rightarrow \tau \nu\right)}}
\newc{\mueg}{\ensuremath{\BR\left(\mu^+\to e^+\gamma\right)}}
\newc{\taueg}{\ensuremath{\BR\left(\tau^{\pm}\to e^\pm\gamma\right)}}
\newc{\taumug}{\ensuremath{\BR\left(\tau^{\pm}\to \mu^\pm\gamma\right)}}
\newc{\mue}{\ensuremath{\BR\left(\mu^{+}\to e^+e^+e^-\right)}}
\newc{\taue}{\ensuremath{\BR\left(\tau^{\pm}\to e^{\pm}e^+e^-\right)}}
\newc{\taumu}{\ensuremath{\BR\left(\tau^{\pm}\to \mu^{\pm}\mu^+\mu^-\right)}}

\newcommand*{\reffig}[1]{Fig.~\ref{#1}}
 
        \newcommand*{\refeq}[1]{Eq.~(\ref{#1})}
     \newcommand*{\refsec}[1]{Sec.~\ref{#1}}

\newcommand*{\neutone}{\ensuremath{\tilde{\chi}^0_1}}
\newcommand*{\neuttwo}{\ensuremath{\tilde{{\chi}}^0_2}}

\newcommand*{\charone}{\ensuremath{\tilde{{\chi}}^{\pm}_1}}

\newcommand*{\eight}{\ensuremath{\sqrt{s}=8\tev}}
\newcommand*{\four}{\ensuremath{\sqrt{s}=14\tev}}

\newcommand*{\dsusy}{DarkSUSY}
\newcommand*{\higgsbounds}{H{\scriptsize IGGS}B{\scriptsize OUNDS}}
\newcommand*{\higgssignals}{H{\scriptsize IGGS}S{\scriptsize IGNALS}}
\newcommand*{\feynhiggs}{\texttt{FeynHiggs}}
\newcommand*{\spheno}{\texttt{SPheno}}

\newcommand*{\multinest}{MultiNest}

\newcommand*{\susyflav}{\texttt{SUSY\textunderscore FLAVOR}}

\let\oldcite\cite
\renewcommand*{\cite}{~\oldcite}

\newcommand*{\hl}{\ensuremath{h}}

\newc{\Yb}{\ensuremath{Y_b}}
\newc{\Ys}{\ensuremath{Y_s}}
\newc{\Ym}{\ensuremath{Y_{\mu}}} 
\newc{\mtpole}{\ensuremath{m_t^{\rm pole}}}

\begin{document}
\begin{titlepage}
\vspace*{2cm}
\begin{center} 
{\bf\Large Exact {\boldmath $SU(5)$} Yukawa matrix unification in the\\[2mm]
           General Flavour Violating MSSM}\\[1cm]
{\large Mateusz Iskrzy\'nski${^a}$ and Kamila Kowalska${^b}$}\\[1cm]
{$^a$\small\it Institute of Theoretical Physics, University of Warsaw, Pasteura 5, 02-093 Warsaw, Poland}\\
{$^b$\small\it National Centre for Nuclear Research, Ho$\dot{z}$a 69, 00-681 Warsaw, Poland}\\[1cm]
{\bf Abstract}\\[5mm] 
\end{center}
We investigate the possibility of satisfying the $SU(5)$ boundary condition
$\mathbf{Y}^d=\mathbf{Y}^{e\,T}$ at the GUT scale within the renormalizable
$R$-parity conserving Minimal Supersymmetric Standard Model (MSSM). Working in
the super-CKM basis, we consider non-zero flavour off-diagonal entries in the
soft SUSY-breaking mass matrices and the $A$-terms. At the same time, the
diagonal $A$-terms are assumed to be suppressed by the respective Yukawa
couplings.  We show that a non-trivial flavour structure of the soft
SUSY-breaking sector can contribute to achieving precise Yukawa coupling
unification for all the three families. However, large non-zero values of the
flavour-violating parameter $(m^2_{dl})_{12}$ lead to a strong tension
with the Lepton Flavour Violating (LFV) observables.  Nevertheless, the LFV
problem does not arise when the Yukawa coupling unification requirement is
restricted to the second and third families only. We
demonstrate that such a scenario is consistent with a wide set of experimental
constraints, including flavour and electroweak observables, Higgs physics and
the LHC bounds.  We also point out that in order to provide a proper value for
the relic density of dark matter, the lightest neutralino needs to be almost
purely bino-like and with mass in the range of $200\div 600\gev$. Such a clear
experimental prediction makes the flavour-violating $SU(5)$ Yukawa unification
scenario fully testable at the LHC \four.

\end{titlepage}
\setcounter{page}{1}

\section{Introduction}

One of the virtues of supersymmetry (SUSY) is that it allows unification of
the gauge couplings. Such a feature suggests that above the unification scale
the Minimal Supersymmetric Standard Model (MSSM) should be replaced with a
more general theory. The simplest realization of a Grand Unified Theory (GUT)
is based on the $SU(5)$ gauge symmetry\cite{Dimopoulos:1981zb}, and its
straightforward consequence is equality of the Yukawa couplings of down-type
quarks and charged leptons at the \gut\ scale. While such a condition is easy
to satisfy for the third family of fermions, obtaining
$Y_s(\mgut)=Y_\mu(\mgut)$ and $Y_d(\mgut)=Y_e(\mgut)$ turns out to be quite a
non-trivial task in the minimal $SU(5)$, given the experimentally measured
values of the fermion masses. Several solutions to this problem have been
proposed in the literature, which either considered an extended Higgs sector
above the \gut\ scale\cite{Georgi:1979df}, or employed higher-dimensional
operators\cite{EmmanuelCosta:2003pu,Antusch:2009gu,Antusch:2013rxa}.

Another approach to the Yukawa coupling unification is based on an observation
that SUSY threshold corrections at the superpartner decoupling scale can
considerably alter or even generate masses of the light
fermions\cite{Buchmuller:1982ye}.  Such a possibility was first investigated
in the context of SUSY grand unification in Ref.\cite{Hall:1985dx} where the
presence of general flavour-violating interactions in the soft SUSY-breaking
Lagrangian was assumed. More recently, several studies have been devoted to a
possibility of using the threshold corrections to facilitate unification of
the first and second family Yukawa couplings in the framework of
supersymmetric $SU(5)$. In Ref.\cite{DiazCruz:2000mn} trilinear soft terms
proportional to the corresponding Yukawa matrices were considered. In such a
case, it turned out impossible to obtain simultaneous unification for more
than two families when the scalar masses were universal. On the other hand,
when large off-diagonal trilinear terms were allowed, a strong tension between
the unification requirement and the experimental limits on Flavour Changing
Neutral Current (FCNC) processes appeared. To avoid these problems, the
proportionality assumption was abandoned in Ref.\cite{Enkhbat:2009jt} where
general diagonal $A$-terms were considered. Consequently, the threshold
corrections were driven by large values of the corresponding trilinear
couplings, leading to a successful Yukawa unification for all the three
families. A similar scenario was considered in Ref.\cite{Iskrzynski:2014zla}
which updated the previous analysis in view of the Higgs boson discovery and
strengthened experimental limits on the superpartner masses. With the SUSY
particles getting heavier, tensions with flavour physics become weaker. It was
confirmed that the $SU(5)$ Yukawa coupling unification of all three families
is phenomenologically viable and attainable for a wide range of
\tanb. However, is comes at a price of having a long-lived but metastable MSSM
vacuum.

In the present paper, we explore an alternative scenario. We assume that the
diagonal entries of the trilinear terms have the same hierarchy as the Yukawa
couplings. However, we allow for non-zero off-diagonal entries both in the
trilinear terms and in the sfermion mass matrices.  We employ the chirally
enhanced MSSM threshold corrections to fermion masses as collected in
Ref.\cite{Crivellin:2011jt} and previously calculated in
Refs.\cite{Crivellin:2008mq,Crivellin:2010er,Crivellin:2010gw}.  The most
important feature of this type of corrections is that they allow to
``transmit'' a large $Y_b$-driven threshold correction to the bottom quark
mass to the strange quark mass as well, provided off-diagonal
flavour-violating entries in the soft mass matrices are present. We show that
achieving a satisfactory and phenomenologically viable Yukawa coupling unification for the second
and third generations is facilitated by a non-zero off-diagonal element $(2,3)$ in the
down-squark mass matrix. On the other hand, we find Yukawa unification for
all the three families only when all the off-diagonal elements in the
down-squark mass matrix are non-zero, as well as the $(1,2)$ and $(2,1)$
entries in the down-sector trilinear term. However, such flavour-violating soft 
terms that involve the first family lead to strong tensions
with Lepton Flavour Violating (LFV) observables like ${\mathcal B}(\mu \to e \gamma)$,
at least when the superpartners are not much heavier than a few TeV. Nevertheless, just the
second and third generation cases alone provide evidence that a
more general treatment of the flavour structure of the renormalizable
$R$-parity conserving MSSM can contribute to fulfilling the $SU(5)$ GUT-scale
boundary conditions.

The phenomenology of models with $SU(5)$ symmetry at the \gut\ scale and
General Flavour Violation (GFV) in the squark mass matrices has been studied
in various contexts. Ref.\cite{Guasch:1999jp,Cao:2007dk,Fichet:2014vha}
analysed possible signatures of their spectra at the
LHC. Ref.\cite{Herrmann:2011xe} investigated properties of the dark matter
(DM) candidate, while the consequences for the Higgs mass, $B$-physics and
electroweak (EW) observables were discussed in
Refs.\cite{Heinemeyer:2004by,Cao:2006xb,AranaCatania:2011ak,Arana-Catania:2014ooa,Kowalska:2014opa}.

In the present study, we perform a full phenomenological analysis of the GFV
$SU(5)$ model in which the Yukawa matrix unification constraint is 
imposed. We take into account the mass and the rates of the lightest Higgs
boson, EW precision tests, flavour observables in the quark and lepton
sectors, relic density of the neutralino dark matter, spin-independent
proton-neutralino scattering cross-section, as well as the 8\tev\ LHC
exclusion bounds from the direct SUSY searches. If the unification
condition is imposed on the second and third generations only, the
model is consistent at $3\sigma$ with all the considered experimental
constraints.

The phenomenological features of the scenario discussed in our study make it
an attractive alternative to models assuming Minimal Flavour Violation. First
of all, the FCNC processes triggered by the chirality-preserving mixing
between the second and the third generation of down-type squarks, which plays
a crucial role in the successful Yukawa coupling unification, are less
constrained than those driven by other off-diagonal entries. Additionally, in
the considered scenario, SUSY contributions to flavour observables in the
quark sector remain relatively small thanks to moderate values of \tanb\
and heaviness of the squarks. We also show that the lightest neutralino needs
to be almost purely bino-like, and to have mass in the range of
$200\div 600$\gev. Interestingly, SUSY spectra of this kind have started to be
probed by the LHC at \eight, and will be completely tested at \four.

The article is organised as follows. In \refsec{analysis}, we discuss SUSY
threshold corrections to the Yukawa couplings in the presence of
flavour-violating soft squark mass matrices. In \refsec{sec:impact}, the
impact of off-diagonal soft entries on the Yukawa unification is analysed
numerically, and the parameter space favoured by successful unification is
determined. In \refsec{pheno}, we study thoroughly the phenomenology of our
Yukawa unification scenario in the light of available experimental data. We
summarize our findings in \refsec{concl}.

\section{Anatomy of the minimal {\boldmath $SU(5)$} Yukawa unification}\label{analysis}

Our analysis is performed in a setting that can shortly be summarized in terms
of renormalization scales. The low-energy Yukawa couplings are fixed in the
Standard Model, below a scale where it is matched with the MSSM (henceforth
named $\mu_{\rm sp}$), and at which the supersymmetric threshold corrections are
calculated. Validity of the MSSM extends up to \mgut, where the minimal
$SU(5)$ boundary conditions are imposed on its parameters.
         
The simplest realization of the grand unification idea is based on $SU(5)$, as
this is the smallest symmetry to encompass the SM gauge group. In its
supersymmetric version, the MSSM superfields $Q$, $U$, $D$, $L$, $E$ are
embedded into the 5- and 10-dimensional representations of $SU(5)$ as
\begin{align}
 \underbrace{(\mathbf{\bar 3},\mathbf{1},\tfrac{1}{3})}_{D} \oplus \underbrace{(\mathbf{1},\mathbf{2}, 
  -\tfrac{1}{2})}_L &= \underbrace{\mathbf{\bar 5}}_{\Psi_{\bar 5}} \\ 
\underbrace{(\mathbf{3},\mathbf{2},\tfrac{1}{6})}_Q \oplus \underbrace{(\mathbf{\bar 3},
   \mathbf{1},-\tfrac{2}{3})}_{U} \oplus \underbrace{(\mathbf{1},\mathbf{1},1)}_E &= 
\underbrace{\mathbf{10}}_{\Psi_{10}}, 
\end{align}
where we use the conventional SM normalization for the hypercharges.  The
Yukawa terms for $SU(5)$ GUT read\cite{Dimopoulos:1981zb}
\begin{equation}\label{yuksu5}
{\cal W} \ni   \Psi_{10} \mathbf{Y}^{de} \Psi_{\bar 5} H_{\bar 5} 
             + \Psi_{10} \mathbf{Y}^u \Psi_{10} H_5,
\end{equation}
where $H_{\bar 5}$ and $H_5$ denote two Higgs multiplets that are coupled to
 matter.  From Eq.~(\ref{yuksu5}) it follows that all parameters that allow to
 determine the masses of the SM fermions are encoded in two independent
 $3\times 3$ matrices $\mathbf{Y}^{de}$ and $\mathbf{Y}^{u}$.  Below the GUT
 scale $M_{\rm GUT} \simeq 2\times 10^{16}\,$GeV, the $SU(5)$ model is
 replaced with the MSSM, and the effective superpotential is given by
\begin{equation}
{\cal W}_{MSSM}=  Q \mathbf{Y}^u U H_u +  Q  \mathbf{Y}^d D H_d +  L \mathbf{Y}^e E H_d + \mu H_d H_u.
\end{equation}
 A straightforward consequence of the GUT symmetry is the equality of the
 matrices $\mathbf{Y}^d$ and $\mathbf{Y}^{e\,T}$ at $M_{\rm GUT}$. This
 condition is true up to a basis redefinition and possible one-loop threshold
 corrections at this scale.  In the present study, we are mainly interested in
 low-energy properties of successful unification scenarios rather than in the
 exact realization of their high-energy UV completion. For this reason, in the
 subsequent analysis, we are going to allow for moderate threshold corrections
 at the GUT scale (see Eq.~(\ref{TresholdGUT})) without investigating their
 origin.

Unification conditions for the Yukawa couplings of down-type quarks and
 charged leptons take the simplest form in a basis where the superpotential
 flavour mixing is entirely included in $\mathbf{Y}^u$, while $\mathbf{Y}^d$
 and $\mathbf{Y}^e$ are real and diagonal. In such a case, it is enough to
 require equality of the diagonal entries at the GUT scale,
\begin{equation}\label{yukunif}
Y^d_{ii}=Y^e_{ii}, \hspace{20pt} i=1,2,3.
\end{equation}

\subsection{Flavour-violating threshold corrections to the Yukawa matrices 
         at the decoupling scale}\label{corrthr}
         
Diagonal entries of the Yukawa couplings are constrained by measurements of
the quark and lepton masses that are performed at or below the electroweak
scale. Consequently, these entries are most easily fixed within the SM. One
needs, however, to determine their renormalized values within the MSSM, which
is assumed to be an underlying effective theory that connects the electroweak
scale with \mgut. This is done by calculating threshold corrections
$\Sigma_{ii}^f$ at the matching scale $\mu_{\rm sp}$. Such corrections depend on
values of the soft SUSY-breaking terms,
\be
v_f Y_{ii}^{f\,MSSM}=v_f Y_{ii}^{f\,SM}-\Sigma_{ii}^f((m^2_{\tilde{f}})_{ij},A^f_{ij},m_{H_i},M_i).
\ee
The Yukawa coupling values at \mgut\ are then determined by solving their MSSM
renormalization group equations (RGEs) which do not depend on the soft
parameters. Once this is done, the Yukawa unification quality for a given set
of parameters can be tested.

The experimentally measured values of fermion masses can give a qualitative
feeling about the problems encountered in achieving the full Yukawa matrix
unification. As is well known, the constraint $Y_b(M_{\rm GUT})=Y_\tau(M_{\rm
GUT})$ can be satisfied without large threshold corrections at $\mu_{\rm sp}$,
at least for moderate $\tan\beta$. On the other hand, achieving strict
unification of the Yukawa couplings for the remaining families ($Y_s(M_{\rm
GUT})=Y_\mu(M_{\rm GUT})$ and $Y_d(M_{\rm GUT})=Y_e(M_{\rm GUT})$) requires
the threshold corrections to be of the same order as the leading terms. To
satisfy the minimal $SU(5)$ boundary conditions on Yukawas, the MSSM strange
quark mass has to be raised w.r.t.\ the SM one, whereas the down-quark has to
be made lighter by the threshold corrections. That does not contradict
perturbativity of the model, as the corresponding leading terms are small
enough, $Y^{(0)}_{s,d} \ll 1$.

The dominant supersymmetric threshold corrections to the Yukawa couplings
beyond the small \tanb\ limit have been calculated in
Ref.\cite{Crivellin:2011jt}.  It was shown that in the SUSY-decoupling limit,
the chirality-flipping parts of the renormalized quark (lepton) self energies
$\Sigma$ are linear functions of the Yukawa couplings, with a proportionality
factor $\epsilon$ and an additive term $\Sigma_{\slashed{Y}}$,
\be
m^{d(\ell)\,SM}_i-v_d Y^{d(\ell)MSSM}_{ii} \;=\;
\Sigma_{ii}^{d(\ell)\,LR} \;=\;
\Sigma_{ii\,{\cancel{Y}}}^{d(\ell)\,LR} \, + \,
\epsilon_i^{d(\ell)}\,v_u\,\,Y^{d(\ell)(0)}_{ii} ~+~ O(\tfrac{v^2}{M_{SUSY}}), 
\label{eq:epsilon_b}
\ee
where \msusy\ is defined as $\msusy=\sqrt{m_{\tilde{t}_1}m_{\tilde{t}_2}}$.
In this approximation, the relation can easily be inverted, and the corrected
MSSM Yukawa couplings in the super-CKM basis read
\be
\label{TCstructEq}
Y^{d(\ell)MSSM}_{ii}=\frac{m^{d(\ell)\,SM}_i-\Sigma^{d(\ell)\,LR}_{ii\,{\cancel{Y}}}}{v_d 
(1+ \tan\beta \cdot\epsilon_{i}^{d(\ell)})}.
\ee
Various contributions to $\Sigma_{ii}^{d(\ell)\,LR}$ from sfermion-gluino,
sfermion-neutralino and sfermion-chargino diagrams, as well as the threshold
corrections to the CKM matrix can be found in
Ref.\cite{Crivellin:2011jt}. Their non-trivial dependence on the GFV
parameters is encoded in the term $\Sigma_{\slashed{Y}}$. The off-diagonal
soft mass elements enter \refeq{eq:epsilon_b} through rotation matrices that
diagonalise the sfermion mass matrices. On the other hand, the off-diagonal
trilinears appear explicitly but give no contribution when the rotation
matrices are diagonal.

\begin{figure}[t]
\centering
\subfloat[]{
\includegraphics[width=0.39\textwidth]{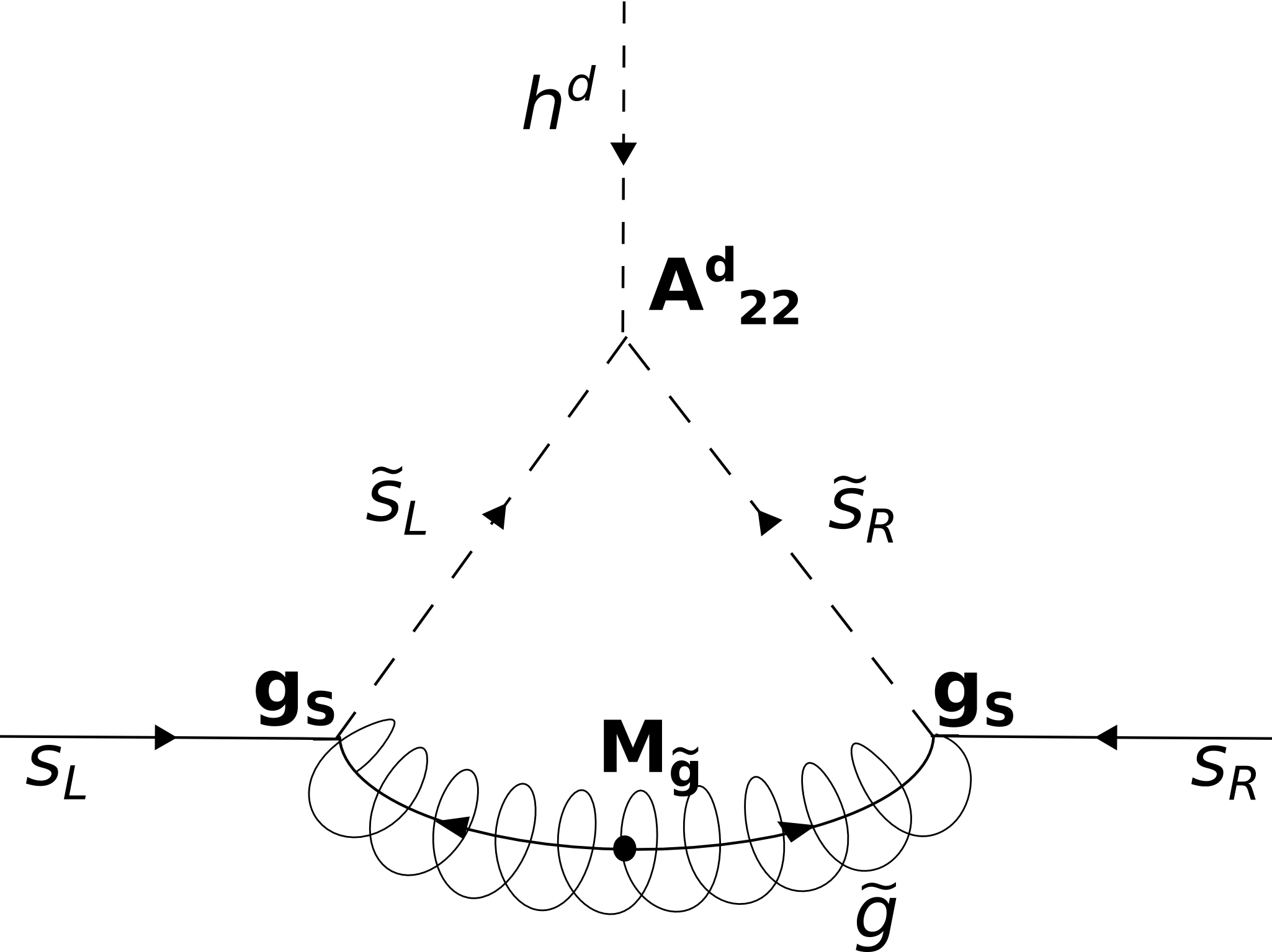}}
\subfloat[]{
\includegraphics[width=0.39\textwidth]{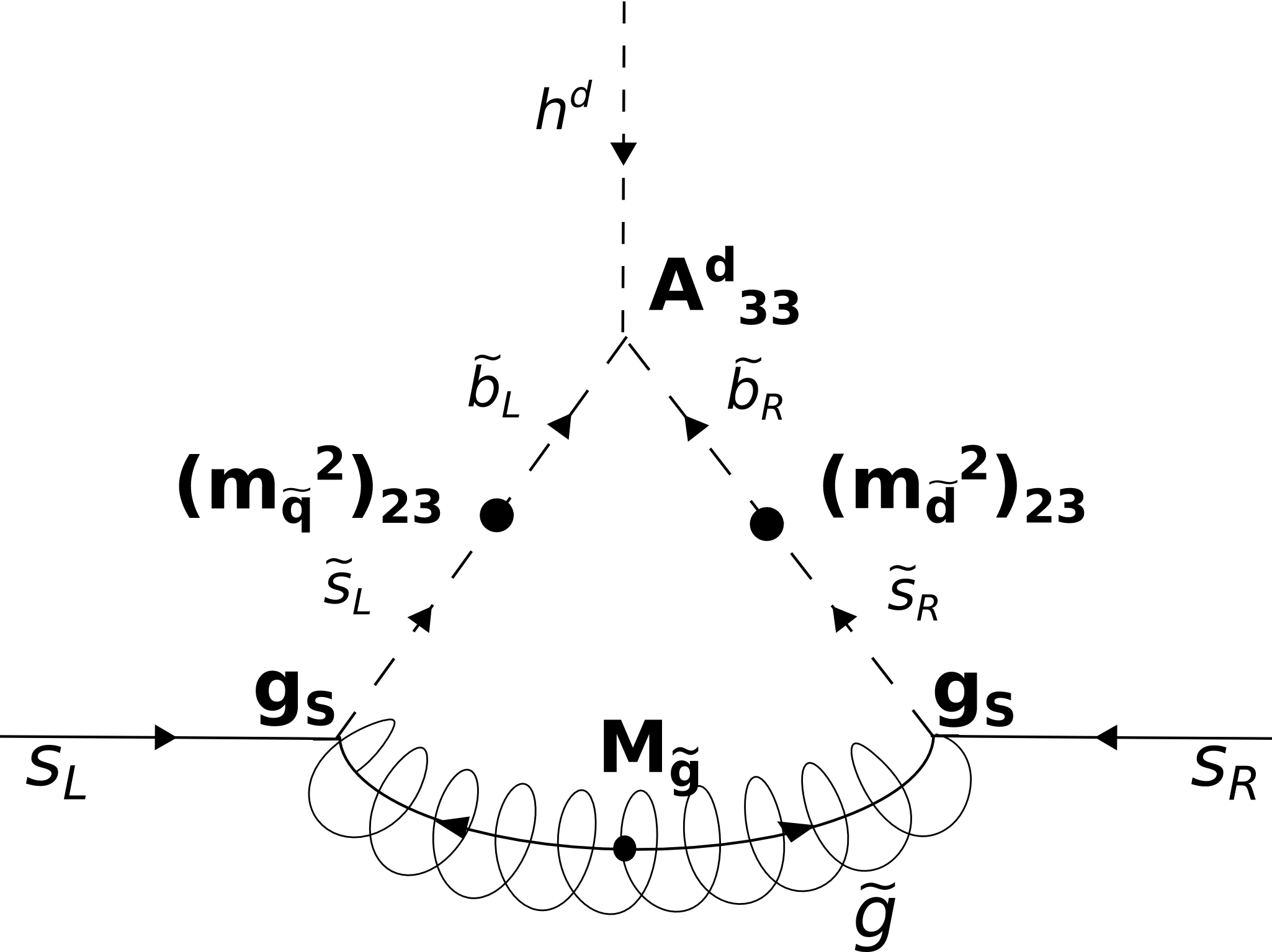}}
\caption{Examples of diagrams that describe threshold corrections to the strange
Yukawa coupling at $\mu_{\rm sp}$. (a) the flavour-diagonal leading gluino
diagram; (b) possibly an even bigger contribution arising in the case of
flavour mixing in the soft-mass matrix.}
\label{diags}
\end{figure}

To get some intuition about functional dependence of \refeq{eq:epsilon_b} on
various entries in the sfermion mass matrices, let us consider an example of
threshold corrections to the self-energy of down-type quarks. The most
significant contribution comes from a gluino loop, as it is enhanced by a
large value of the strong gauge coupling.  If all the flavour-violating mass
matrix entries were zero (the relevant diagram in the strange quark case is
depicted in the left panel of Fig.~\ref{diags}), threshold corrections to the
Yukawa couplings from a gluino loop would be given by the corresponding
diagonal entries of the trilinear terms (see Ref.\cite{Crivellin:2011jt}),
\be\label{thres:mfv}
(\Sigma^d_{ii})^{\tilde{g}}=\frac{2\alpha_s}{3\pi}\mgluino v_d(A_{ii}^d-Y^d_{ii}\mu\tanb) 
C_0(\mgluino^{\!\!\! 2},m^2_{\tilde{q}^L_i},m^2_{\tilde{d}^R_i}),
\ee
where the loop function $C_0(\mgluino^{\!\!\!
2},m^2_{\tilde{q}^L_i},m^2_{\tilde{d}^R_i})$ of dimension mass$^{-2}$ can be
found, e.g., in the appendix of Ref.\cite{Crivellin:2011jt}. Interestingly,
for the third family, the expression given by Eq.~(\ref{thres:mfv}) tends to
cancel with a contribution mediated by the higgsino exchange, which makes the
ratio $Y_b/Y_{\tau}$ quite stable with respect to the SUSY threshold
corrections. On the contrary, for the first and second generation, the gluino
contribution is dominant and can be used to fix the ratios of the
corresponding Yukawa couplings at $\mu_{\rm sp}$. Such a possibility was
considered in Refs.\cite{DiazCruz:2000mn,Enkhbat:2009jt,Iskrzynski:2014zla}.

Since $C_0$ is always negative, it results immediately from
Eq.~(\ref{thres:mfv}) that $A_{22}^d$ should be positive and $\mu$ negative to
maximize the necessary positive correction to $m_s^{SM}$ in the numerator on
the r.h.s.\ of Eq.~(\ref{TCstructEq}).  In the first family case, a
negative $A_{11}^d$ is sufficient to generate a correction to $Y_d$ of the
right sign.

The situation changes when non-zero off-diagonal soft mass entries are allowed
(see the right panel of Fig.~\ref{diags}).  In such a case, the dominant GFV
contribution to the strange quark self-energy associated with a gluino 
loop\cite{Crivellin:2011jt} can be written as
\be\label{sigma22}
(\Sigma^d_{22})^{\tilde{g}}=\frac{\alpha_s}{6\pi}m_{\tilde{g}}v_{d}(A^d_{33}-Y_b\mu\tan\beta)\,
S\,\sin 2\theta_{\tilde{q}}\,\sin 2\theta_{\tilde{d}}\!
\sum_{m,n=2,3} (-1)^{n+m} C_0(\mgluino^{\!\!\!2},m^2_{\tilde{q}^L_m},m^2_{\tilde{d}^R_n}),
\ee
where $S={\rm sgn}\left[\left(m^2_{\tilde{q}^L_2} - m^2_{\tilde{q}^L_3}\right)
                            \left(m^2_{\tilde{d}^R_2} - m^2_{\tilde{d}^R_3}\right)\right]$,
and
\begin{displaymath}
\sin 2\theta_{\tilde{q}}\,=\,
\frac{2(m^2_{\tilde{q}})_{23}}{
\sqrt{\left[(m_{\tilde{q}}^2)_{22}-(m^2_{\tilde{q}})_{33}\right]^2+4((m^2_{\tilde{q}})_{23})^2}},
\hspace{5mm}
\sin 2\theta_{\tilde{d}}\,=\,
\frac{2(m^2_{\tilde{d}})_{23}}{\sqrt{\left[(m^2_{\tilde{d}})_{22}-
(m^2_{\tilde{d}})_{33}\right]^2+4((m^2_{\tilde{d}})_{23})^2}}.
\end{displaymath}
Here, we have assumed that $(m^2_{\tilde{q}})_{23}$ and
$(m^2_{\tilde{d}})_{23}$ are the only non-zero off-diagonal elements of the
down-squark mass matrix, and that they are real. It follows from
\refeq{sigma22} that chirality-conserving GFV interactions~
$[(m^2_{\tilde{q}})_{23}\, \tilde{s}_L^* \tilde{b}_L + {\rm h.c.}]$~ 
and~
$[(m^2_{\tilde{d}})_{23}\, \tilde{s}_R^* \tilde{b}_R + {\rm h.c.}]$~ 
generate a threshold correction to \Ys\ of the order of $\Delta \Ys\sim
\alpha_s A_{33}^d/M_{\rm SUSY}$, which in general can be large enough to
facilitate Yukawa coupling unification for the second family, even when the
coupling $A_{22}^d$ is small.

The above description, however, should be treated only as a simplified
qualitative illustration. In a general case, also other off-diagonal elements
of the squark mass matrix can significantly differ from zero, which renders
mixing among all the three generations important. To make sure that all such
effects are properly taken into account, in the following we turn to a full
numerical analysis.

\section{Impact of the GFV threshold corrections on Yukawa unification}\label{sec:impact}

Let us start this section with briefly describing the numerical tools and
procedures employed to identify the GFV $SU(5)$ GUT parameter space regions
where the Yukawa coupling unification becomes possible. Next, we shall
perform a numerical scan and use its results to determine those SUSY
parameters whose non-zero values are indispensable from the point of view of
the considered unification.

\subsection{Numerical tools and scanning methodology}\label{sec:tool}

Points satisfying the Yukawa coupling unification conditions at the \gut\
scale are found by scanning the parameter space of the model defined in
\refsec{input}. To this end, we use the numerical package BayesFITSv3.2 that
was first developed in Ref.\cite{Fowlie:2012im} and modified to incorporate
the full GFV structure of the soft SUSY-breaking sector in
Ref.\cite{Kowalska:2014opa}. The package uses for sampling \texttt{\multinest\
v2.7}\cite{Feroz:2008xx} which allows a fast and efficient scanning according
to a predefined likelihood function. The likelihood corresponding to the
$SU(5)$ boundary condition (\ref{yukunif}) is modeled with a Gaussian
distribution as follows,
\be \label{TresholdGUT}
\like_{\rm Yuk}=\sum_{i=1,2,3}\exp\left[-(1-Y^e_{ii}(\mgut)/Y^d_{ii}(\mgut))^2/2\sigma_{\rm Yuk}^2\right],
\ee
and the allowed deviation from the exact unification condition,
$\sigma_{\rm Yuk}$, is set to 5\%.
\begin{table}[t]\footnotesize
\centering
\renewcommand{\arraystretch}{1.4}
\begin{tabular}{| c | c | c | c | c | c | c | c |}
\hline
\multicolumn{2}{|c|}{\mtpole} & \multicolumn{2}{c|}{\mbmbmsbar} & \multicolumn{2}{c|}{\alphasmzms} & 
\multicolumn{2}{c|}{\alphaemmz} \\
\multicolumn{2}{|c|}{$173.34 \pm 0.76$ GeV} & \multicolumn{2}{c|}{$4.18 \pm 0.03$ GeV} & 
\multicolumn{2}{c|}{$0.1184 \pm 0.0007$} & \multicolumn{2}{c|}{$127.944 \pm 0.015$} \\
\hline
\mumbmsbar & \mdmbmsbar & \msmbmsbar & \mcmbmsbar & \mepole & \mmupole & \mtaupole & \mzpole \\
2.3 MeV & 4.8 MeV & 95 MeV & 1.275 GeV & 511 keV & 106 MeV & 1.777 GeV & 91.19 GeV \\
\hline
\multicolumn{2}{|c|}{$\bar{\rho}$} & \multicolumn{2}{c|}{$\bar{\eta}$} & \multicolumn{2}{c|}{$A$} & 
\multicolumn{2}{c|}{$\lambda$} \\
\multicolumn{2}{|c|}{$0.159 \pm 0.045$} & \multicolumn{2}{c|}{$0.363 \pm 0.049$} & 
\multicolumn{2}{c|}{$0.802 \pm 0.020$} & \multicolumn{2}{c|}{$0.22535 \pm 0.00065$} \\
\hline
\end{tabular}
\caption{Standard Model parameters\cite{Agashe:2014kda,utfit} used in
our numerical calculations. The light ($u$, $d$, $s$) quark masses are
$\overline{\rm MS}$-renormalized at 2$\,$GeV.\label{SMinput}}
\end{table}

Mass spectra are calculated with \texttt{\spheno\
v3.3.3}\cite{Porod:2011nf}. The choice is dictated by the fact that at the
moment \spheno\ is the only SUSY spectrum generator where the full flavour
structure of threshold corrections to the Yukawa couplings, as given in
Ref.\cite{Crivellin:2011jt}, is implemented.  As is the case of other tools of
this kind, the renormalization group equations of the MSSM are solved by an
iterative algorithm that interpolates between different scales at which the
parameters are defined. The boundary with the SM (i.e. the scale $\mu_{\rm
sp}$) is set at $M_Z$.

Four SM parameters (\mtpole, \mbmbmsbar, \alphaemmz\ and \alphasmzms) are
treated as nuisance parameters and randomly drawn from a Gaussian distribution
centred around their experimentally measured central values\cite{Agashe:2014kda}. The
elements of the CKM matrix in the Wolfenstein parametrisation ($\bar{\rho}$,
$\bar{\eta}$, $A$, $\lambda$) are scanned as well, with central values and
errors given by the UTfit Collaboration for the scenario allowing new physics
effects in loop observables\cite{utfit}. The other SM parameters which are
passed as an input to \spheno\ (\msmbmsbar, \mcmbmsbar, \mdmbmsbar,
\mumbmsbar, \mtaupole, \mmupole, \mepole, \mzpole) are fixed at their
experimentally measured values. Our SM input is collected in
Table~\ref{SMinput}.

\subsection{Input SUSY parameters}\label{input}

We start our analysis by defining a general set of SUSY parameters at the GUT
scale.  Since {\it a priori} we do not know which GFV parameters in the
down-squark sector are indispensable to achieve the Yukawa coupling
unification and which can be skipped, initially we allow all of them to assume
non-zero values. The aim of the numerical scan is to identify those parameters
that are essential.

We assume for simplicity that all the soft SUSY-breaking parameters are real,
therefore neglecting the possibility of new SUSY sources of CP violation. The
GUT-scale $SU(5)$ boundary conditions for the soft-masses read
\bea\label{softmass:su5}
\mllij=\mdrij\equiv \mdlij, \qquad \mqlij=\murij=\merij\equiv (m^2_{ue})_{ij}. 
\eea
We do not impose any additional conditions on the relative magnitudes of the
diagonal entries. In particular, both normal and inverted hierarchies for the
elements $(m^2_{dl})_{ii}$ and $(m^2_{ue})_{ii}$ are allowed. The off-diagonal
elements of the down-squark matrix are normalized to the entry
$(m^2_{dl})_{33}$, and are required to satisfy the upper limit
$(m^2_{dl})_{ij}/(m^2_{dl})_{33}\le 1$.

We further assume that
\be
(m^2_{ue})_{ij}=0,\;\;i\neq j.
\ee
Such an assumption is not expected to cause any significant loss of generality
because relatively large off-diagonal elements of $\mqlij$ are generated
radiatively at \msusy\ due to the RGE running in the super-CKM basis.  We
restrict our study to the case $(m^2_{dl})_{ij}>0$, as it is a desired
property for the Yukawa unification. At this point we are allowed to
introduce a short-hand notation
\be
m^{dl}_{ij}\equiv\sqrt{(m^2_{dl})_{ij}},\qquad m^{ue}_{ij}\equiv\sqrt{(m^2_{ue})_{ij}}.
\ee

The GUT-scale $SU(5)$ boundary conditions for the trilinear terms are given by
\bea\label{trili:su5}
A_{ij}^{d}=A_{ji}^{e}\equiv A_{ij}^{de}.
\eea
We constrain the relative magnitude of the diagonal entries by the
corresponding Yukawa couplings
\begin{equation}
\frac{|A^f_{ii}|}{|A^f_{33}|} < \frac{Y^f_{ii}}{Y^f_{33}}.
\end{equation}
We do so because we aim at relaxing the strong tension between the EW vacuum
stability condition and Yukawa unification that has been
observed\cite{Iskrzynski:2014zla} in the case of large diagonal $A$-terms.  We
also impose that
\be
A^u_{ij}=0,\;\;i\neq j.
\ee
On the other hand, the off-diagonal entries in the down-sector trilinear
matrix are not constrained in our initial scan to scale proportionally to the corresponding Yukawa
matrix entries. The are only required to satisfy
$|(A^{de}_{ij})/(A^{de}_{33})|\le 0.5$.

Finally, we assume that the gaugino mass parameters are universal at $M_{\rm
GUT}$,
\be
M_1=M_2=M_3\equiv M_{1/2},
\ee
which is the simplest among relations that naturally arise in the framework of
SUSY $SU(5)$ GUTs. The sign of the parameter $\mu$ is chosen to be negative to
facilitate the second family unification, as explained in \refsec{analysis}.
\begin{table}[t]
\centering
\renewcommand{\arraystretch}{1.1}
\begin{tabular}{|c|c|c|}
\hline 
Parameter &  Scanning Ranges:  $GFV_{23}$  &  Scanning Ranges: $GFV_{123}$ \\
\hline 
$M_{1/2}$	& [$100$, $4000$] \gev\ & [$100$, $4000$] \gev\ \\
\mhu            & [$100$, $8000$] \gev\  & [$100$, $8000$] \gev\ \\
\mhd            & [$100$, $8000$] \gev\  & [$100$, $8000$] \gev\ \\
\tanb	        & [$3$, $45$]& [$3$, $45$] \\
\signmu		& $-1$& $-1$ \\
\hline 
$A_{33}^{de}$   & [$0$, $5000$] \gev\ & [$0$, $5000$] \gev\ \\
$A_{33}^{u}$    & [$-9000$, $9000$] \gev\ & [$-9000$, $9000$] \gev\ \\
$A^{de}_{11}/A^{de}_{33}$   & [$-0.00028$, $0.00028$] & [$-0.00028$, $0.00028$]\\
$A^{de}_{22}/A^{de}_{33}$  & [$-0.065$, $0.065$] & [$-0.065$, $0.065$]\\
$A^{u}_{22}/A^{u}_{33}$    & [$-0.005$, $0.005$] & [$-0.005$, $0.005$] \\
$A^{de}_{ij}/A^{de}_{33},\; i\neq j$    & $0$ & [$-0.5$, $0.5$]\\
\hline 
$m_{ii}^{dl},\; i=1,2,3$   & [$100$, $10000$] \gev\ & [$100$, $7000$] \gev\ \\
$m^{dl}_{23}/m^{dl}_{33}$  & [$0$, $1$]  & [$0$, $1$]\\
$m^{dl}_{13}/m^{dl}_{33}$  & $0$ & [$0$, $1$] \\
$m^{dl}_{12}/m^{dl}_{33}$  & $0$ & [$0$, $1$] \\
$m_{ii}^{ue},\; i=1,2,3$   & [$100$, $7000$] \gev\  & [$100$, $7000$] \gev\ \\
\hline 
\end{tabular}
\caption{Ranges of the input SUSY parameters used in our initial scan. Several omitted
soft SUSY-breaking parameters at the GUT scale (namely $A^{u}_{11}$ as well as
$A^{u}_{ij}$ and $m^{ue}_{ij}$ for $i\neq j$) have been set to zero.}
\label{tab:priors$SU(5)$_unif}
\end{table}

In our study, we consider two different scenarios for the Yukawa matrix unification:\\[2mm]
{\bf $GFV_{23}$:} Only the third and the second generation Yukawa couplings
are unified at the \gut\ scale. The only relevant GFV parameter in this case
is $m^{dl}_{23}$.\\[2mm]
{\bf $GFV_{123}$:} Yukawa couplings of all the three families are unified at
the \gut\ scale. All the GFV parameters in the down-squark sector can assume
non-zero values.\\[2mm]
In Table~\ref{tab:priors$SU(5)$_unif}, we collect the scanning ranges for our
initial set of the input SUSY parameters. In the next subsection, we shall
identify those being particularly relevant from the point of view of the
Yukawa coupling unification.

\subsection{Yukawa unification through flavour-violating terms}\label{sec:unif}

\subsubsection{$\boldsymbol{GFV_{23}}$: unification of the third and second family}\label{gfv23}

The performed scan has returned the MSSM spectrum for 121986
parameter-space points, among which 34758 yield the Yukawa couplings of
leptons and down-type quarks at \mgut{} equal within 10\% for the second and
third generations.

\begin{figure}[h]
\centering
\subfloat[]{
\includegraphics[width=0.41\textwidth]{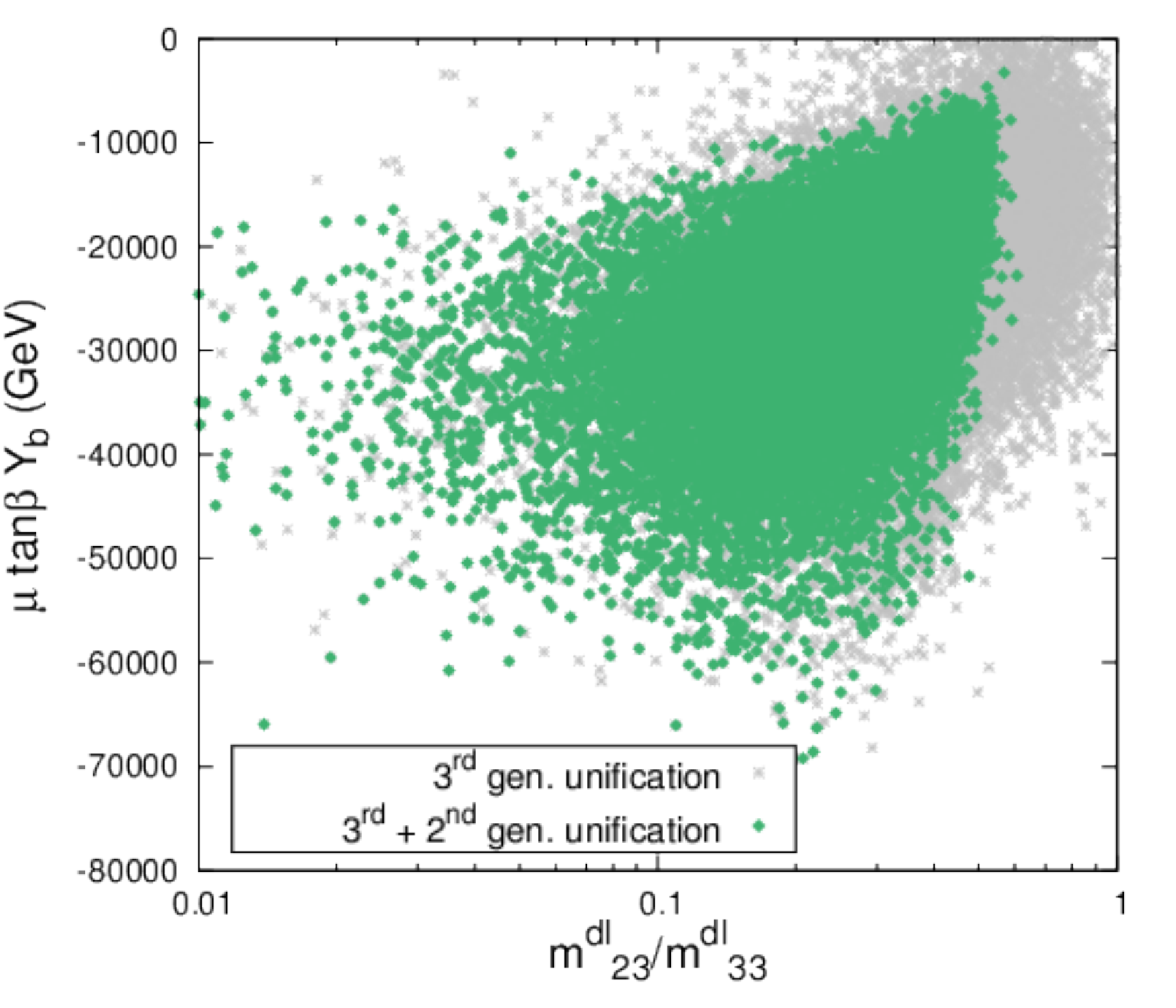}}
\subfloat[]{
\includegraphics[width=0.41\textwidth]{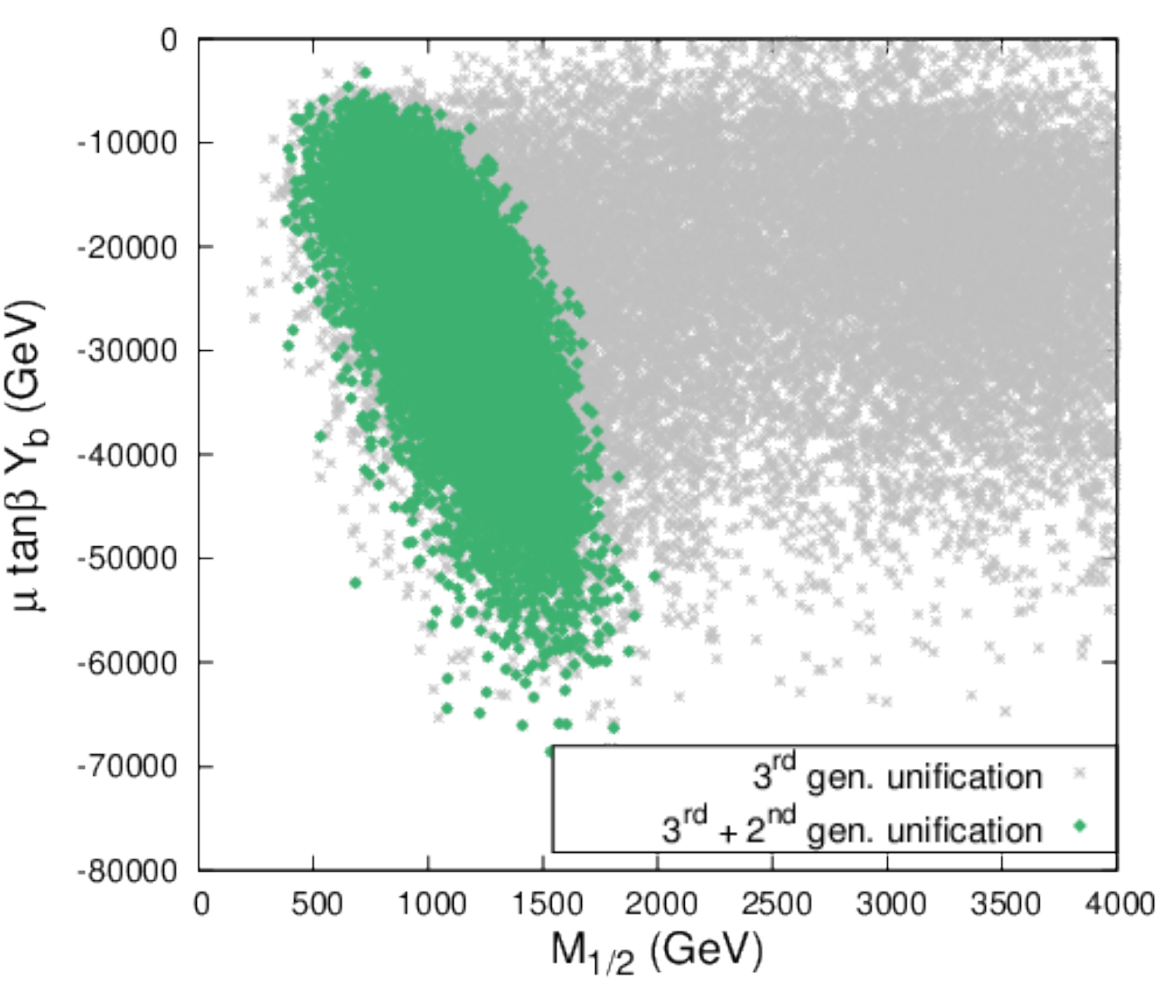}}\\
\subfloat[]{
\includegraphics[width=0.40\textwidth]{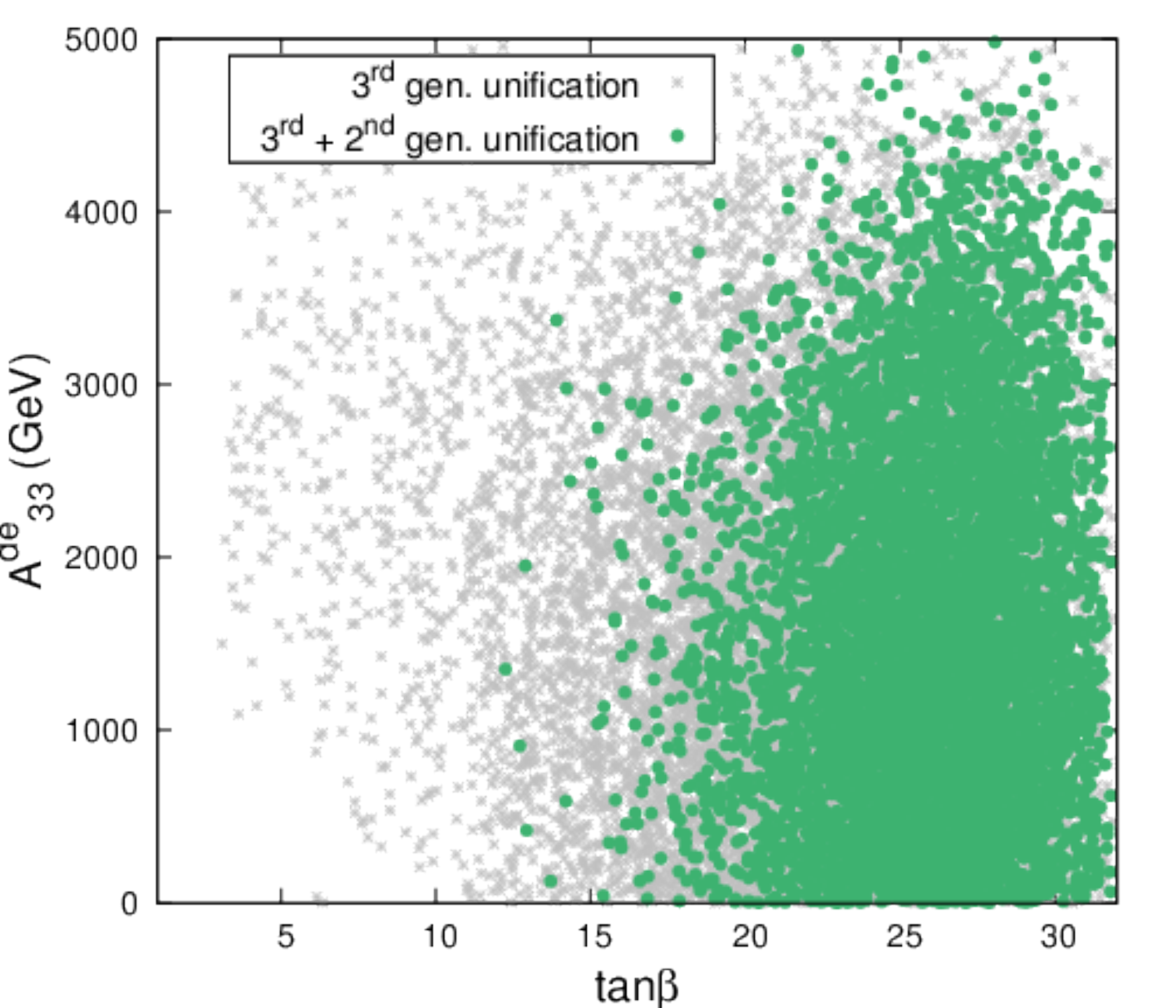}}
\caption{Scatter plot of the $GFV_{23}$ points in the planes
($m^{dl}_{23}/m^{dl}_{33}$, $Y_b\mu\tanb$) (a), ($\mhalf$, $Y_b\mu\tanb$) (b), and
($\tanb$, $A^{de}_{33}$) (c). Gray stars: all
points satisfying the Yukawa unification condition for the third generation at
$2\sigma$; green dots: points for which both heavier generations are unified
at $2\sigma$.}
\label{m23ad22}
\end{figure}

In \reffig{m23ad22}, we present distributions of the collected points in the
planes of ($m^{dl}_{23}/m^{dl}_{33}$, $Y_b\mu\tanb$) (a), ($\mhalf$, $Y_b\mu\tanb$)
(b), and ($\tanb$, $A^{de}_{33}$) (c). All the
points that satisfy the Yukawa unification condition for the third generation
at $2\sigma$ ($0.9<Y_b/Y_{\tau}<1.1$) are represented as gray stars. Those for
which both third and second generations are unified at $2\sigma$ are shown as
green dots.

The most characteristic feature of the $GFV_{23}$ scenario is the lack of
points consistent with the Yukawa unification and with a large universal
gaugino mass parameter, $\mhalf > 2000\gev$. This is a straightforward 
consequence of Eq.~(\ref{sigma22}) in which the gluino mass appears both as a multiplicative 
factor and through the loop function $C_0$. The impact of the latter is of particular importance since for 
$\mhalf > 2000\gev$, and for the fixed sfermion masses,
it leads to strong suppression of the threshold correction $(\Sigma^d_{22})^{\tilde{g}}$.
 At the same time, the correction is somewhat enhanced when the mass of sfermions is decreased with a fixed $\mhalf$, although this effect is less prominent.

Secondly, the strange-muon unification occurs for both large and small
values of $A^{de}_{33}$, which is of relevance for the minimisation of the
scalar potential, as further discussed in \refsec{OffDiagEWSBsec}.  
This is confirmed by panels (a) and (b) of \reffig{m23ad22} where the values of another relevant factor $Y_b\mu\tanb$, which appears in Eq.~(\ref{sigma22}), are depicted for all the model points satisfying the Yukawa unification condition. One can observe, by comparing with panel (c) of \reffig{m23ad22}, that $|Y_b\mu\tanb| \gg A^{de}_{33}$ and the impact of $A^{de}_{33}$ in Eq.~(\ref{sigma22}) is indeed negligible.

This means that the threshold correction $(\Sigma^d_{22})^{\tilde{g}}$ essentially depends on 
three parameters: the gluino mass, the off-diagonal element $m^{dl}_{23}$, and the higgsino mass term $\mu$.  
As a consequence, the strange-muon Yukawa coupling unification can be achieved through an interplay of several mechanisms.
For small gluino masses (that correspond to the upper-left corner of \reffig{m23ad22}(b)), the threshold correction $(\Sigma^d_{22})^{\tilde{g}}$ can be enhanced either by large values 
of the GFV parameter $m^{dl}_{23}$ (upper-right corner of \reffig{m23ad22}(a)) or by large values of the factor $Y_b\mu\tanb$ with $m^{dl}_{23}$ as small as $0.01$. 
The latter effect is additionally enhanced for moderate values of \tanb{} which are preferred
in the $GFV_{23}$ scenario, as shown in panel (c) of \reffig{m23ad22}.

For large gluino masses, that correspond to $\mhalf \geq 1500$ \gev, the threshold correction of Eq.~(\ref{sigma22}) becomes
suppressed by the loop function $C_0$. Therefore, both large $m^{dl}_{23}$ and an even larger $\mu$-term are needed to achieve 
Yukawa unification for the second family.

Finally,
in the squark and slepton sector, the $GFV_{23}$ scenario favours a spectrum
with the third generation much heavier than the first two ones. It
has important phenomenological consequences for the dark matter and flavour
observables, as discussed in \refsec{phenoGFV23}.
\subsubsection{$\boldsymbol{GFV_{123}}$: unification of all three families}

We collected about $1.5\times10^5$ points through our scanning procedure of
the parameter space defined in Table~\ref{tab:priors$SU(5)$_unif}.

\begin{figure}[h]
\centering
\subfloat[]{
\includegraphics[width=0.41\textwidth]{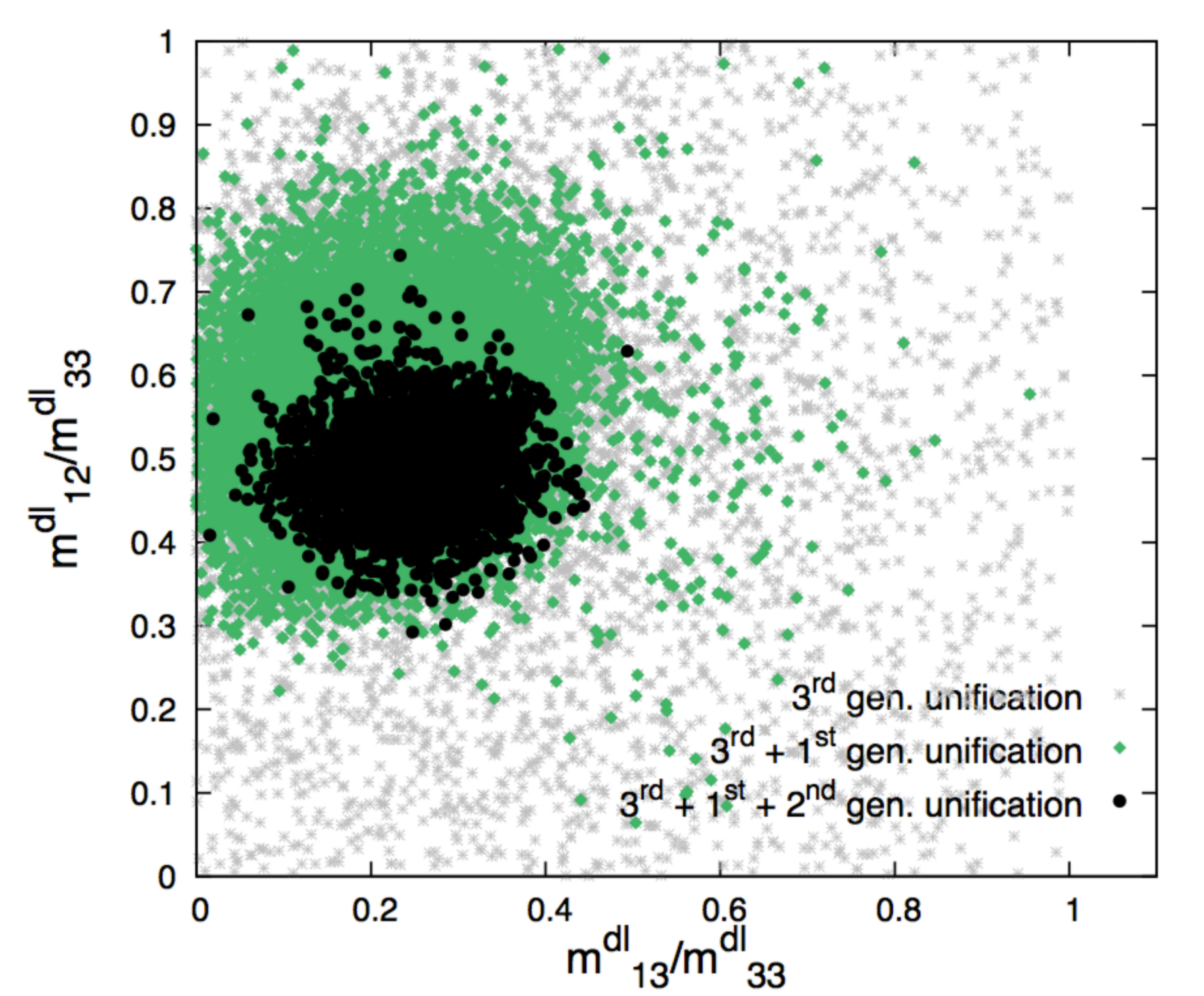}}
\subfloat[]{
\includegraphics[width=0.41\textwidth]{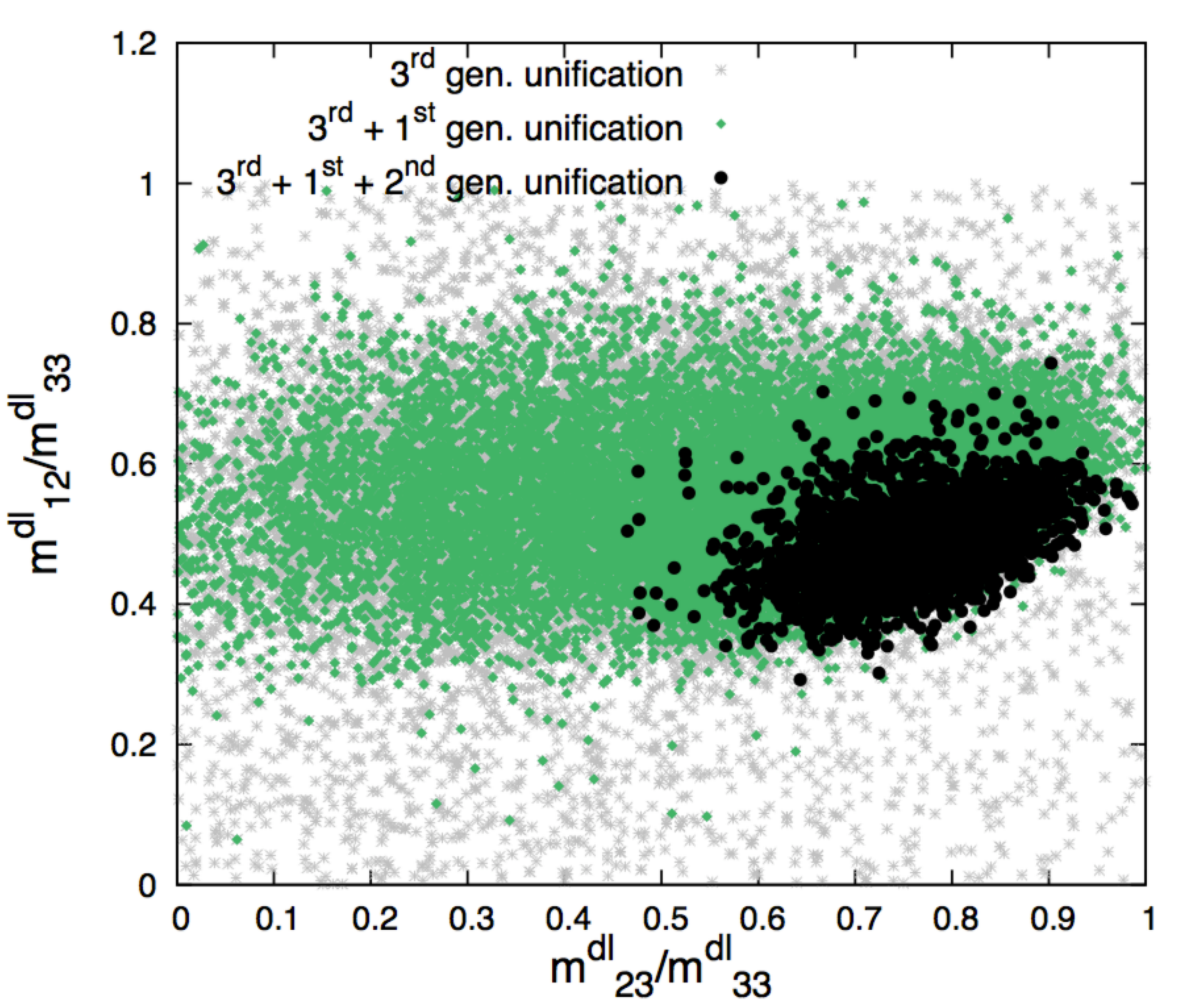}}
\caption{Scatter plot of the $GFV_{123}$ points in the planes
($m^{dl}_{13}/m^{dl}_{33}$,
$m^{dl}_{12}/m^{dl}_{33}$) (a), and
($m^{dl}_{23}/m^{dl}_{33}$,
$m^{dl}_{12}/m^{dl}_{33}$) (b). Gray stars: all the points
satisfying the Yukawa unification condition for the third generation at
$2\sigma$; green diamonds: points additionally requiring $2\sigma$ unification
of the first family; black dots: points for which all the three generations are
unified at $2\sigma$.}
\label{fig:m12m13m23}
\end{figure}
\begin{figure}[t]
\centering
\subfloat[]{
\includegraphics[width=0.41\textwidth]{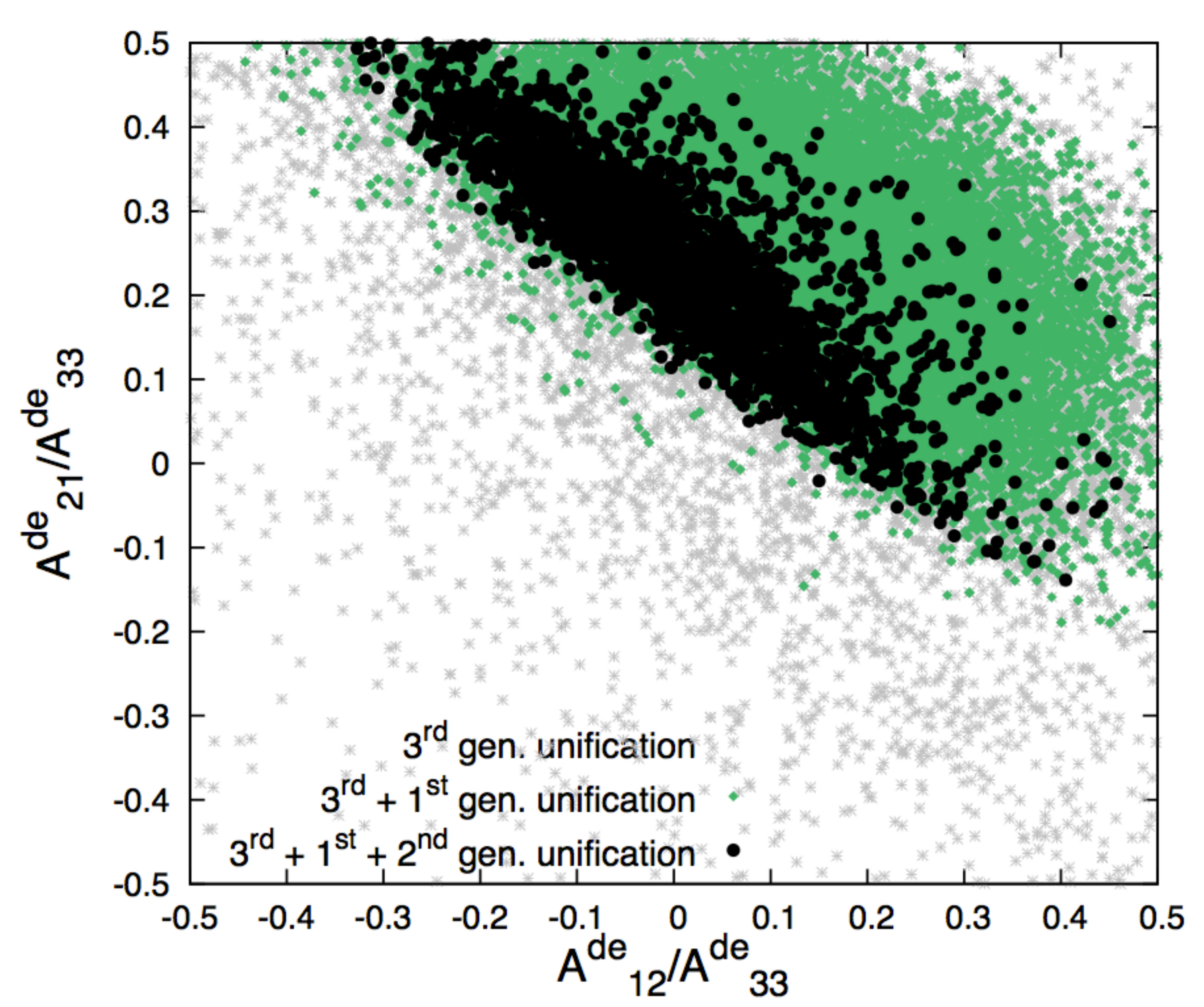}}
\subfloat[]{
\includegraphics[width=0.41\textwidth]{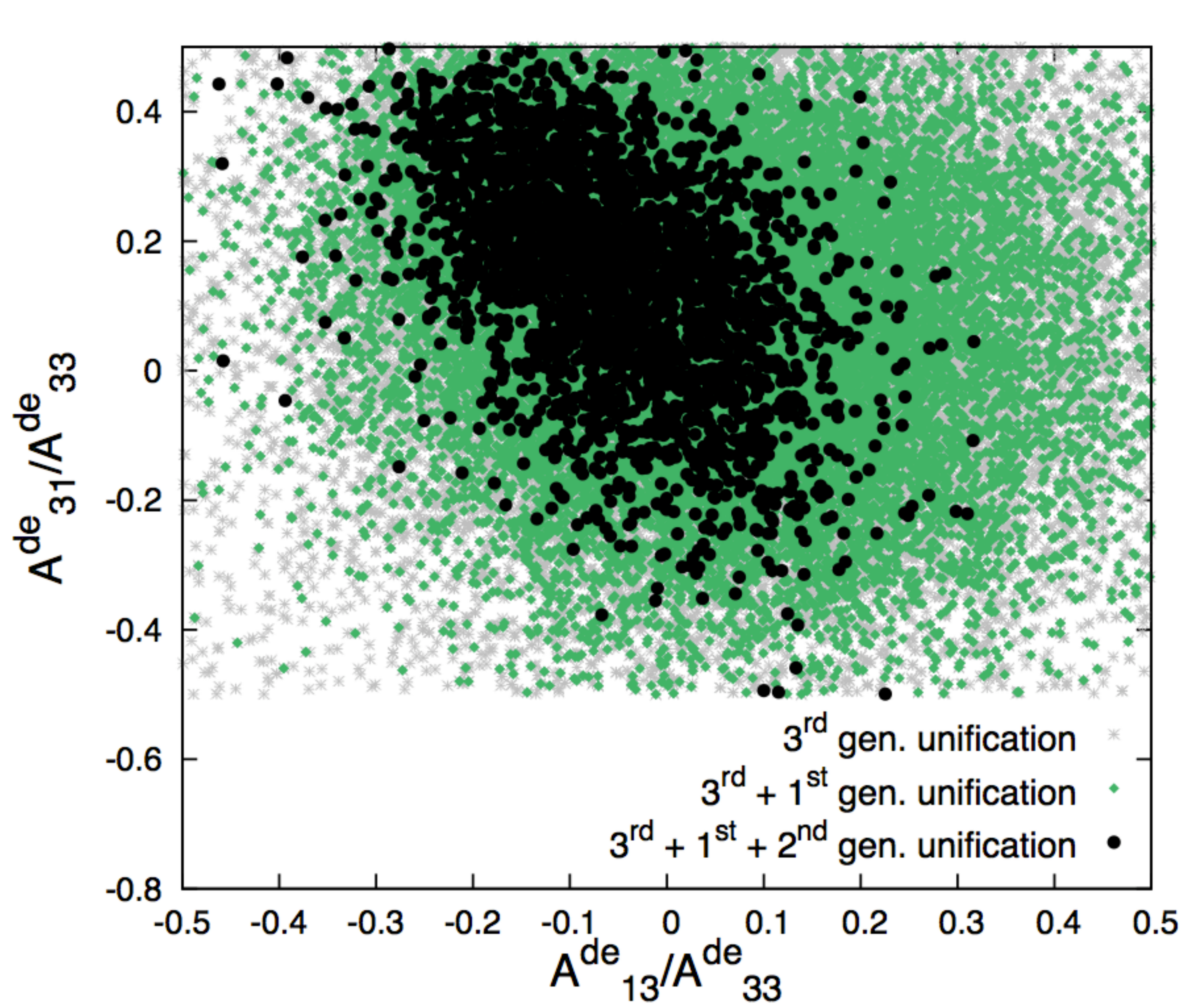}}\\
\subfloat[]{
\includegraphics[width=0.41\textwidth]{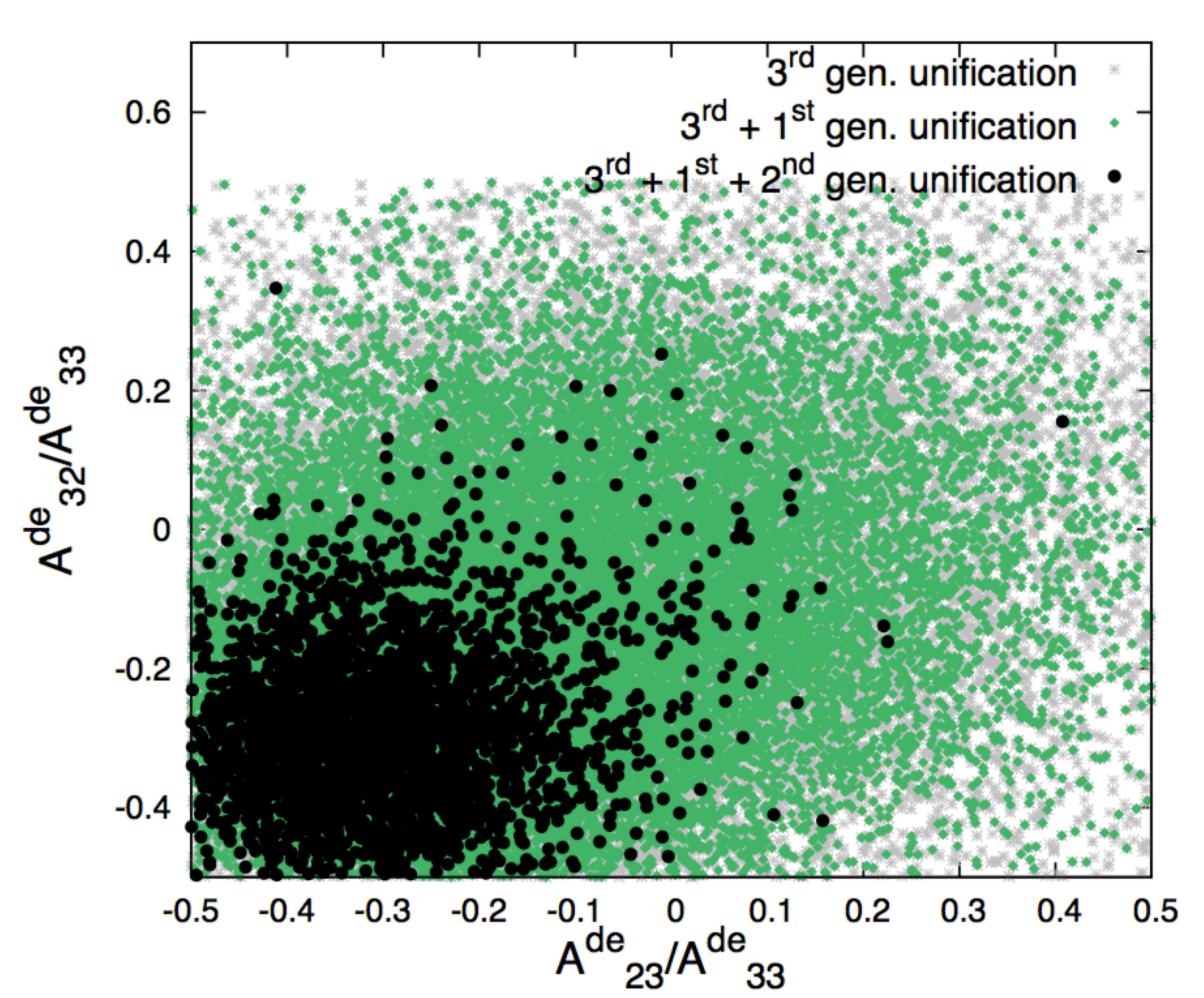}}
\caption{Scatter plot of the $GFV_{123}$ points in the planes
($A^{de}_{12}/A^{de}_{33}$,
$A^{de}_{21}/A^{de}_{33}$) (a),
($A^{de}_{13}/A^{de}_{33}$,
$A^{de}_{31}/A^{de}_{33}$) (b), and
($A^{de}_{23}/A^{de}_{33}$,
$A^{de}_{32}/A^{de}_{33}$) (c). The colour code is the same
as in \reffig{fig:m12m13m23}.}
\label{fig:t12t13t23}
\end{figure}

In \reffig{fig:m12m13m23}, we present distributions of points in the planes
($m^{dl}_{13}/m^{dl}_{33}$,
$m^{dl}_{12}/m^{dl}_{33}$) (a) and
($m^{dl}_{23}/m^{dl}_{33}$,
$m^{dl}_{13}/m^{dl}_{33}$) (b). 
All the points that satisfy the Yukawa unification condition for the third
generation at $2\sigma$ are depicted as gray stars (they account for $49\%$ of
all the points), while those that additionally fulfil $2\sigma$ unification of
the first family as green diamonds ($29\%$ of all the points). Finally, black
dots correspond to those points for which all three generations are unified at
$2\sigma$ ($1.7\%$ of all the points collected by the scan). In
\reffig{fig:t12t13t23}, similar distributions are shown for the
flavour-violating entries of the trilinear down-sector matrix, in the planes
corresponding to
($A^{de}_{12}/A^{de}_{33}$,
$A^{de}_{21}/A^{de}_{33}$) (a),
($A^{de}_{13}/A^{de}_{33}$,
$A^{de}_{31}/A^{de}_{33}$) (b), and
($A^{de}_{23}/A^{de}_{33}$,
$A^{de}_{32}/A^{de}_{33}$) (c).

Several observations can now be made. First of all, it is known that
satisfactory unification of the third family Yukawa couplings is quite easy to
achieve in the Minimal Flavour Violating $SU(5)$ for moderate values of
\tanb. This is confirmed by both \reffig{fig:m12m13m23} and
\reffig{fig:t12t13t23} where gray points can easily be found for vanishing
flavour-violation in the GUT-scale soft parameters.

Secondly, the functional form of the threshold correction in
Eq.~(\ref{sigma22}) might suggest that non-zero soft-mass elements
$m^{dl}_{23}$ and $m^{dl}_{13}$ are sufficient to allow the Yukawa unification
in both the second and first family cases. Such a simplistic picture, however,
is not true, as can be seen from the panel (a) of \reffig{fig:m12m13m23} where
large $m^{dl}_{12}$ is clearly favoured. To understand what happens, let us
note that the GFV corrections $(\Sigma^d_{22})^{\tilde{g}}$ and
$(\Sigma^d_{11})^{\tilde{g}}$ (obtained from Eq.~(\ref{sigma22}) by replacing
indices ``2'' with ``1'') are determined by overlapping sets of parameters, in
particular \mhalf\ and $A_{33}^{de}$. On the other hand, sizes of those
corrections as required by the Yukawa coupling unification differ by two
orders of magnitude. Let us now assume that $(\Sigma^d_{11})^{\tilde{g}}$ is
fixed by the unification condition for the first family. Thus \mhalf\ and
$A_{33}^{de}$, already constrained by unification of the third family, are
even more limited. With such a choice of parameters, however, the correction
$(\Sigma^d_{22})^{\tilde{g}}$ is still too small to allow unification of the
second family, and needs to be further enhanced by another contribution. Such
a contribution comes from a diagram like the one shown in Fig.~\ref{diags}(b),
but with the trilinear term $A_{21}^{de}$ in the vertex and $m^{dl}_{12}$
mixing in the right-handed sector. However, a similar diagram also exists for
the first family, and the corresponding contribution should be added to the
one driven by $m^{dl}_{13}$. That explains why all the five parameters
$m^{dl}_{12}$, $m^{dl}_{13}$, $m^{dl}_{23}$, $A_{12}^{de}$ and $A_{21}^{de}$
must be adjusted simultaneously. Note also that $A_{12/21}^{de}$ can be kept
relatively low, as this contribution is always enhanced by a large value of
$m^{dl}_{12}$.
\begin{figure}[t]
\centering
\subfloat[]{
\includegraphics[width=0.41\textwidth]{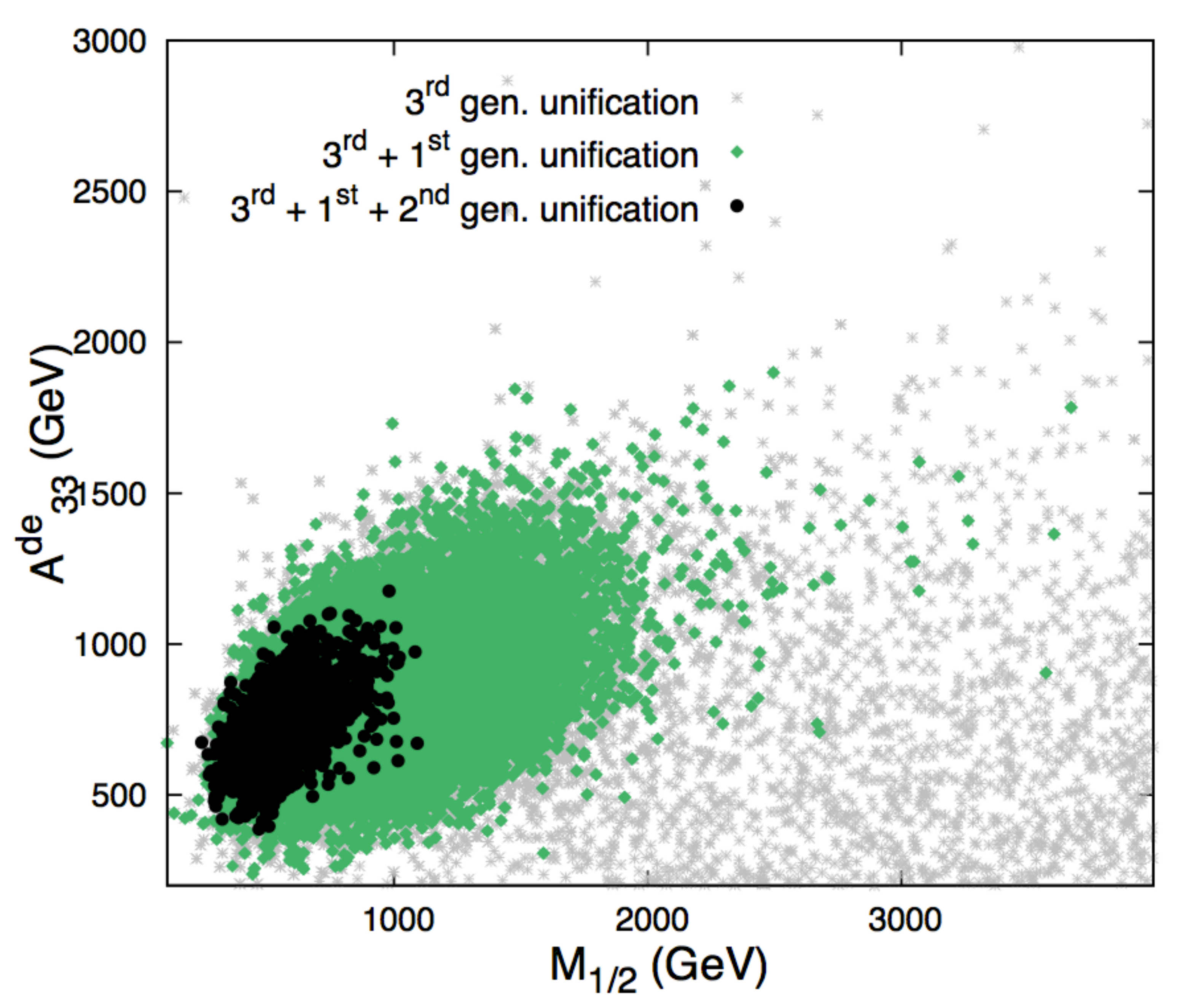}}
\subfloat[]{
\includegraphics[width=0.41\textwidth]{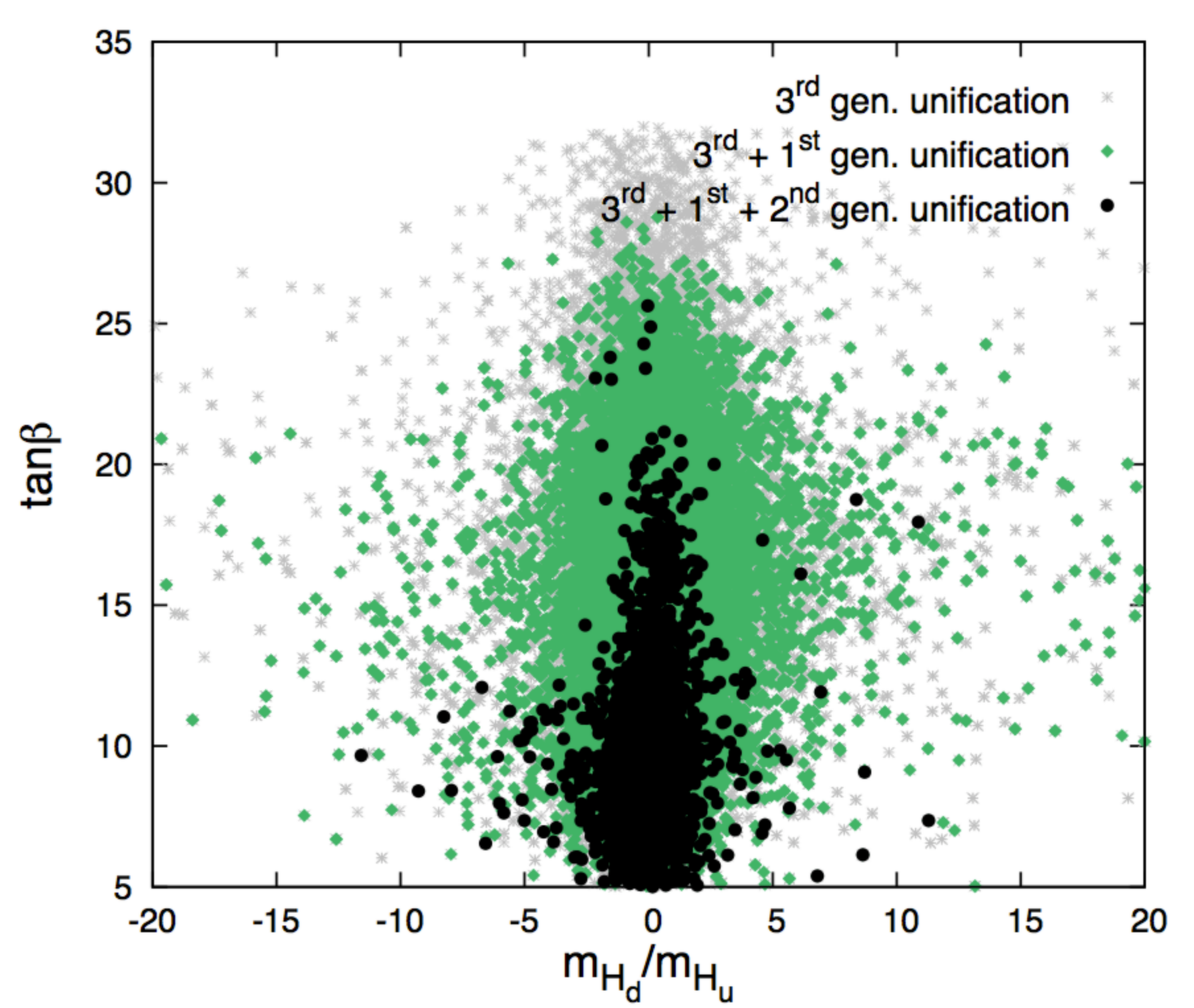}}\\
\subfloat[]{
\includegraphics[width=0.41\textwidth]{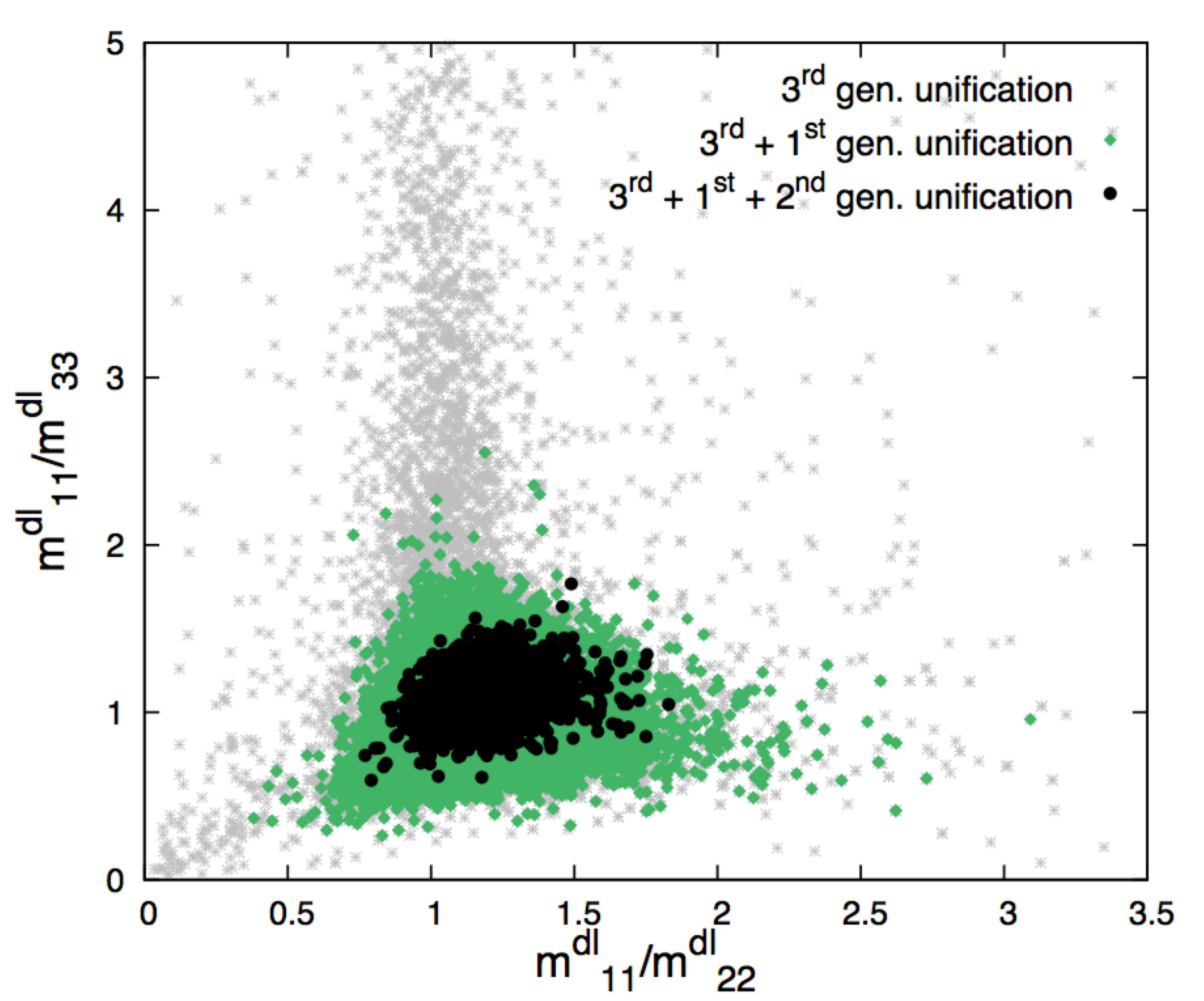}}
\caption{Scatter plot of the $GFV_{123}$ points in the planes
($\mhalf$, $A^d_{33}$) (a), ($\mhd/\mhu$, $\tanb$) (b), and
($m^{dl}_{11}/m^{dl}_{22}$,
$m^{dl}_{11}/m^{dl}_{33}$) (c). The colour code is the same
as in \reffig{fig:m12m13m23}.}
\label{fig:m12a33tanb}
\end{figure}

Flavour-violating parameters are not the only ones constrained by the Yukawa
unification condition. In \reffig{fig:m12a33tanb}, we present distributions of
points in the planes ($\mhalf$, $A^{de}_{33}$) (a), ($\mhd/\mhu$, $\tanb$)
(b), and ($m^{dl}_{11}/m^{dl}_{22}$, $m^{dl}_{11}/m^{dl}_{33}$) (c). The
colour code is the same as in \reffig{fig:m12m13m23}. One can observe that
values of both \mhalf\ and $A^d_{33}$ need to be very limited in order to
facilitate unification in the first and second family cases, as they directly
enter Eq.~(\ref{sigma22}). The ratio $\mhd/\mhu$ in the range $[0-2]$ allows
the unification of the second family for larger values of \tanb,
$\tanb\in[15-25]$. Finally, large mass splittings between the diagonal entries
of the down-squark mass matrix are disfavoured because they would lead to a
strong suppression of SUSY threshold corrections, as can be deduced from
Eq.~(\ref{sigma22}).

We conclude this section with summarising the allowed ranges of the non-zero
GFV parameters that characterise the $SU(5)$ GUT scenario with the full Yukawa
coupling unification:
\bea
0.5<m^{dl}_{23}/m^{dl}_{33}<1,&\quad 0 < m^{dl}_{13}/m^{dl}_{33} < 0.5,&\quad 0.3 < m^{dl}_{12}/m^{dl}_{33} < 0.7, \nonumber
\eea
\bea
0 <A^{d}_{12}/A^{d}_{33} < 0.2, &\quad 0 <A^{d}_{21}/A^{d}_{33} < 0.2.
\eea

\section{Phenomenology of the unification scenarios}\label{pheno}

So far we have discussed the possibility of satisfying the $SU(5)$ boundary
conditions for the Yukawa couplings by allowing a non-trivial flavour
structure of the soft terms at the \gut\ scale. In this section, we will
discuss compatibility of the results obtained in \refsec{sec:impact} with
various phenomenological constraints. First, we are going to consider a set of
experimental measurements from the dark matter searches, Higgs physics,
flavour physics and EW precision measurements. Next, theoretical limits
provided by the EW vacuum stability requirement will be taken into account. 
In this exposition, we do not aim at finding the best possible fit of the MSSM
to the experimental data. Rather, we show that the Yukawa unification
requirement does not contradict consistency of the MSSM with observations.

\subsection{Experimental constraints}

In order to find points satisfying both the Yukawa coupling unification
conditions at the GUT scale and the experimental constraints, we use the tools
described in \refsec{sec:tool}, and scan the parameter space of the
unification scenarios given in Table~\ref{tab:priorsYuk_pheno}. The
experimental constraints applied in the analysis are listed in
Table~\ref{tab:exp_constraints}.
\begin{table}[t]
\centering
\renewcommand{\arraystretch}{1.1}
\begin{tabular}{|c|c|c|}
\hline 
Parameter &  Scanning Ranges: $GFV_{23}$ &  Scanning Ranges: $GFV_{123}$\\
\hline 
$M_{1/2}$	& [$100$, $2000$] \gev\ & [$200$, $1100$] \gev\ \\
\mhu, \mhd  & [$100$, $8000$] \gev\   & [$100$, $8000$] \gev\ \\
\tanb	    & [$3$, $35$] & [$3$, $25$] \\
\signmu		& $-1$ & $-1$ \\
\hline 
$A_{33}^{de}$ & [$0$, $5000$] \gev\  & [$400$, $1100$] \gev\ \\
$A_{33}^{u}$    & [$-9000$, $9000$] \gev\ & [$-9000$, $9000$] \gev\ \\
$A^{de}_{11}/A^{de}_{33}$   & [$-0.00028$, $0.00028$]  & [$-0.00028$, $0.00028$]\\
$A^{de}_{22}/A^{de}_{33}$  & [$-0.065$, $0.065$]  & [$-0.065$, $0.065$] \\
$A^{u}_{22}/A^{u}_{33}$    & [$-0.005$, $0.005$]  & [$-0.005$, $0.005$]\\
$A^{de}_{12}/A^{de}_{33}$,\; $A^{de}_{21}/A^{de}_{33}$ & $0$ & [$-0.2$, $0.2$] \\
\hline 
$m_{ii}^{dl},\; i=1,2,3$   & [$100$, $10000$] \gev\ & [$100$, $7000$] \gev\ \\
$m^{dl}_{23}/m^{dl}_{33}$  & [$0.0$, $0.4$] & [$0.5$, $1.0$] \\
$m^{dl}_{13}/m^{dl}_{33}$  & $0$ & [$0.0$, $0.5$] \\
$m^{dl}_{12}/m^{dl}_{33}$  & $0$ & [$0.3$, $0.7$] \\
$m_{ii}^{ue},\; i=1,2,3$   & [$100$, $7000$] \gev\ & [$100$, $7000$] \gev\ \\
\hline 
\end{tabular}
\caption{Ranges of the input SUSY parameters in the GFV $SU(5)$ scenario with
Yukawa coupling unification. Several parameters that are not explicitly listed in
the table (namely $A^{u}_{11}$, $A^{u}_{ij}$ and $m^{ue}_{ij}$ for $i\neq j$, $A^{de}_{23/32}$,
$A^{de}_{13/31}$) have been set to zero.}
\label{tab:priorsYuk_pheno}
\end{table}

All the flavour observables have been evaluated with the code
\texttt{\susyflav\ v2.10}\cite{Crivellin:2012jv}. It calculates the
renormalized Yukawa matrices in the MSSM and derives the correct CKM matrix as
described in Ref.\cite{Crivellin:2011jt}. Apart from the $B$-meson mass
differences $\Delta M_{B_d}$ and $\Delta M_{B_s}$, we also consider their
ratio $\Delta M_{B_d}/\Delta M_{B_s}$ for which more precise lattice inputs
are available\cite{Aoki:2013ldr,Amhis:2014hma}. Theory uncertainties in the
cases of $\brbxsgamma$, ${\rm BR}(B_{s(d)} \to \mu^+ \mu^-)$ and
$|\epsilon_K|$ have been estimated following Refs.\cite{Misiak:2006zs},
\cite{Bobeth:2013uxa}, and\cite{Brod:2011ty}, respectively.  In the latter
case, they arise mainly from poor convergence of the QCD perturbation series
for double charm contributions. Errors stemming from the Wolfenstein
parameters $\rho$ and $\eta$ are not taken into account in the third column of
Table~\ref{tab:exp_constraints}. Each point of our scan corresponds to a
particular value of $(\rho,\eta)$, and it is tested against SUSY-sensitive
loop observables that matter for determining allowed regions in the
$(\rho,\eta)$ plane. At the same time, the considered values of all the CKM
parameters are within the ranges allowed by tree-level observables. We use the
same code to determine the values of the Lepton Flavour Violating observables.

The relic density and the spin-independent neutralino-proton cross section
\sigsip\ have been calculated with the help of \texttt{\dsusy\
v5.0.6}\cite{Gondolo:2004sc}. For the EW precision constraints
\texttt{\feynhiggs\
v2.10.0}\cite{Hahn:2013ria,Frank:2006yh,Degrassi:2002fi,Heinemeyer:1998yj} has
been used. To include the exclusion limits from Higgs boson searches at LEP,
Tevatron, and LHC, as well as the \chisq\ contributions from the Higgs boson
signal rates from Tevatron and LHC, we applied \texttt{\higgsbounds\
v4.0.0}\cite{Bechtle:2008jh,Bechtle:2011sb,Bechtle:2013wla} interfaced with
\texttt{\higgssignals\ v1.0.0}\cite{Bechtle:2013xfa}.
\begin{table}[t]
\begin{center}
\renewcommand{\arraystretch}{1.1}
\begin{tabular}{|l|l|l|l|l|l|}
\hline
Measurement & Mean or range & Error [~exp.,~th.] & Reference\\
\hline
\abundchi      & $0.1199$ 	& [$0.0027$,~$10\%$]		& \cite{Ade:2013zuv}\\
LUX (2013) & See Sec.~3 of\cite{Kowalska:2014hza} 	& See Sec.~3 of\cite{Kowalska:2014hza} & \cite{Akerib:2013tjd}\\ 
\hline
\mhl\ (by CMS) & $125.7\gev$ & [$0.4$, $3.0$] \gev & \cite{CMS:yva} \\
LHC \eight\ & \refsec{sec:lhc} & \refsec{sec:lhc} & \cite{CMS-PAS-SUS-13-018,Aad:2014wea,Aad:2014vma}\\
\hline
\sinsqeff 			& $0.23155$     & [$0.00012$, $0.00015$] &  \cite{Agashe:2014kda}\\
$M_W$                     	& $80.385\gev$  & [$0.015$, $0.015$] \gev &  \cite{Agashe:2014kda}\\
\hline
\brbxsgamma $\times 10^{4}$ & $3.43$   	& [$0.22$, $0.23$] & \cite{Amhis:2014hma} \\ 
$\brbsmumu\times 10^9$	  & $2.8$ & [$0.7$, $0.23$] & \cite{CMS:2014xfa}\\ 
$\brbdmumu\times 10^{10}$ & $3.9$ & [$1.6$, $0.09$]  & \cite{CMS:2014xfa}\\ 
$\Delta M_{B_s}\times 10^{11}$ & $1.1691\gev$ & [$0.0014$, $0.1580$] \gev & \cite{Agashe:2014kda}\\
$\Delta M_{B_d}\times 10^{13}$ & $3.357\gev$ & [$0.033$, $0.340$] \gev & \cite{Agashe:2014kda} \\
$\Delta M_{B_d}/\Delta M_{B_s}\times 10^{2}$ & $2.87$ & [$0.02$, $0.14$] & \cite{Amhis:2014hma}\\ 
$\sin(2\beta)_{\rm exp}$ & $0.682$ & [$0.019$, $0.003$] & \cite{Agashe:2014kda}\\ 
\brbutaunu $\times 10^{4}$  & $1.14$ & [$0.27$, $0.07$] & \cite{Agashe:2014kda}\\ 
$\textrm{BR}(K^+ \to \pi^+ \nu \bar{\nu})\times 10^{10}$  &$1.73$  
            & [$1.15, 0.04$] & \cite{Agashe:2014kda}\\ 
$\epsilon_K\times 10^3$  & $2.228$ & [$0.011$, $0.17$] &  \cite{Agashe:2014kda}  \\
\hline
$|d_n|\times 10^{26}$  & < $2.9$ $e$ cm & [$0$, $30\%$] & \cite{Baker:2006ts} \\
\hline
\mueg $\times 10^{13}$ & < $5.7$ & [0,0] & \cite{Adam:2013mnn} \\
\taueg $\times 10^{8}$ & < $3.3$ & [0,0] & \cite{Aubert:2009ag} \\
\taumug $\times 10^{8}$ & < $4.4$ & [0,0] & \cite{Aubert:2009ag} \\
\mue $\times 10^{12}$ & < $1.0$ & [0,0] & \cite{Bellgardt:1987du} \\
\taue $\times 10^{8}$ & < $2.7$ & [0,0] & \cite{Hayasaka:2010np} \\
\taumu $\times 10^{8}$ & < $2.1$ & [0,0] & \cite{Hayasaka:2010np} \\

\hline
\end{tabular}
\caption{
The experimental constraints applied in the analysis.} 
\label{tab:exp_constraints}
\end{center}
\end{table}

In all the cases, the likelihood for positive experimental measurements was
modeled with Gaussian distribution, while for upper bounds -- with
half-Gaussian one. The likelihood relative to the LUX
results\cite{Akerib:2013tjd} was calculated by closely following the procedure
first developed in Ref.\cite{Cheung:2012xb} and described in Sec.~3 of
Ref.\cite{Kowalska:2014hza}. The likelihood for the LHC direct SUSY searches
was not included in the global likelihood used by \texttt{\multinest}. Our
treatment of this constraint will be described in Sec.~\ref{sec:lhc}.  We
collected in total about $2.1\times10^5$ points.  

\subsection{$\boldsymbol{GFV_{23}}$: unification of the third and second family}\label{phenoGFV23}

\subsubsection{Dark matter}\label{sec:dm}

Relic abundance of the dark matter in GUT-constrained SUSY scenarios is
usually the most stringent constraint. It is well known that properties of the
DM candidate strongly depend on its composition. In the almost purely bino
case, the relic density is generally too large, and the annihilation
cross-section needs to be enhanced by some mechanism, usually by
co-annihilation with the lightest sfermion, or resonance annihilation through
one of the Higgs bosons. On the other hand, a significant higgsino component
of the lightest neutralino opens a possibility of efficient annihilation into
gauge bosons through a $t$-channel exchange of higgsino-like \charone\ and
\neuttwo. In fact, the annihilation cross-section in such a case is usually
too large, which leads to DM underabundance.

\begin{figure}[t]
\centering
\subfloat[]{
\includegraphics[width=0.41\textwidth]{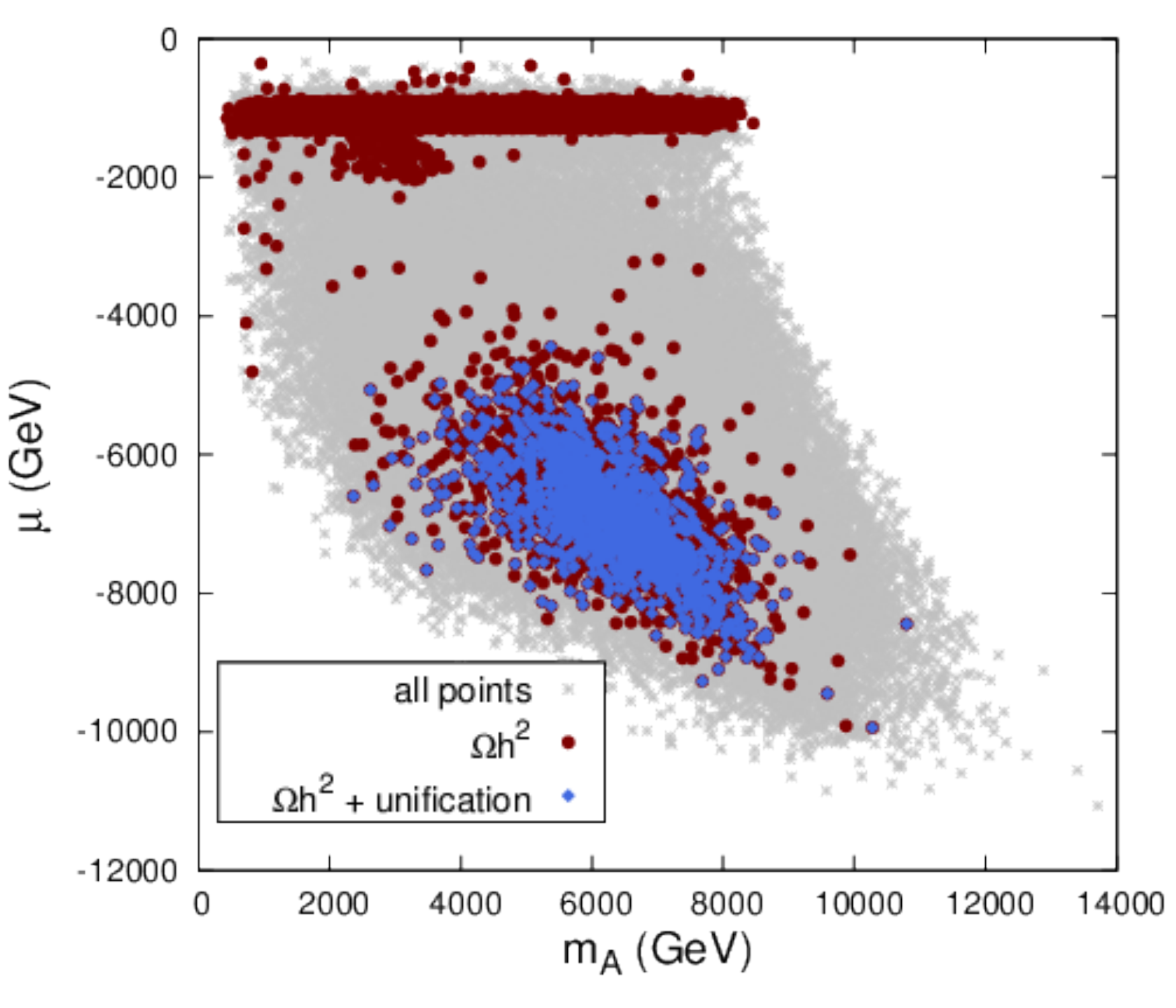}}
\subfloat[]{
\includegraphics[width=0.41\textwidth]{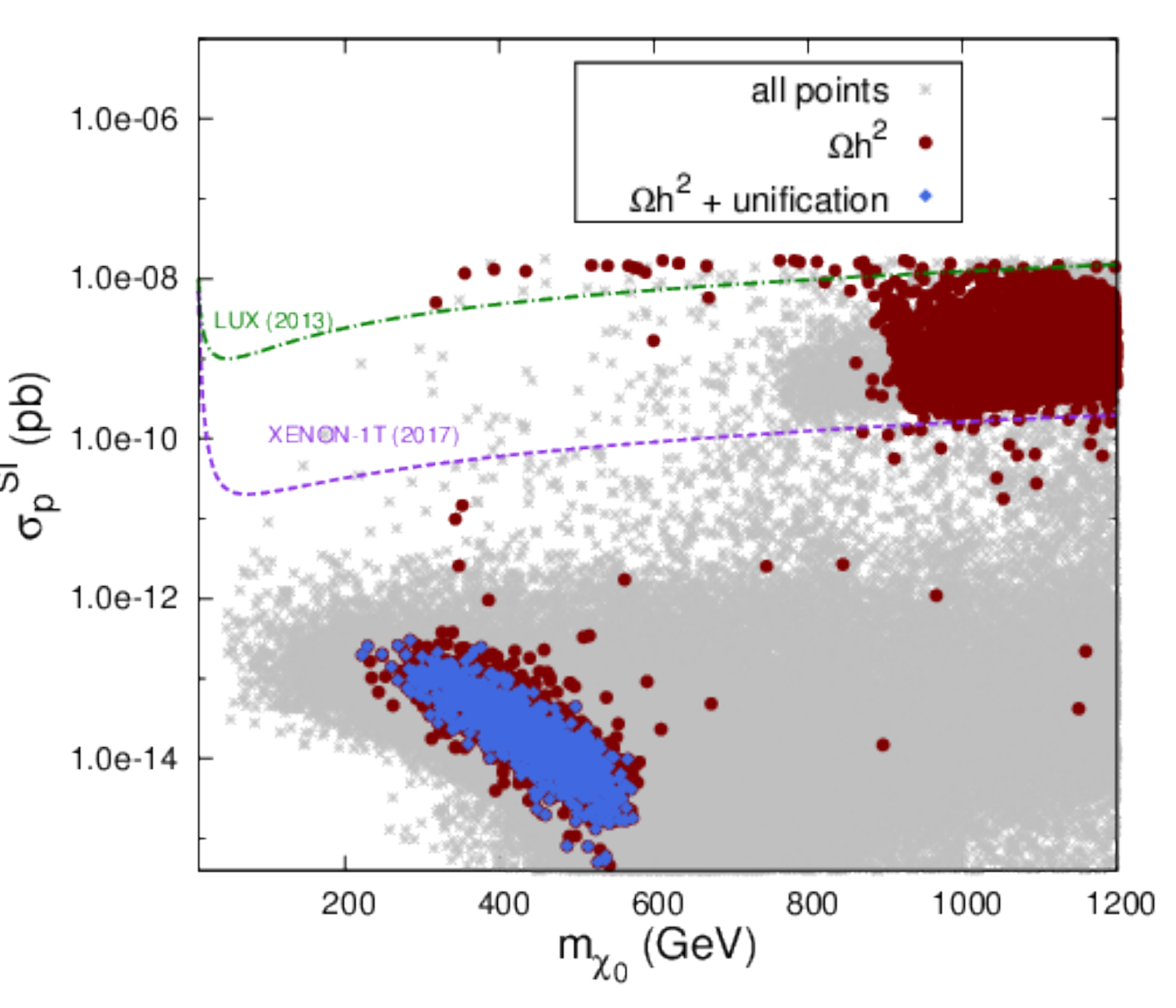}}
\caption{Scatter plot of the $GFV_{23}$ points in the planes ($\ma$, $\mu$)
(a) and ($m_{\neutone}$, \sigsip) (b). Gray stars: all
points collected by the scan; brown dots: points satisfying the DM relic
density constraint at $3\sigma$; blue diamonds: points for which additionally
the Yukawa unification of the second an third family holds. The meaning of dashed lines is described in the
text.}
\label{pheno:CDM_111}
\end{figure}

In \reffig{pheno:CDM_111}, we present distributions of points found by our
scanning procedure in the planes ($\ma$, $\mu$) (a) and ($m_{\neutone}$,
\sigsip) (b). All the points collected by the scan are depicted as gray
stars, while those that satisfy at $3\sigma$ the experimental constraint on
the DM relic density appear as brown dots. Blue diamonds correspond to those
scenarios for which additionally the Yukawa coupling unification holds. The
green dashed line indicates the $90\%$ \cl\ exclusion bound on the \sigsip\
based on the 85-day measurement by the LUX collaboration\cite{Akerib:2013tjd},
while the purple dashed line is a projection of XENON1T
sensitivity\cite{Aprile:2012zx}. By comparing panels (a) and (b) of
\reffig{pheno:CDM_111} one can see that in the region where Yukawa unification
is achieved, the neutralino LSP is bino-like, which corresponds to a
relatively low spin-independent proton-neutralino cross-section. In other
words, the condition of Yukawa coupling unification strongly disfavours purely
or partly higgsino-like neutralino. This is due to the fact that only the
points with $|\mu| \lsim 1\tev$ in \reffig{pheno:CDM_111}(a) correspond to
a significant higgsino component of the LSP. In such a case, the
$\mu$-dependent contribution in Eq.~(\ref{sigma22}) is too small to allow for
unification of the second family Yukawa couplings. Thus, it is an
important phenomenological signature of the GFV $SU(5)$ Yukawa unification
scenario with the universal gaugino masses that only bino-like neutralino is
allowed.

A unique mechanism that makes the effective increase of the DM annihilation
cross-section possible in this case is the neutralino co-annihilation with the
lightest sneutrino. The pseudoscalar is too heavy to allow for resonant
\neutone\ annihilation (as can be read from the panel (a) of
\reffig{pheno:CDM_111}), while the masses of the coloured sfermions in the
GUT-constrained unification scenarios are always larger than those of the
sleptons. It is due to a renormalization effect, as their RGE running is
strongly driven by the gluino. Such a property of the spectrum, however, has
important consequences for experimental testability. In
\reffig{pheno:CDM_111}(b) the dashed lines indicate the present reach and the
expected sensitivity for LUX and XENON1T. The region favoured by the relic
density constraint in the Yukawa unification scenario remains far beyond the
reach for any of them. On the other hand, spectra characterized by the
presence of a light bino-like \neutone\ and a sneutrino close in mass are
already being tested by the LHC \eight, as will be discussed in
\refsec{sec:lhc}. It is an interesting aspect of the $GFV_{23}$ Yukawa
unification scenario that the requirement of satisfying the DM relic density
constraint makes it testable by the LHC \four\ run.

\begin{figure}[p]
\centering
\subfloat[]{
\includegraphics[width=0.41\textwidth]{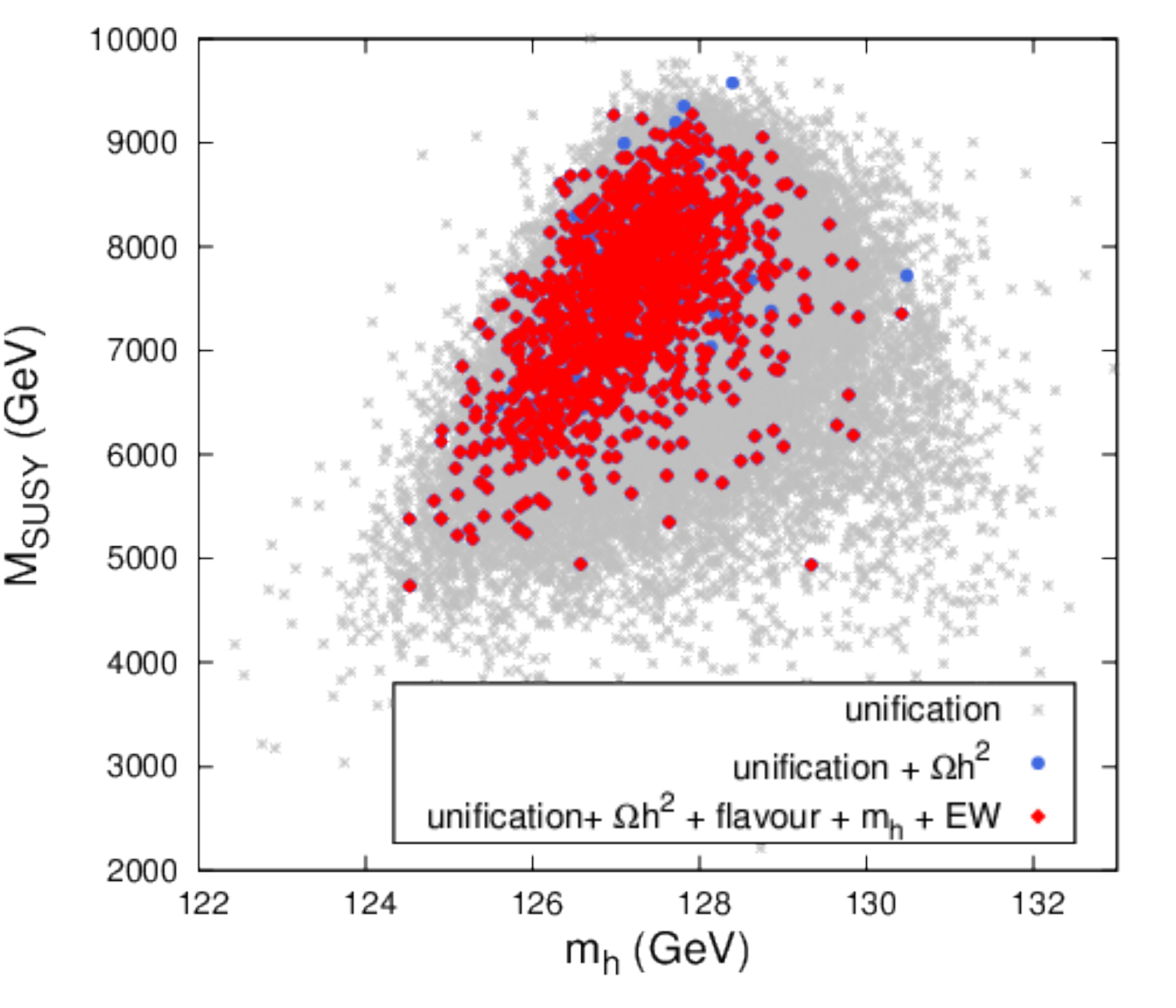}}
\subfloat[]{
\includegraphics[width=0.41\textwidth]{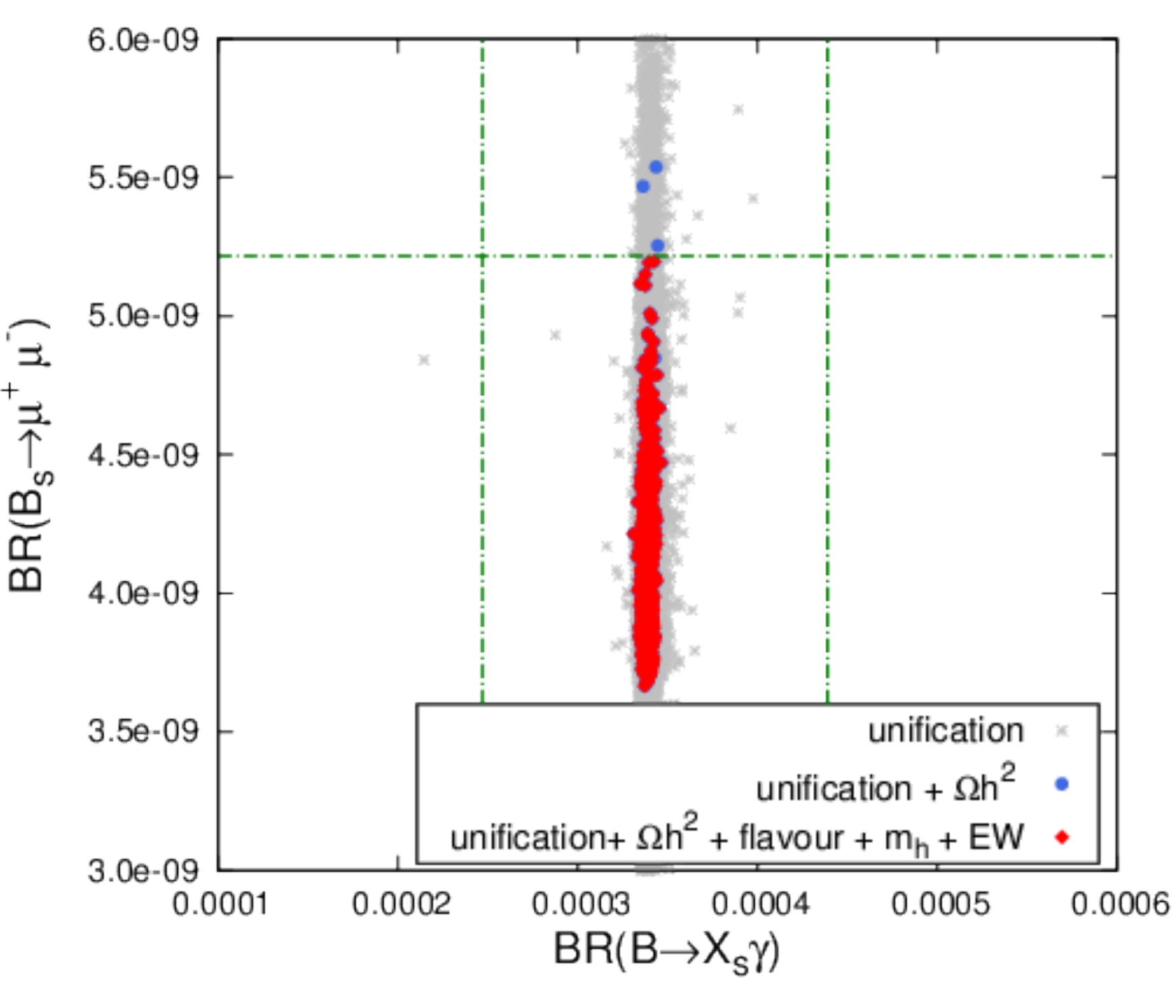}}\\
\subfloat[]{
\includegraphics[width=0.41\textwidth]{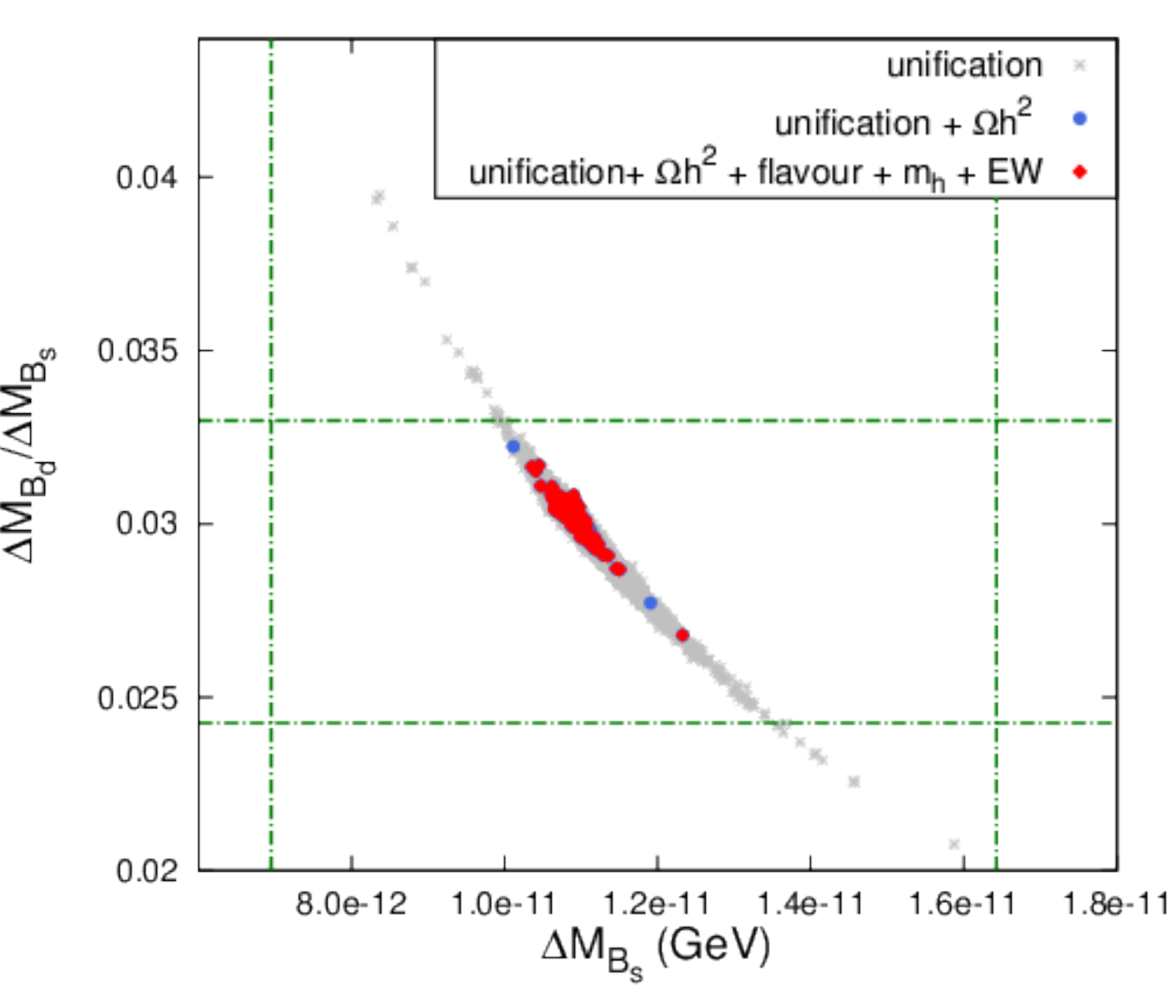}}
\subfloat[]{
\includegraphics[width=0.41\textwidth]{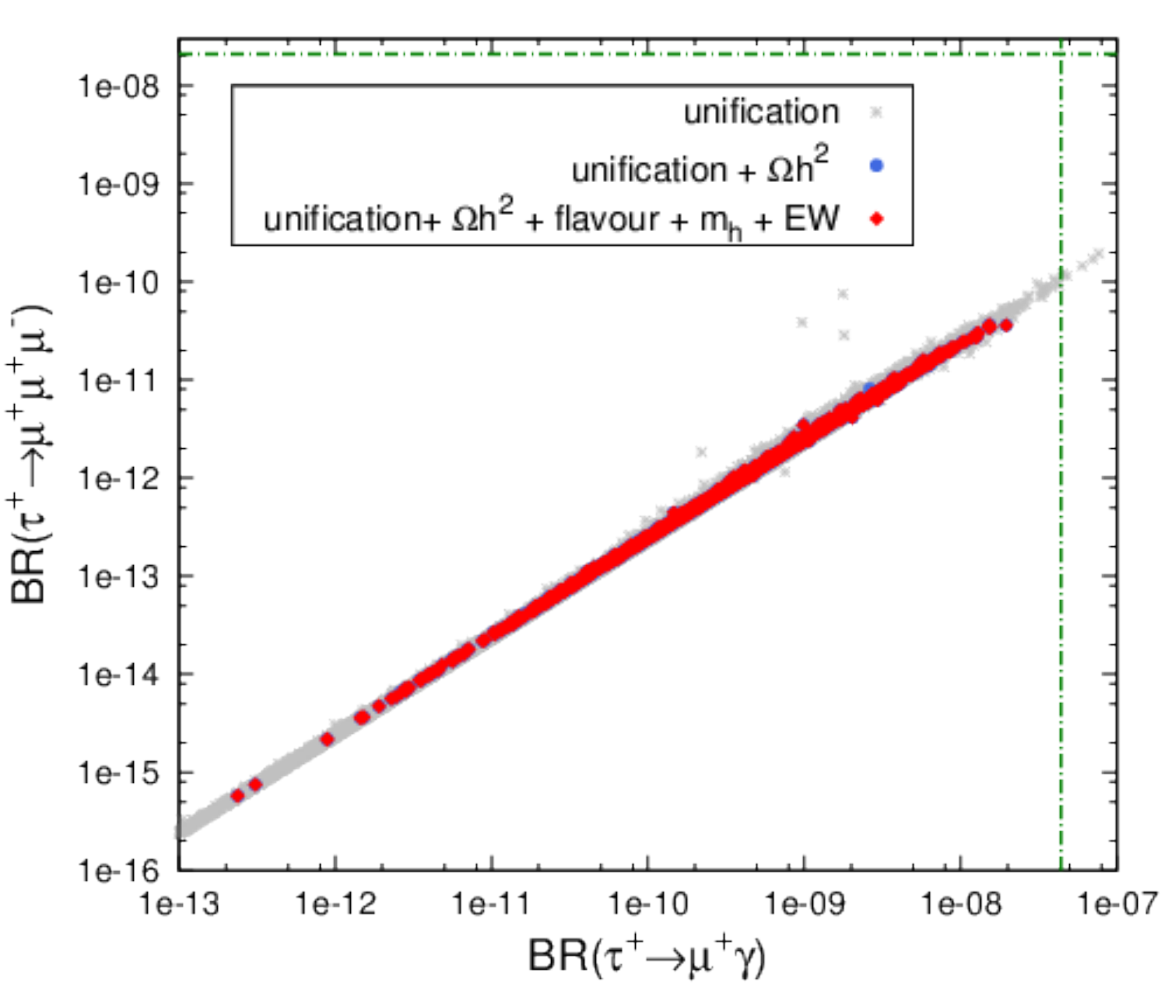}}
\caption{Scatter plot of the $GFV_{23}$ points in the planes (\mhl, \msusy)
(a), (\brbxsgamma, \brbsmumu) (b), ($\Delta M_{B_s}$, $\Delta M_{B_d}/\Delta
M_{B_s}$) (c), and (\taumug, \taumu) (d). Gray stars: all the points that allow for the Yukawa coupling
unification; blue dots: points satisfying additionally the DM relic density
constraint at $3\sigma$; red diamonds: points with good Yukawa coupling
unification and all the constraints listed in Table~\ref{tab:exp_constraints}
satisfied at $3\sigma$ (except the LHC). Dashed lines correspond to $3\sigma$
experimental limits on the corresponding observables.}
\label{flavour}
\end{figure}

\subsubsection{Higgs, flavour and electroweak observables}\label{sec:flav}

In \reffig{flavour}, we present distributions of points for several relevant
observables: (\mhl, \msusy) (a), (\brbxsgamma, \brbsmumu) (b),($\Delta
M_{B_s}$, $\Delta M_{B_d}/\Delta M_{B_s}$) (c), and (\taumug, \taumu)
(d). Gray stars indicate all the points for which it is possible to achieve
the Yukawa coupling unification for the third and second generations. Points
that satisfy the relic density constraint at $3\sigma$ are shown as blue dots,
while red diamonds correspond to those cases where additionally all the other
constraints listed in Table~\ref{tab:exp_constraints} are met at $3\sigma$
(except the LHC bounds from direct SUSY searches that will be discussed in
\refsec{sec:lhc}).

The Higgs boson mass dependence on the GFV parameters has been discussed in
Ref.\cite{Cao:2006xb,AranaCatania:2011ak,Arana-Catania:2014ooa,Kowalska:2014opa}. It
was shown that while \mhl\ can be enhanced by non-zero $(2,3)$ entries of the
trilinear down-squark matrix, its dependence on the off-diagonal soft-mass
elements is negligible. Therefore, in the scenario considered in this study,
the only parameters relevant for the Higgs physics remain $A^u_{33}$ and
\msusy. That is confirmed by the panel (a) of \reffig{flavour} where no
tension between the correct value of the Higgs boson mass and the Yukawa
unification constraint is observed. The EW observables are not affected either
because the dominant GFV contribution to $M_W$ and $\sinsqeff$ would be driven
by the element $m^{ue}_{23}$\cite{Heinemeyer:2004by}.

On the other hand, the presence of sizeable off-diagonal entries in the squark
mass matrices might lead to disastrously high SUSY contributions to FCNC
processes. It turns out, however, that in the considered scenario most of the
flavour constraints in the quark sector are quite easily satisfied for the
points that have survived imposing the DM experimental limit.  This is mainly
due to the fact that the coloured sfermions, in particular those of the third
generation, are relatively heavy in our setup, while \tanb\ needs to be low or
moderate in order to facilitate the Yukawa coupling unification of the third
and second family.

In consequence, SUSY contributions to the FCNC processes in the quark
sector are suppressed in our $GFV_{23}$ Yukawa unification
scenario. Only $\Delta M_{B_d}$, $\Delta M_{B_s}$, \brbsmumu{} and
\brbxsgamma{} vary non-negligibly compared to the present bounds. Note also
that even if the theoretical uncertainties in determination of $\Delta
M_{B_s}$ and $\Delta M_{B_d}/\Delta M_{B_s}$, as well as the experimental error
in \brbsmumu{} were further reduced, no tension would arise,
as the accepted points in \reffig{flavour}(c) are distributed quite uniformly
over the $3\sigma$ region.

A potential threat to the $GFV_{23}$ scenario is posed by the LFV observables
that severely constrain any non-zero mixing among
sleptons\cite{Arana-Catania:2013nha}. However, in the case of $m^{dl}_{23}$,
the current constraints on the relevant processes are still easily satisfied,
as can be seen in panel (d) of \reffig{flavour}.

\subsubsection{LHC direct SUSY searches}\label{sec:lhc}

\begin{figure}[t]
\centering
\includegraphics[width=0.65\textwidth]{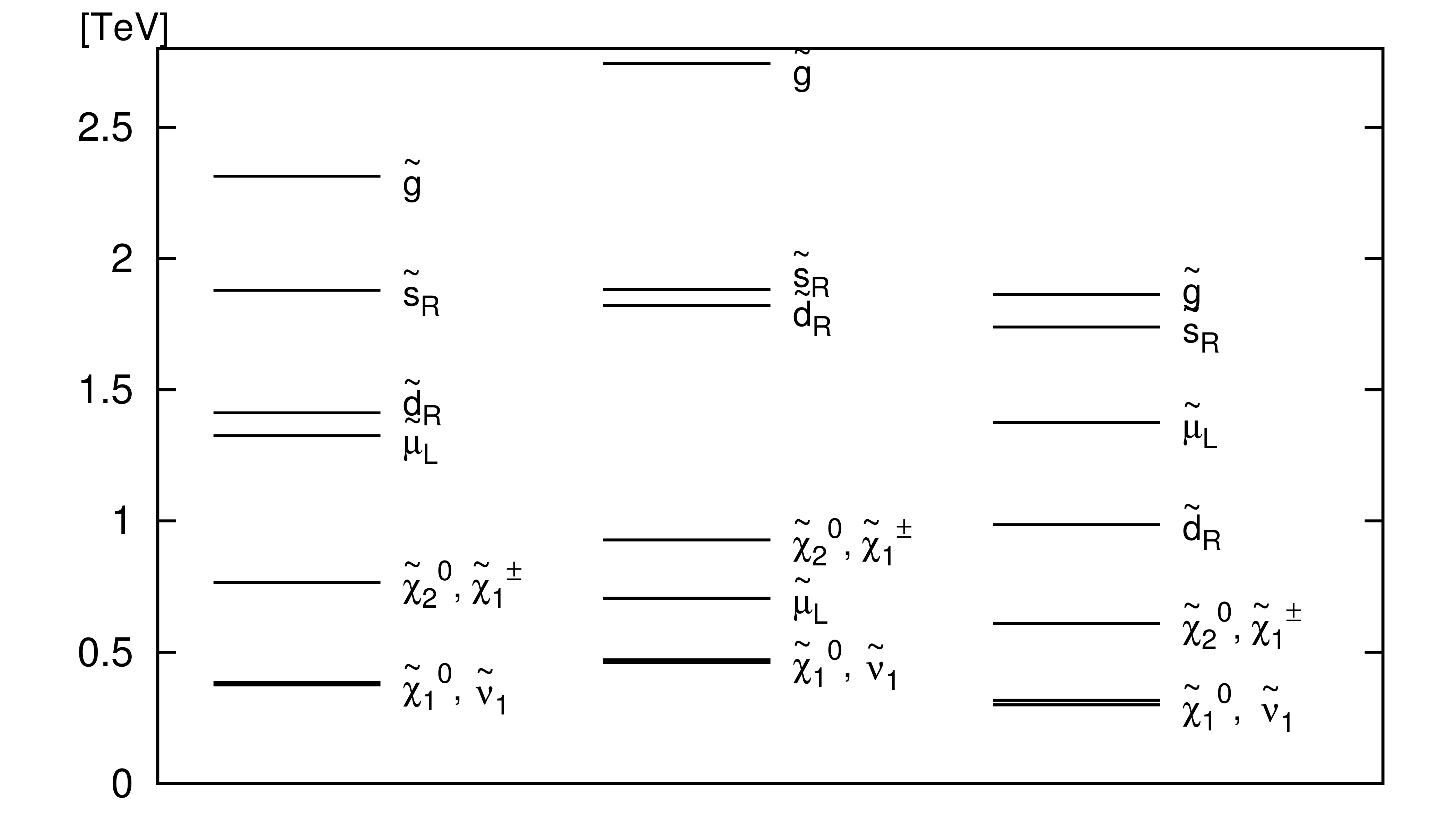}
\caption{Examples of spectra characteristic of the $GFV_{23}$ Yukawa unification scenario.}
\label{spectrum}
\end{figure}

All the points from our scan that demonstrate the Yukawa unification and
satisfy the experimental constraints listed in Table~\ref{tab:exp_constraints}
share the same characteristic pattern of the light part of the spectrum.
Three examples are shown in \reffig{spectrum}.  The Next-to-Lightest SUSY
particle (NLSP) is the lightest sneutrino, while one charged slepton,
neutralino \neuttwo\ and chargino \charone\ are slightly heavier. The presence
of light sleptons in the spectrum is very important as it leads to a
characteristic 3-lepton signature at the LHC. The next particle on the mass
ladder is the lightest down-type squark which is followed by the gluino. All
the other coloured particles, the remaining sleptons and heavy Higgses are
much heavier and effectively decoupled. Comparison of the production
cross-sections reveals that the dominant production channels at the LHC would
be direct $\tilde{d}_1\tilde{d}_1$, \charone\neuttwo\ and \charone\charone\
production. The cross-section for gluino pair production is one order of
magnitude lower, and practically all of the produced gluinos decay via
$\gluino\to\tilde{d}_1 q$. Therefore, the Yukawa unification scenario is
characterized by two distinct LHC signatures: 3 leptons plus missing energy
(MET) signature for the electroweakino production, and 0 leptons, jets plus
MET signature for the coloured particles production.

The strongest limit on the gluino mass comes from the ATLAS 0 lepton, 2-6 jets
plus MET inclusive search\cite{Aad:2014wea} that provides a stringent 95\%
\cl\ exclusion bound $m_{\gluino} \gtrsim 1400\gev$ for neutralino LSP lighter
than $300\gev$. The strongest bound on the lightest sbottom mass comes from
the CMS 0 leptons, 2 jets and MET search\cite{CMS-PAS-SUS-13-018}, while in
the electroweakino sector the most stringent experimental exclusion limits are
obtained using the 3-lepton plus MET CMS search\cite{Aad:2014vma}.

However, one needs to keep in mind that the bounds provided by the
experimental collaborations are interpreted in the Simplified Model Scenarios
(SMS) that make strong assumptions about the hierarchy of the spectrum and the
decay branching ratios. Usually it is assumed that there is only one light
SUSY particle apart from the neutralino LSP, and only one decay channel of the
NLSP with the branching ratio set to 100\% is considered. In a more general
case, however, the presence of other light particles in the spectrum may alter
the decay chain, and the assumption concerning the branching ratio may not
hold either. In such a case, the exclusion limits for the SMS should be
treated with care, and the actual limits are expected to be weaker.

\begin{figure}[t]
\centering
\subfloat[]{
\includegraphics[width=0.41\textwidth]{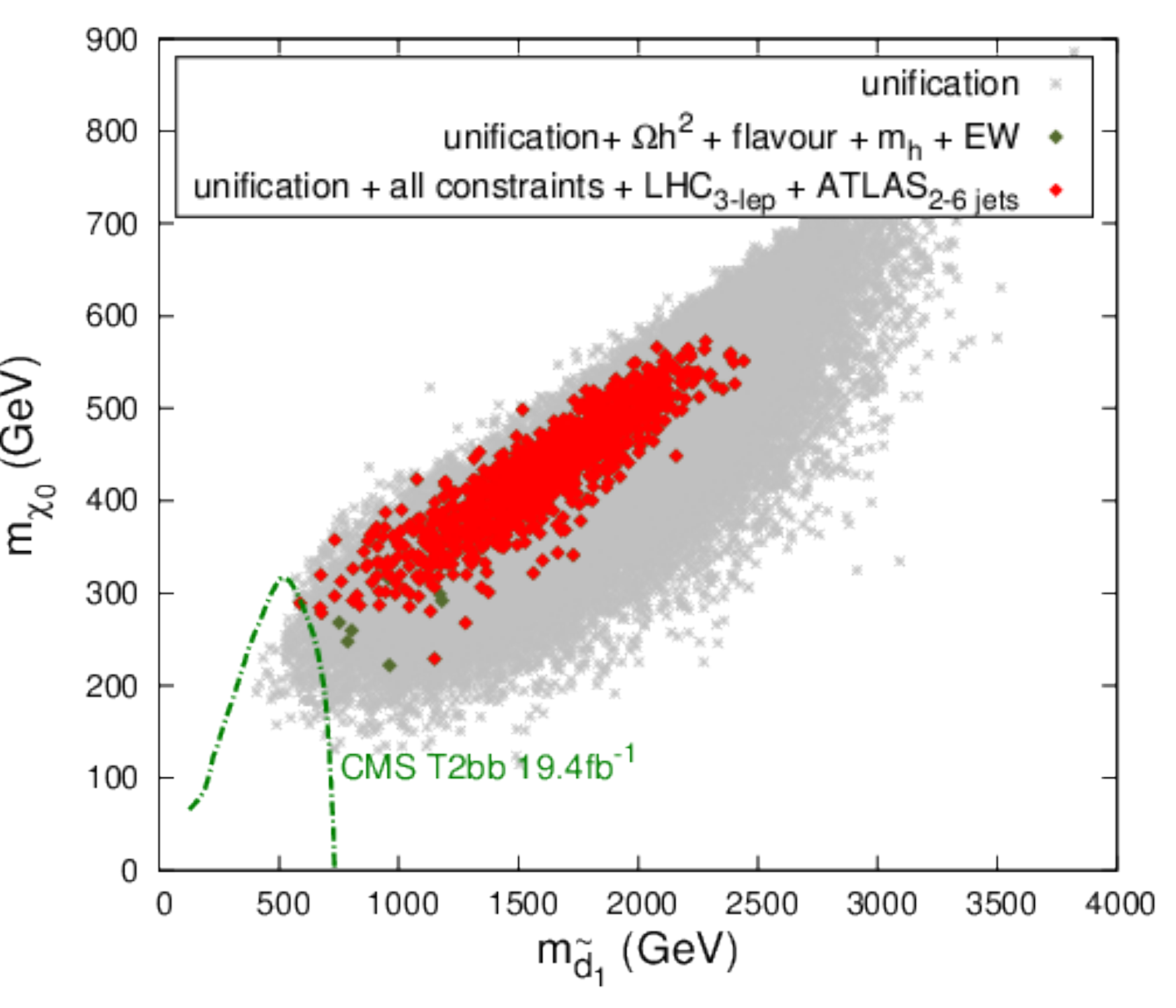}}
\subfloat[]{
\includegraphics[width=0.41\textwidth]{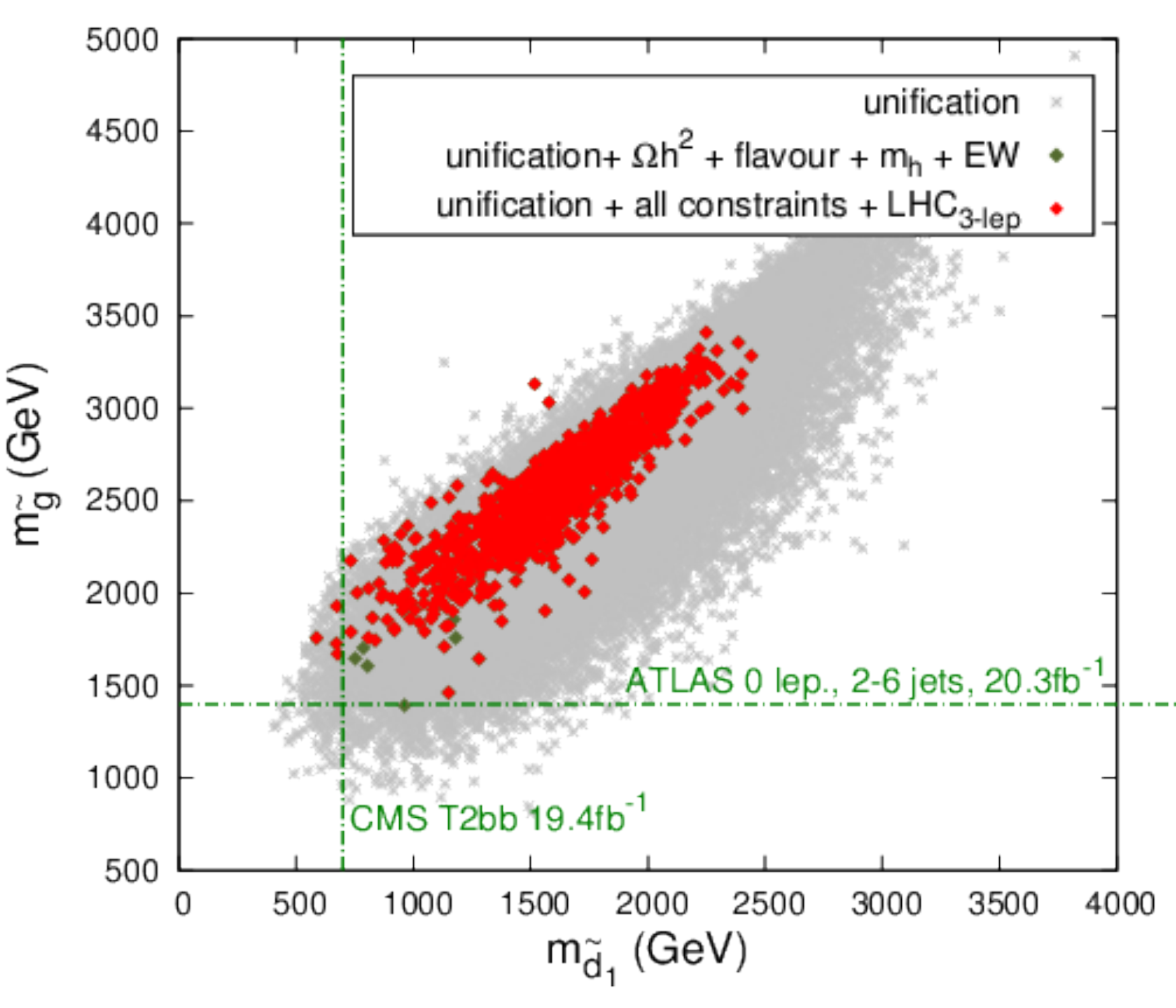}}
\caption{Scatter plot of the $GFV_{23}$ points in the planes
($m_{\tilde{d}_1}$, $m_{\neutone}$) (a), and ($m_{\tilde{d}_1}$,
$m_{\gluino}$) (b). Gray stars: all the points that allow for the Yukawa coupling
unification; dark green dots: points with good Yukawa coupling unification and
all the constraints listed in Table~\ref{tab:exp_constraints} (except the LHC)
satisfied at $3\sigma$; red diamonds: points additionally surviving the
CMS 3-lepton search at $3\sigma$. Dashed lines correspond to 95\% \cl\ limits
provided by other LHC SUSY searches discussed in the text.}
\label{lhc}
\end{figure}

In our analysis, we have used the experimental exclusion bound for the gluino
mass\cite{Aad:2014wea} at face value because this search is inclusive and
therefore tests any gluino-driven multijet signature, irrespectively of a
particular decay chain. We have also decided to use a direct 95\% \cl\ limit
from the CMS sbottom production search\cite{CMS-PAS-SUS-13-018}. For the SMS
T2bb, it reads $m_{\tilde{b}_1} \gtrsim 700\gev$ for $m_{\neutone}\simeq
150\gev$, and $m_{\tilde{b}_1} \gtrsim 640\gev$ for $m_{\neutone}\simeq
250\gev$.  In our scenario, the sbottom decay corresponds exactly to the SMS
T2bb, i.e. $\textrm{BR}(\tilde{b}\to b\neutone)=100\%$.  We neglect here a
possibility that the actual limit can be weakened due to significant mixing
between the sbottoms and other down-type squarks, and we leave a detailed
analysis of GFV effects in such a case for future studies. We will see,
however, that this simplifying assumption is justified by the fact that the
limits derived from Ref.\cite{CMS-PAS-SUS-13-018} are not the dominant ones.

On the other hand, interpretation of the CMS 3-lepton search strongly depends
on hierarchy in the considered spectrum, as well as on actual branching ratios
for neutralino and chargino decays. Therefore, in order to correctly quantify
the effect of this search in the $GFV_{23}$ scenario, we perform a
full reinterpretation of the experimental analysis using the tools developed
first in Ref.\cite{Fowlie:2012im}, and modified to recast the limits from SMS
in Ref.\cite{Kowalska:2013ica}. For the purpose of the present study, we have
updated the previously implemented\cite{CMS-PAS-SUS-12-022} CMS 3-lepton
search to include the full set of data with integrated luminosity of
$19.5\invfb$\cite{Aad:2014vma}.

In \reffig{lhc}, we present a distribution of the model points in the
($m_{\tilde{d}_1}$, $m_{\neutone}$) plane (a), and in the ($m_{\tilde{d}_1}$,
$m_{\gluino}$) plane (b). All the points for which the Yukawa coupling
unification is possible are shown as gray stars, and those that additionally
satisfy at $3\sigma$ the experimental constraints listed in
Table~\ref{tab:exp_constraints} (except the LHC) as dark green dots. Finally,
red diamonds depict the points that survive (at $3\sigma$) the CMS 3-lepton
plus MET search. Dashed lines correspond to the 95\% \cl\ exclusion bounds
from the CMS and ATLAS multijet searches described above, and should be
interpreted as lower bounds on sbottom and gluino masses.

One can see that already at the LHC \eight, the 3-lepton search provides a
stronger constraint on the Yukawa unification scenario than the limits from
Refs.\cite{CMS-PAS-SUS-13-018} and\cite{Aad:2014wea}, although even in this
case only a small part of the parameter space is tested.  The efficiency of
the search is weakened with respect to the corresponding SMS, for which the
interpretations are provided in the experimental analysis\cite{Aad:2014vma},
since only one generation of light sleptons is present.

On the other hand, $GFV_{23}$ unification scenario has a chance to be fully
tested during the second run of the LHC.  Gluino masses up to $2.5\tev$ will
be probed at the LHC\four\ with the luminosity of
$300\invfb$\cite{ATL-PHYS-PUB-2014-010} as long as neutralino LSP is lighter
than $700\gev$. The limits on the sbottom mass will be improved up to
$1400\gev$ for neutralinos lighter than $600\gev$. Those limits can be further
extended up to $3000\gev$ and $1600\gev$, respectively, for the final LHC
luminosity of $3000\invfb$.
      
\subsubsection{Electroweak symmetry breaking}\label{OffDiagEWSBsec}

\begin{figure}[t]
\centering
\subfloat[]{
\includegraphics[width=0.48\textwidth]{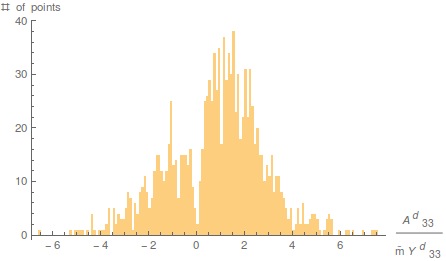}}
\subfloat[]{
\includegraphics[width=0.48\textwidth]{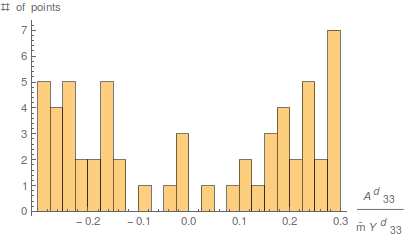}}
\caption{Histograms of points gathered by the $GFV_{23}$ scan 
as a function of $A^d_{33}/(Y^d_{33} \widetilde{m})$
at the scale $M_{SUSY}$: (a) all the points gathered by the scan; (b) enlarged vicinity of zero.}
\label{fig:EWSB_GFV_23}
\end{figure}     

As mentioned in \refsec{gfv23}, the $GFV_{23}$ unification scenario allows
large values of $A^{de}_{33} / M_{1/2}$ at \mgut. This may lead to a global
charge/colour breaking (CCB) minimum of the MSSM scalar potential. Indeed,
large values of $A^{d}_{33} (M_{SUSY})$ make it easier to tune the Yukawa
threshold corrections to the size required by bottom-tau and strange-muon
Yukawa coupling unification, and most of our points that satisfy
this unification condition correspond to $\frac{A^d_{33}}{Y^d_{33}
\widetilde{m}}(M_{SUSY}) \sim O(1)$, where
$\widetilde{m}=\sqrt{(m_{\tilde{d}_{3}}^2+m_{\tilde{e}_3}^2+m_{H_d}^2)/3}$. For
some points however, this ratio is close to zero, as can be seen in the panel
(a) of Fig.~\ref{fig:EWSB_GFV_23}.  Therefore, the $GFV_{23}$ scenario does
not necessarily lead to a metastable vacuum, as the relevant factor
$\frac{A^d_{33}}{Y^d_{33} \widetilde{m}}(M_{SUSY})$ can always be fitted to
satisfy the coarse bound $\frac{A_{ii}}{Y_{ii} \widetilde{m}_i} < 1$.
     
\subsection{$\boldsymbol{GFV_{123}}$: unification of all three families}

\subsubsection{Lepton Flavour Violating observables}

\begin{figure}[t]
\centering
\subfloat[]{
\includegraphics[width=0.41\textwidth]{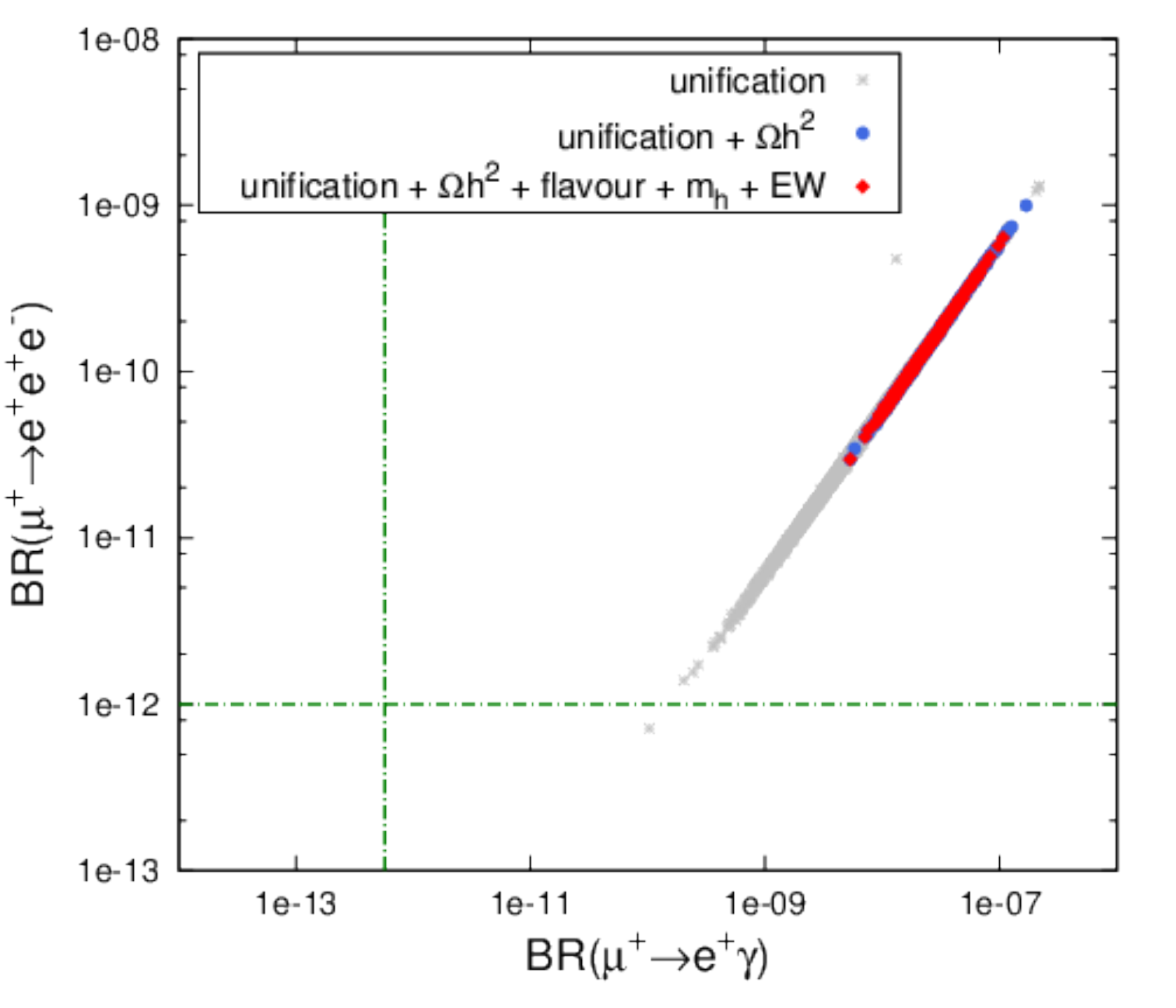}}
\subfloat[]{
\includegraphics[width=0.41\textwidth]{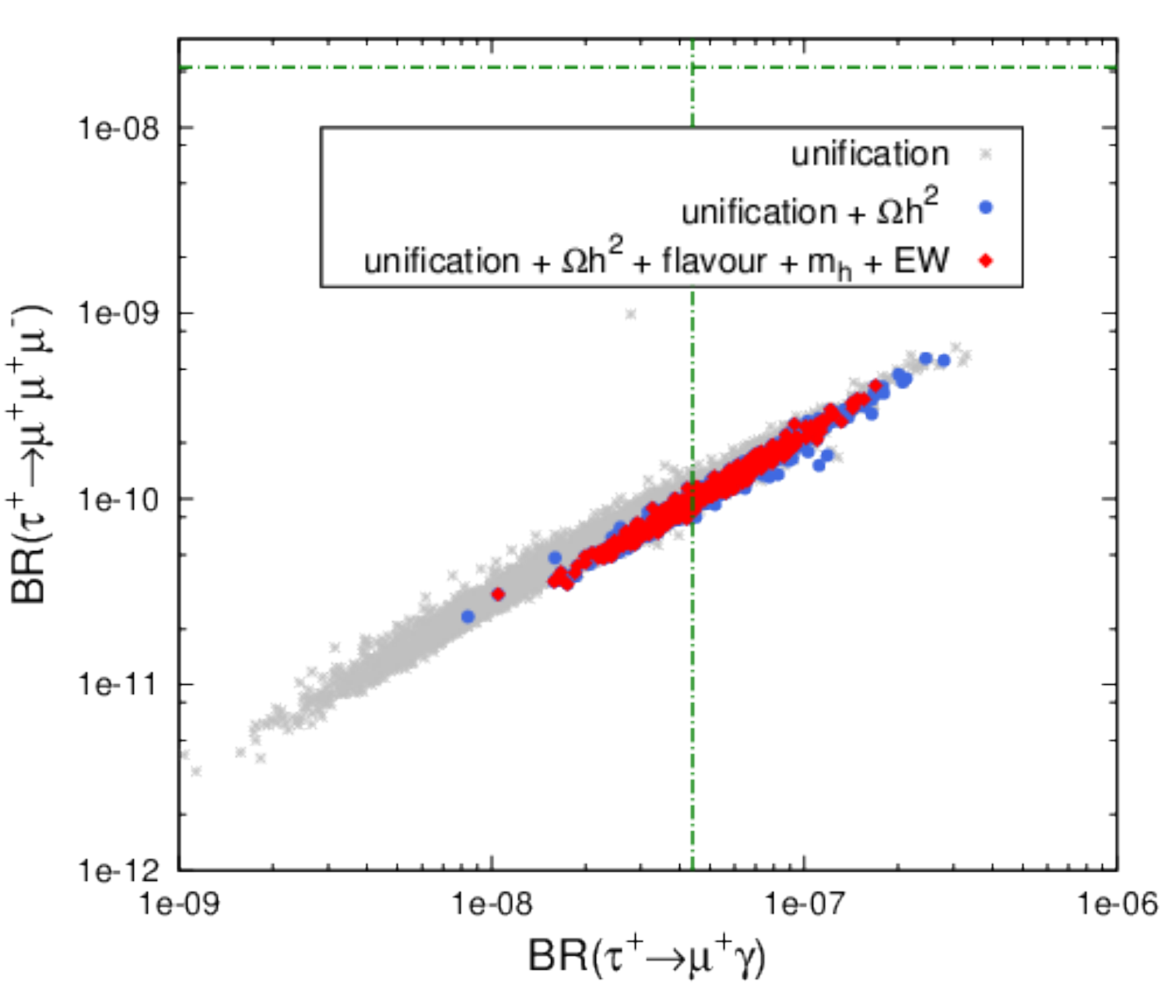}}
\caption{Scatter plot of the $GFV_{123}$ points in the planes
(\mueg, \mue) (a), and (\taueg, \taue) (b). The colour code is the same
as in \reffig{flavour}.}
\label{lfv:123}
\end{figure}

In the $GFV_{123}$ scenario, the muon-electron conversion observables might be
strongly enhanced, as they are influenced by large non-zero values of the
parameters $m^{dl}_{12}$, $A^{de}_{12}$ and $A^{de}_{21}$, required by the
Yukawa unification of the first family. In \reffig{lfv:123}, we present 
distributions of model points in the planes (\mueg, \mue) (a), and (\taueg,
\taue) (b). The colour code is the same as in \reffig{flavour}. It is clear
that the 90\% \cl\ upper bound on \mueg{} reported in Ref.\cite{Adam:2013mnn}
is violated in the $GFV_{123}$ scenario by about five orders of magnitude.

On the other hand, it is theoretically possible to evade this constraint while
unifying the electron and down quark Yukawa couplings by raising
the overall scale of the superpartner masses. However, such a scenario is
difficult to test with our present numerical tools which assume
$\mu_{\rm sp}=M_Z$ and henceforth are not reliable when \msusy\ is very
large.

\subsubsection{EW vacuum stability}\label{vacSect}

There are also important non-decoupling effects of phenomenological importance
that characterise the $GFV_{123}$ scenario, namely the vacuum
(meta)stability problem. As discussed in \refsec{sec:impact}, non-zero
elements $A^{de}_{12/21}$ are required to achieve the Yukawa coupling
unification for both the first and second families. However, off-diagonal
entries of the trilinear couplings (as well as the diagonal ones) are strongly
constrained by the requirement of EW vacuum stability. When the
flavour-violating entries are too large, a CCB minimum may appear in the MSSM
scalar potential, and it may become deeper than the standard EW one. The
potential may also become unbounded from below
(UFB)\cite{Frere:1983ag,AlvarezGaume:1983gj,Derendinger:1983bz,Kounnas:1983td,Casas:1995pd,
Casas:1996de}. An important feature of all such constraints is that, unlike
the FCNC ones, they do not become weaker when the scale \msusy\ is increased.

\begin{figure}[t]
\centering
\subfloat[]{
\includegraphics[width=0.41\textwidth]{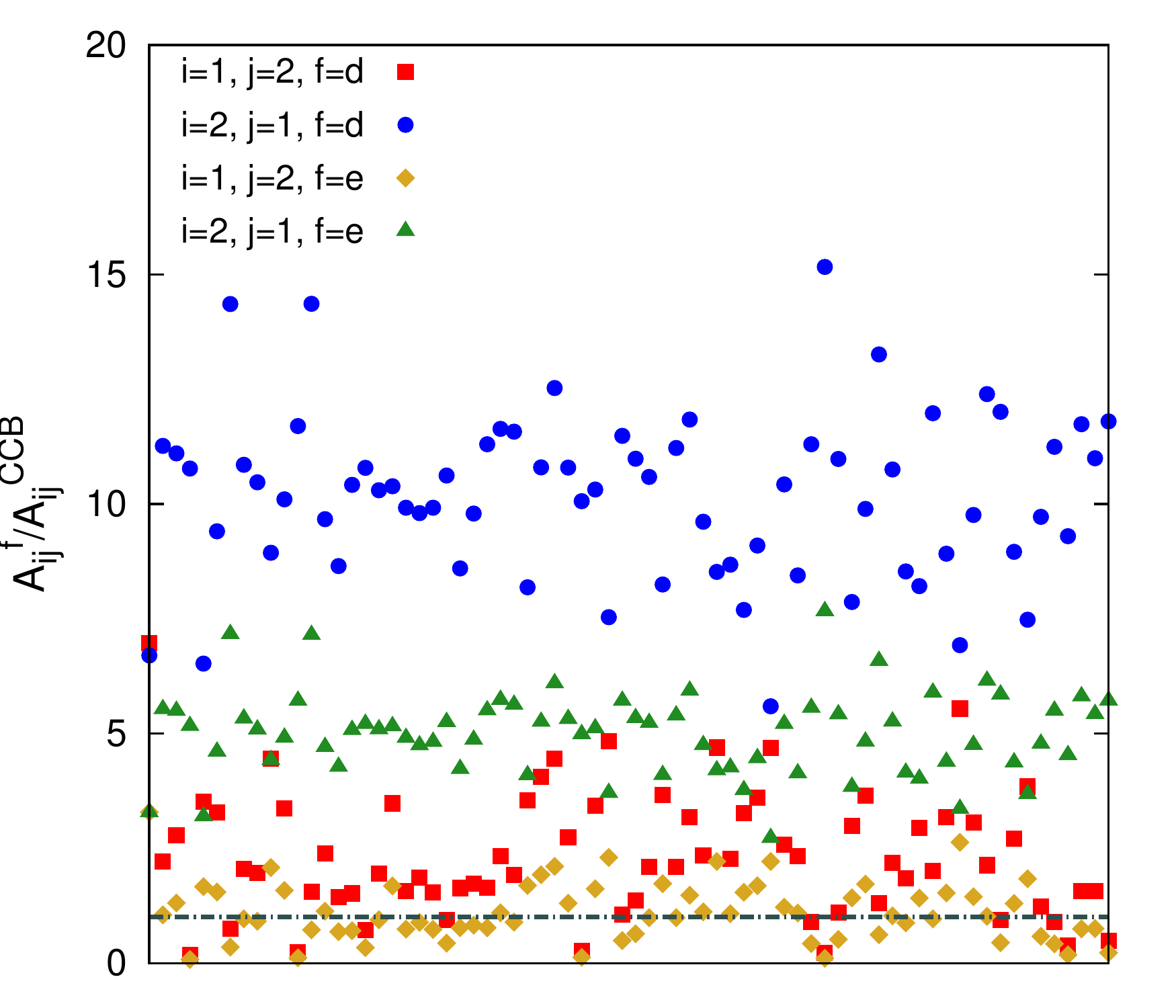}}
\subfloat[]{
\includegraphics[width=0.41\textwidth]{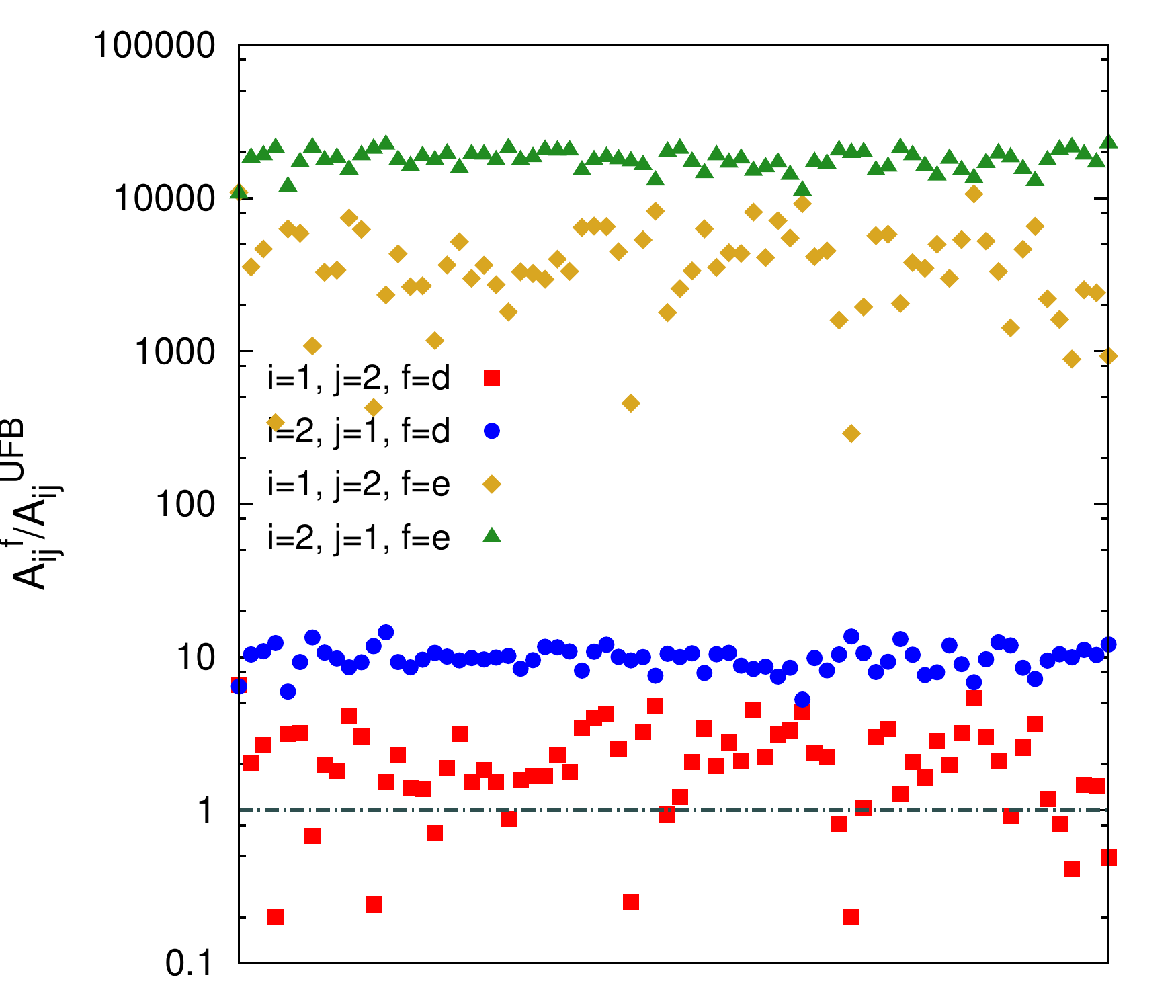}}
\caption{EW vacuum CCB (a) and UFB (b) stability bounds on the elements
$A^{d}_{12/21}$ and $A^{e}_{12/21}$. Dashed line indicates the upper limit on
the allowed size of the off-diagonal trilinear terms.}
\label{vacuum}
\end{figure}

In the down-squark sector, tree-level formulae for the CCB bounds are given
by\cite{Casas:1996de}
\bea\label{ccb}
(v_d/\sqrt{2})A^{d}_{ij}&\leq& m^d_{k}[(m^2_{\tilde{q}})_{ii}+(m^2_{\tilde{d}})_{jj}+\mhd^2+\mu^2]^{1/2},\quad k=\textrm{Max}(i,j).
\eea
The limits on $A^{e}_{ij}$ have an analogous form, up to replacing the matrix
$m^2_{\tilde{d}}$ by $m^2_{\tilde{e}}$. Similarly, the UFB bounds
read\cite{Casas:1996de}
\bea\label{ufb}
(v_d/\sqrt{2})A^{d}_{ij}&\leq& m^d_{k}[(m^2_{\tilde{q}})_{ii}+(m^2_{\tilde{d}})_{jj}+(\mll^2)_{ii}+(m^2_{\tilde{e}})_{jj}]^{1/2},\nonumber\\
(v_d/\sqrt{2})A^{e}_{ij}&\leq& \sqrt{3}m^l_{k},\quad k=\textrm{Max}(i,j).
\eea
In \reffig{vacuum}, we show to what extent the CCB and UFB limits are
satisfied for the points that allow the Yukawa coupling unification and
satisfy at $3\sigma$ all the experimental constraints listed in
Table~\ref{tab:exp_constraints}. Dashed line indicates the upper limit on the
allowed size of the off-diagonal trilinear terms. One can see that the CCB
stability bounds are violated by around an order of magnitude. The situation
is even worse in the case of the UFB bounds where the size of the elements
$A^e_{ij}$ are around four orders of magnitude larger than it is allowed by
the stability constraint. It results from the fact that the UFB limit on
$A^e_{ij}$ is of the order of the muon mass. Therefore, we conclude that the
EW MSSM vacuum is not stable in the Yukawa unification scenario considered in
our study.

On the other hand, even if a CCB minimum appears, it does not imply that a
considered model is not valid, as long as the standard EW vacuum lives longer
than the age of the universe. Moreover, in such a case, the UFB bound becomes
irrelevant because the probability of a tunnelling process along the CCB
direction is much higher\cite{Park:2010wf}. To derive metastability bounds,
the bounce action for a given scalar potential should be calculated
numerically, which is beyond the scope of our paper. To evaluate the impact of
metastability on the validity of the unification scenario, we will use instead
the results of the analysis performed in Ref.\cite{Park:2010wf}. The derived
metastability bounds do not depend on the Yukawa couplings, and therefore are
in general much less stringent than the CCB ones. The effect is the strongest
for the $A_{12/21}$ elements, in which case the CCB limit can be weakened by
three to four orders of magnitude, depending on a particular choice of the
model parameters. We can therefore conclude that the scenario considered in
our study leads to an unstable, but a long-lived vacuum.

It should be stressed, though, that the tension between the Yukawa unification
condition and the EW vacuum stability is less severe in our present scenario
than in the case of Yukawa unification through large diagonal $A$-terms where
the CCB bounds were violated by two orders of
magnitude\cite{Iskrzynski:2014zla}. On the other hand, theoretical
calculations of the stability conditions are still marred with many
uncertainties. This leaves a possibility that future improvements might
further reduce (or even eliminate) the tension between the vacuum stability
and the Yukawa unification.

\section{Conclusions}\label{concl}

In this study, we provided evidence that $SU(5)$ boundary conditions for the
Yukawa matrices at the GUT-scale can be satisfied within the renormalizable
$R$-parity-conserving MSSM if a more general flavour structure of the soft
SUSY-breaking sector is allowed. In particular, we found that a non-zero
$(2,3)$ element of the down-type squark mass matrix helps to achieve an
approximate Yukawa unification for the second family, as the SUSY threshold
correction driven by this entry is proportional to the combination
$(A^d_{33}-Y_b\mu\tan\beta$) that involves large third-generation
couplings. Moreover, simultaneous unification of the Yukawa couplings for
the second and first family additionally requires the elements
$A_{12/21}^{de}$, $m^{dl}_{13}$ and $m^{dl}_{12}$ to assume non-zero values.
The latter, however, leads to unacceptably large SUSY corrections to some
Lepton Flavour Violating processes, in particular to $\mu\to e\gamma$ decay.

On the other hand, the scenario with Yukawa unification for the second and
third family only, is consistent with a wide set of experimental measurements,
including those from the FCNC processes, which are often raised as an argument
against the GFV MSSM. The consistency holds even when theoretical
uncertainties in determination of the flavour observables are reduced. We
showed that a non-trivial flavour structure of the down-squark soft
terms does not pose a threat for the quark-sector flavour
observables for two reasons. First, the value of \tanb\ required by both the
Yukawa unification condition and the DM relic density measurement must stay in
the range of $\tanb\in(10-30)$, so no large $\tan\beta$-enhancement occurs in
supersymmetric loop corrections to rare $B$-meson decays. Secondly, squarks
remain relatively heavy.  This may suggest that the principle of Minimal
Flavour Violation as a unique way to remain in agreement with the flavour
observables is not always well motivated.

Another interesting feature of the $GFV_{23}$ unification scenario are
the properties of dark matter. It turns out that the points with the Yukawa
matrix unification are characterised by a neutralino LSP which is almost
purely bino-like, and with the mass in the range of $200-600\gev$. This, in
turn, enforces a particular hierarchy of masses in the corresponding SUSY
spectrum, with sneutrino NLSP and one light charged slepton. Spectra of this
kind, that lead to a characteristic 3-lepton collider signature, have started
to be tested in the direct SUSY searches at the LHC \eight. In the upcoming
LHC Run II, they are going to be tested in a complete manner.

Let us conclude with one more remark. The aim of our study was to check
whether large GUT-scale threshold corrections to the Yukawa couplings or
modifications of the Yukawa boundary conditions can be avoided in the
$R$-parity conserving MSSM. We did not try, however, to construct a full and
self-consistent model valid above the GUT-scale. For this reason some issues
related to the minimal $SU(5)$ GUT, like the proton lifetime, remained
unaddressed. On the other hand, ways of avoiding too fast proton decay in the
framework of minimal $SU(5)$ have been proposed in the literature, for example
by employing higher-dimensional operators\cite{EmmanuelCosta:2003pu}. It
seems, therefore, feasible to combine two complementary approaches to the
Yukawa coupling unification, the one using flavour-violating low-scale
threshold corrections, and the one introducing higher-dimensional operators,
to construct a correct $SU(5)$ UV-completion of the MSSM.

\bigskip\bigskip
\begin{center}
\textbf{ACKNOWLEDGMENTS}
\end{center}

We would like to thank Werner Porod for great help with issues related to
\texttt{\spheno}. We also would like to thank Andreas Crivellin, Christophe
Grojean, Gian Giudice, Mikołaj Misiak, Ulrich Nierste and Enrico Maria Sessolo
for many useful comments and discussions. M.I. was supported in part by the
Foundation for Polish Science International PhD Projects Programme co-financed
by the EU European Regional Development Fund, by the Karlsruhe Institute of
Technology, and by the National Science Centre (Poland) research project,
decision DEC-2014/13/B/ST2/03969. K.K. was supported by the EU and MSHE Grant
No. POIG.02.03.00-00-013/09. The use of the CIS computer cluster at the
National Centre for Nuclear Research is gratefully acknowledged.


\bigskip

\begin{thebibliography}{99}

\bibitem{Dimopoulos:1981zb}
  S.~Dimopoulos and H.~Georgi,
  Nucl.\ Phys.\ B {\bf 193} (1981) 150.

\bibitem{Georgi:1979df}
  H.~Georgi and C.~Jarlskog,
  Phys.\ Lett.\ B {\bf 86} (1979) 297.

\bibitem{EmmanuelCosta:2003pu}
  D.~Emmanuel-Costa and S.~Wiesenfeldt,
  Nucl.\ Phys.\ B {\bf 661} (2003) 62
  [hep-ph/0302272].
  
\bibitem{Antusch:2009gu}
  S.~Antusch and M.~Spinrath,
  Phys.\ Rev.\ D {\bf 79} (2009) 095004
  [arXiv:0902.4644 [hep-ph]].

\bibitem{Antusch:2013rxa}
  S.~Antusch, S.~F.~King and M.~Spinrath,
  Phys.\ Rev.\ D {\bf 89} (2014) 055027
  [arXiv:1311.0877 [hep-ph]].
    
\bibitem{Buchmuller:1982ye}
  W.~Buchmuller and D.~Wyler,
  Phys.\ Lett.\ B {\bf 121} (1983) 321.

\bibitem{Hall:1985dx}
  L.~J.~Hall, V.~A.~Kostelecky and S.~Raby,
  Nucl.\ Phys.\ B {\bf 267} (1986) 415.

\bibitem{DiazCruz:2000mn}
  J.~L.~Diaz-Cruz, H.~Murayama and A.~Pierce,
  Phys.\ Rev.\ D {\bf 65} (2002) 075011
  [hep-ph/0012275].
  
\bibitem{Enkhbat:2009jt}
  T.Enkhbat,
  arXiv:0909.5597 [hep-ph].

\bibitem{Iskrzynski:2014zla}
  M.~Iskrzy\'nski,
  Eur.\ Phys.\ J.\ C {\bf 75} (2015) 2,  51
  [arXiv:1408.2165 [hep-ph]].
  
\bibitem{Crivellin:2011jt}
  A.~Crivellin, L.~Hofer and J.~Rosiek,
  JHEP {\bf 1107} (2011) 017
  [arXiv:1103.4272 [hep-ph]].

\bibitem{Crivellin:2008mq}
  A.~Crivellin and U.~Nierste,
  Phys.\ Rev.\ D {\bf 79} (2009) 035018
  [arXiv:0810.1613 [hep-ph]].

\bibitem{Crivellin:2010er}
  A.~Crivellin,
  Phys.\ Rev.\ D {\bf 83} (2011) 056001
  [arXiv:1012.4840 [hep-ph]].

\bibitem{Crivellin:2010gw}
  A.~Crivellin and J.~Girrbach,
  Phys.\ Rev.\ D {\bf 81} (2010) 076001
  [arXiv:1002.0227 [hep-ph]].
  
\bibitem{Guasch:1999jp}
  J.~Guasch and J.~Sola,
  Nucl.\ Phys.\ B {\bf 562} (1999) 3
  [hep-ph/9906268].

\bibitem{Cao:2007dk}
  J.~J.~Cao, G.~Eilam, M.~Frank, K.~Hikasa, G.~L.~Liu, I.~Turan and J.~M.~Yang,
  Phys.\ Rev.\ D {\bf 75} (2007) 075021
  [hep-ph/0702264].

\bibitem{Fichet:2014vha}
  S.~Fichet, B.~Herrmann and Y.~Stoll,
  Phys.\ Lett.\ B {\bf 742} (2015) 69
  [arXiv:1403.3397 [hep-ph]].
  
\bibitem{Herrmann:2011xe}
  B.~Herrmann, M.~Klasen and Q.~Le Boulc'h,
  Phys.\ Rev.\ D {\bf 84} (2011) 095007
  [arXiv:1106.6229 [hep-ph]].

\bibitem{Heinemeyer:2004by}
  S.~Heinemeyer, W.~Hollik, F.~Merz and S.~Penaranda,
  Eur.\ Phys.\ J.\ C {\bf 37} (2004) 481
  [hep-ph/0403228].

\bibitem{Cao:2006xb}
  J.~Cao, G.~Eilam, K.~i.~Hikasa and J.~M.~Yang,
  Phys.\ Rev.\ D {\bf 74} (2006) 031701
  [hep-ph/0604163].

\bibitem{AranaCatania:2011ak}
  M.~Arana-Catania, S.~Heinemeyer, M.~J.~Herrero and S.~Penaranda,
  JHEP {\bf 1205} (2012) 015
  [arXiv:1109.6232 [hep-ph]].

\bibitem{Arana-Catania:2014ooa}
  M.~Arana-Catania, S.~Heinemeyer and M.~J.~Herrero,
  Phys.\ Rev.\ D {\bf 90} (2014) 075003
  [arXiv:1405.6960 [hep-ph]]. 

\bibitem{Kowalska:2014opa}
  K.~Kowalska,
  JHEP {\bf 1409} (2014) 139
  [arXiv:1406.0710 [hep-ph]].
  
  \bibitem{Arana-Catania:2013nha}
  M.~Arana-Catania, S.~Heinemeyer and M.~J.~Herrero,
  Phys.\ Rev.\ D {\bf 88} (2013) 1,  015026
  [arXiv:1304.2783 [hep-ph]].

\bibitem{Fowlie:2012im}
  A.~Fowlie, M.~Kazana, K.~Kowalska, S.~Munir, L.~Roszkowski, E.~M.~Sessolo, S.~Trojanowski and Y.~L.~S.~Tsai,
  Phys.\ Rev.\ D {\bf 86} (2012) 075010
  [arXiv:1206.0264 [hep-ph

\bibitem{Feroz:2008xx}
  F.~Feroz, M.~P.~Hobson and M.~Bridges,
  Mon.\ Not.\ Roy.\ Astron.\ Soc.\  {\bf 398} (2009) 1601
  [arXiv:0809.3437 [astro-ph]].

 \bibitem{Porod:2011nf}
  W.~Porod and F.~Staub,
  Comput.\ Phys.\ Commun.\  {\bf 183} (2012) 2458
  [arXiv:1104.1573 [hep-ph]].
    
\bibitem{Agashe:2014kda}
  K.~A.~Olive {\it et al.}  [Particle Data Group Collaboration],
  Chin.\ Phys.\ C {\bf 38} (2014) 090001.

\bibitem{utfit}
    http://www.utfit.org/UTfit/ResultsSummer2014PostMoriondNP

\bibitem{Crivellin:2012jv}
  A.~Crivellin, J.~Rosiek, P.~H.~Chankowski, A.~Dedes, S.~Jaeger and P.~Tanedo,
  Comput.\ Phys.\ Commun.\  {\bf 184} (2013) 1004
  [arXiv:1203.5023 [hep-ph]].

\bibitem{Aoki:2013ldr}
  S.~Aoki, Y.~Aoki, C.~Bernard, T.~Blum, G.~Colangelo, M.~Della Morte, S.~Dürr and A.~X.~El Khadra {\it et al.},
  Eur.\ Phys.\ J.\ C {\bf 74} (2014) 9,  2890
  [arXiv:1310.8555 [hep-lat]].

\bibitem{Amhis:2014hma} 
  Y.~Amhis {\it et al.}  [Heavy Flavor Averaging Group Collaboration],
  arXiv:1412.7515 [hep-ex].

\bibitem{Misiak:2006zs}
  M.~Misiak, H.~M.~Asatrian, K.~Bieri, M.~Czakon, A.~Czarnecki, T.~Ewerth, A.~Ferroglia, P.~Gambino {\it et al.},
  Phys.\ Rev.\ Lett.\  {\bf 98} (2007) 022002
  [hep-ph/0609232].

\bibitem{Bobeth:2013uxa}
  C.~Bobeth, M.~Gorbahn, T.~Hermann, M.~Misiak, E.~Stamou and M.~Steinhauser,
  Phys.\ Rev.\ Lett.\  {\bf 112} (2014) 101801
  [arXiv:1311.0903 [hep-ph]].

\bibitem{Brod:2011ty}
  J.~Brod and M.~Gorbahn,
  Phys.\ Rev.\ Lett.\  {\bf 108} (2012) 121801
  [arXiv:1108.2036 [hep-ph]].
        
\bibitem{Gondolo:2004sc}
  P.~Gondolo, J.~Edsjo, P.~Ullio, L.~Bergstrom, M.~Schelke and E.~A.~Baltz,
  JCAP {\bf 0407} (2004) 008
  [astro-ph/0406204].    
  
\bibitem{Hahn:2013ria}
  T.~Hahn, S.~Heinemeyer, W.~Hollik, H.~Rzehak and G.~Weiglein,
  Phys.\ Rev.\ Lett.\  {\bf 112} (2014) 141801
  [arXiv:1312.4937 [hep-ph]].
  
\bibitem{Frank:2006yh}
  M.~Frank, T.~Hahn, S.~Heinemeyer, W.~Hollik, H.~Rzehak and G.~Weiglein,
  JHEP {\bf 0702} (2007) 047
  [hep-ph/0611326].
  
\bibitem{Degrassi:2002fi}
  G.~Degrassi, S.~Heinemeyer, W.~Hollik, P.~Slavich and G.~Weiglein,
  Eur.\ Phys.\ J.\ C {\bf 28} (2003) 133
  [hep-ph/0212020].
  
\bibitem{Heinemeyer:1998yj}
  S.~Heinemeyer, W.~Hollik and G.~Weiglein,
  Comput.\ Phys.\ Commun.\  {\bf 124} (2000) 76
  [hep-ph/9812320].        

\bibitem{Bechtle:2008jh}
  P.~Bechtle, O.~Brein, S.~Heinemeyer, G.~Weiglein and K.~E.~Williams,
  Comput.\ Phys.\ Commun.\  {\bf 181} (2010) 138
  [arXiv:0811.4169 [hep-ph]].
  
\bibitem{Bechtle:2011sb}
  P.~Bechtle, O.~Brein, S.~Heinemeyer, G.~Weiglein and K.~E.~Williams,
  Comput.\ Phys.\ Commun.\  {\bf 182} (2011) 2605
  [arXiv:1102.1898 [hep-ph]].
  
\bibitem{Bechtle:2013wla}
  P.~Bechtle, O.~Brein, S.~Heinemeyer, O.~St{\aa}l, T.~Stefaniak, G.~Weiglein and K.~E.~Williams,
  Eur.\ Phys.\ J.\ C {\bf 74} (2014) 2693
  [arXiv:1311.0055 [hep-ph]].  
  
 \bibitem{Bechtle:2013xfa}
  P.~Bechtle, S.~Heinemeyer, O.~St{\aa}l, T.~Stefaniak and G.~Weiglein,
  Eur.\ Phys.\ J.\ C {\bf 74} (2014) 2711
  [arXiv:1305.1933 [hep-ph]]. 

\bibitem{Akerib:2013tjd}
  D.~S.~Akerib {\it et al.}  [LUX Collaboration],
  Phys.\ Rev.\ Lett.\  {\bf 112} (2014) 9,  091303
  [arXiv:1310.8214 [astro-ph.CO]].

\bibitem{Cheung:2012xb}
  K.~Cheung, Y.~L.~S.~Tsai, P.~Y.~Tseng, T.~C.~Yuan and A.~Zee,
  JCAP {\bf 1210} (2012) 042
  [arXiv:1207.4930 [hep-ph]].
 
\bibitem{Kowalska:2014hza}
  K.~Kowalska, L.~Roszkowski, E.~M.~Sessolo and S.~Trojanowski,
  JHEP {\bf 1404} (2014) 166
  [arXiv:1402.1328 [hep-ph]].

\bibitem{Ade:2013zuv}
  P.~A.~R.~Ade {\it et al.}  [Planck Collaboration],
  Astron.\ Astrophys.\  (2014)
  [arXiv:1303.5076 [astro-ph.CO]].
    
\bibitem{CMS:yva}
  [CMS Collaboration],
  ``Combination of standard model Higgs boson searches and measurements of the
  properties of the new boson with a mass near 125 GeV,''
  CMS-PAS-HIG-13-005.   

\bibitem{CMS:2014xfa}
  V.~Khachatryan {\it et al.} [CMS and LHCb Collaborations],
  Nature {\bf 522} (2015) 68
  [arXiv:1411.4413 [hep-ex]].
     
\bibitem{Baker:2006ts}
  C.~A.~Baker, D.~D.~Doyle, P.~Geltenbort, K.~Green, M.~G.~D.~van der Grinten, P.~G.~Harris, P.~Iaydjiev and S.~N.~Ivanov {\it et al.},
  Phys.\ Rev.\ Lett.\  {\bf 97} (2006) 131801
  [hep-ex/0602020].

\bibitem{CMS-PAS-SUS-13-018}
  CMS Collaboration
  "Search for direct production of bottom squark pairs",
  CMS-PAS-SUS-13-018".
  
\bibitem{Aad:2014wea}
  G.~Aad {\it et al.}  [ATLAS Collaboration],
  JHEP {\bf 1409} (2014) 176
  [arXiv:1405.7875 [hep-ex]].
  
\bibitem{Aad:2014vma}
  G.~Aad {\it et al.}  [ATLAS Collaboration],
  JHEP {\bf 1405} (2014) 071
  [arXiv:1403.5294 [hep-ex]].

\bibitem{Adam:2013mnn}
  J.~Adam {\it et al.}  [MEG Collaboration],
  Phys.\ Rev.\ Lett.\  {\bf 110} (2013) 201801
  [arXiv:1303.0754 [hep-ex]].

\bibitem{Aubert:2009ag}
  B.~Aubert {\it et al.}  [BaBar Collaboration],
  Phys.\ Rev.\ Lett.\  {\bf 104} (2010) 021802
  [arXiv:0908.2381 [hep-ex]]
  
\bibitem{Bellgardt:1987du}
  U.~Bellgardt {\it et al.}  [SINDRUM Collaboration],
  Nucl.\ Phys.\ B {\bf 299} (1988) 1.

\bibitem{Hayasaka:2010np}
  K.~Hayasaka, K.~Inami, Y.~Miyazaki, K.~Arinstein, V.~Aulchenko, T.~Aushev, A.~M.~Bakich and A.~Bay {\it et al.},
  Phys.\ Lett.\ B {\bf 687} (2010) 139
  [arXiv:1001.3221 [hep-ex]].
  
\bibitem{Aprile:2012zx} 
  E.~Aprile [XENON1T Collaboration],
  Springer Proc.\ Phys.\  {\bf 148}, 93 (2013)
  [arXiv:1206.6288 [astro-ph.IM]].
  
\bibitem{Kowalska:2013ica} 
  K.~Kowalska and E.~M.~Sessolo,
  Phys.\ Rev.\ D {\bf 88}, no. 7, 075001 (2013)
  [arXiv:1307.5790 [hep-ph]].
  
\bibitem{CMS-PAS-SUS-12-022}
  CMS Collaboration
  "Search for direct EWK production of SUSY particles in
                       multilepton modes with 8TeV data",
  CMS-PAS-SUS-12-022".

\bibitem{ATL-PHYS-PUB-2014-010}
  ATLAS Collaboration
  "Search for Supersymmetry at the high luminosity LHC with
                       the ATLAS experiment",
  ATL-PHYS-PUB-2014-010".

\bibitem{Frere:1983ag}
  J.~M.~Frere, D.~R.~T.~Jones and S.~Raby,
  Nucl.\ Phys.\ B {\bf 222} (1983) 11. 
 
\bibitem{AlvarezGaume:1983gj}
  L.~Alvarez-Gaume, J.~Polchinski and M.~B.~Wise,
  Nucl.\ Phys.\ B {\bf 221} (1983) 495. 

\bibitem{Derendinger:1983bz}
  J.~P.~Derendinger and C.~A.~Savoy,
  Nucl.\ Phys.\ B {\bf 237} (1984) 307.

\bibitem{Kounnas:1983td}
  C.~Kounnas, A.~B.~Lahanas, D.~V.~Nanopoulos and M.~Quiros,
  Nucl.\ Phys.\ B {\bf 236} (1984) 438

\bibitem{Casas:1995pd}
  J.~A.~Casas, A.~Lleyda and C.~Mu\~noz,
  Nucl.\ Phys.\ B {\bf 471} (1996) 3
  [hep-ph/9507294].
        
\bibitem{Casas:1996de}
  J.~A.~Casas and S.~Dimopoulos,
  Phys.\ Lett.\ B {\bf 387} (1996) 107
  [hep-ph/9606237].  

\bibitem{Park:2010wf}
  J.~h.~Park,
  Phys.\ Rev.\ D {\bf 83} (2011) 055015
  [arXiv:1011.4939 [hep-ph]].

  
\end{thebibliography}
\end{document}